\documentclass[twocolumn,times,tighten,twocolappendix]{aastex631}

\usepackage{times, apjfonts}
\usepackage{amsmath, amsthm, amssymb}
\usepackage{graphicx, epsfig, xcolor}
\usepackage{natbib}
\usepackage{verbatim}
\usepackage{epstopdf} 
\usepackage{CJK} 
\hypersetup{linkcolor=red,citecolor=blue,filecolor=red,urlcolor=blue}
\usepackage{datetime2}
\usepackage{float}

\shorttitle{iron emission template}
\shortauthors{Park et al.}

\journalinfo{accepted for publication in The Astrophysical Journal Supplement Series}

\newcommand{\mbh}{$M_{\rm BH}$}

\newcommand{\msun}{$M_{\odot}$}
\newcommand{\kms}{km~s$^{\rm -1}$}
\newcommand{\ergs}{erg~s$^{\rm -1}$}

\newcommand{\FeII}{\ion{Fe}{2}}
\newcommand{\Hd}{H$\delta$}
\newcommand{\Hg}{H$\gamma$}
\newcommand{\Hb}{H$\beta$}

\newcommand{\CIV}{\ion{C}{4}}
\newcommand{\MgII}{\ion{Mg}{2}}

\begin{document}
\begin{CJK*}{UTF8}{mj}

\title{A New Iron Emission Template for Active Galactic Nuclei. I. Optical Template for the \Hb\ region
\footnote{Based on observations made with the NASA/ESA Hubble Space Telescope obtained from the Space Telescope Science Institute, which is operated by the Association of Universities for Research in Astronomy, Inc., under NASA contract NAS 5-26555. These observations are associated with program GO-14744.}}

\author[0000-0001-9877-1732]{Daeseong Park (박대성)}
\affil{Department of Astronomy and Atmospheric Sciences, Kyungpook National University, Daegu, 41566, Republic of Korea}
\affil{Korea Astronomy and Space Science Institute, Daejeon, 34055, Republic of Korea;
\href{mailto:daeseong.park@gmail.com}{daeseong.park@gmail.com}}

\author[0000-0002-3026-0562]{Aaron J. Barth}
\affil{Department of Physics and Astronomy, 4129 Frederick Reines Hall, University of California, Irvine, CA, 92697-4575, USA; \href{mailto:barth@uci.edu}{barth@uci.edu}}

\author[0000-0001-6947-5846]{Luis C. Ho}
\affil{Kavli Institute for Astronomy and Astrophysics, Peking University, Beijing 100871, China}
\affil{Department of Astronomy, School of Physics, Peking University, Beijing 100871, China; \href{mailto:lho.pku@gmail.com}{lho.pku@gmail.com}}

\author{Ari Laor}  
\affil{Physics Department, Technion, Haifa 32000, Israel; \href{mailto:laor@physics.technion.ac.il}{laor@physics.technion.ac.il}}

\begin{abstract}

We present a new empirical template for iron emission in active galactic nuclei (AGN) covering the $4000-5600$ \AA\ range.
The new template is based on a spectrum  of the narrow-line Seyfert 1 galaxy Mrk 493 obtained with the Hubble Space Telescope. In comparison with the canonical iron template object I~Zw~1, Mrk 493 has narrower broad-line widths, lower reddening, and a less extreme Eddington ratio, making it a superior choice for template construction. We carried out a multicomponent spectral decomposition to produce a template incorporating all permitted and forbidden lines of \ion{Fe}{2} identified in the Mrk 493 spectrum over this wavelength range, as well as lines from \ion{Ti}{2}, \ion{Ni}{2}, and \ion{Cr}{2}.
We tested the template by fitting it to AGN spectra spanning a broad range of iron emission properties, and we present a detailed comparison with fits using other widely used monolithic and multi-component iron emission templates. 
The new template generally provides the best fit (lowest $\chi^2$) compared to other widely used monolithic empirical templates. In addition, the new template yields more accurate spectral measurements including a significantly better match of the derived Balmer line profiles (\Hb, \Hg, \Hd), in contrast with results obtained using the other templates.
Our comparison tests show that the choice of iron template can introduce a systematic bias in measurements of the H$\beta$ line width,
which consequently impacts single-epoch black hole mass estimates by $\sim0.1$ dex on average and possibly up to $\sim0.3-0.5$ dex individually. 

\end{abstract}


\section{Introduction} \label{sec:intro}
\setcounter{footnote}{0}

Iron emission lines are among the most prominent features in the ultraviolet (UV) and optical spectra of many active galactic nuclei (AGN). \ion{Fe}{2} emission is a key coolant in the broad-line region (BLR), accounting for $\sim25\%$ of the total energy output \citep{Wills+1985} in some AGN. Empirical correlations between \ion{Fe}{2} emission and  fundamental AGN properties such as black hole (BH) mass and accretion rate found from the Eigenvector 1 (EV1) sequence \citep{BorosonGreen1992} indicate that \FeII\ emission is a crucial diagnostic for investigation of BLR physics and for understanding the underlying factors responsible for the diversity of AGN spectra. Accurate measurements of \ion{Fe}{2} emission properties, including line fluxes, broadening, and velocity shifts, are required for a broad variety of investigations of AGN physics and phenomenology.

One of the essential components in spectral decomposition analysis of AGN is an iron emission template. Iron emission at UV and optical wavelengths forms a pseudo-continuum consisting of tens of thousands of emission lines  resulting from the extremely complex atomic structure of the Fe$^+$ ion 
\citep[e.g., ][]{Wills+1985,SigutPradhan1998,Verner+1999,Baldwin+2004,Verner+2004}. Since individual iron lines in AGN spectra are so strongly blended together, \ion{Fe}{2} templates are used in analysis of AGN spectra to fit the iron blends and measure the overall properties of the iron emission itself, as well as to deblend the iron emission from other spectral lines and continuum components \citep[e.g.,][]{Marziani2003,GreeneHo2005,Shen+2008,Hu+2008,Ho+2012,Barth+2015,Park+2017}. 
Template fitting is the only practical way to measure the complex blends of numerous overlapping iron emission lines and separate them from other components
\citep[e.g.,][]{BorosonGreen1992,VestergaardWilkes2001}.

Use of \ion{Fe}{2} emission templates has numerous applications in AGN science including investigating broad-line and narrow-line region (BLR and NLR) physics, performing reverberation mapping to understand BLR structure and kinematics \citep{Peterson1993, Cackett+2021},  estimating BH masses from single-epoch spectroscopic measurements \citep{Shen2013}, and deriving metal abundances in AGN BLRs \cite[e.g.,][]{Dietrich+2002,Kurk+2007,DeRosa+2014}. The accuracy of available iron emission templates consequently impacts nearly every study of BH masses that relies on measurement of broad emission-line widths
\cite[e.g.,][]{McLureDunlop2004,Shen+2008,ShenLiu2012,TrakhtenbrotNetzer2012,ShenKelly2012,KellyShen2013,Park+2015,Mejia-Restrepo+2016,Ding+2017,Ding+2020,Schulze+2018}.

In particular, the spectral region of $4000-5600$ \AA,
where the strongest optical \FeII\ emission blends are located,
is widely used for both reverberation mapping of \Hb\ and for single-epoch BH mass determination. Spectral decomposition analysis on this region using iron templates provides significant benefits for AGN reverberation mapping studies for the \Hb\ line, and has recently been used to detect variability and measure reverberation lags of the optical \FeII\ lines themselves \citep[e.g.,][]{Barth+2011:mrk50,Barth+2013,Barth+2015,Park+2012,Pancoast+2014-II,Pancoast+2018,Hu+2015,Pei+2017,Grier+2017,Williams+2018,Li+2018,Rakshit+2019,Wang+2020,Lu+2019,Lu+2021}.

Given the complex energy level structure of the Fe$^+$ ion, the large number of electronic transitions responsible for the UV and optical emission, and the difficulties involved in theoretical modeling of \ion{Fe}{2} line formation and radiative transfer through the BLR, it is an immense challenge for purely theoretical \ion{Fe}{2} emission models to match observed AGN spectra in detail, and theoretically derived iron emission templates  \citep[e.g.,][]{SigutPradhan2003,BruhweilerVerner2008} are still not widely used in fitting observed AGN spectra. Instead, most of the widely adopted emission templates are empirically derived from observations of individual AGN, either by fitting identified iron features to generate a template as the best-fitting iron emission model, or by fitting and removing non-iron features from the data leaving the iron template as a residual spectrum.

\subsection{Currently Used Iron Templates}

The most widely used empirical iron templates have been derived from observations of the bright narrow-line Seyfert 1 (NLS1) galaxy I Zw 1 ($z = 0.060$), which is well known as a prototypical strong \FeII\ emitter \citep{Laor+1997}. With their broad emission-line widths of $<2000$ \kms, NLS1 nuclei having strong \ion{Fe}{2} emission are ideal targets for defining and testing iron emission templates because line blending is minimized, so that individual \FeII\ features can be clearly and reliably distinguished. 

In the optical, \citet{BorosonGreen1992} constructed an empirical template covering $3686-7484$ \AA\ that is perhaps the most widely used optical template. Their approach to template construction was to remove several non-iron emission lines and the underlying continuum from a ground-based spectrum of I Zw 1.
Another empirical optical template covering $3535-7534$ \AA\ was built by \citet{Veron-Cetty2004}\footnote{available at \url{http://cdsarc.u-strasbg.fr/viz-bin/qcat?J/A+A/417/515}}
who adopted a more comprehensive modeling method based on an extensive list of lines. To construct their template, they fit the profiles of all identified iron lines in a ground-based spectrum of I Zw 1, and used the sum of all modeled iron line profiles as the template.
Another approach was used by \citet{Dong+2008}, who constructed an alternate template using the line list from \citet{Veron-Cetty2004}, but allowing different normalizations for lines identified as belonging to the broad-line system and the low-excitation narrow-line system.
This approach has been shown to work well for some studies, e.g., \citet{Dong+2008,Dong+2010,Dong+2011}. Generating a template by combining lines from the \citet{Veron-Cetty2004} line lists with different relative strengths allows for considerably more flexibility in spectral fitting, at the expense of adding additional free parameters to the model. However, in this work, we use only the original, monolithic version of the \citet{Veron-Cetty2004} template (rather than modified versions) for comparison with our new template.

A different approach to template construction was demonstrated by \citet{Kovacevic+2010}\footnote{available at \url{http://servo.aob.rs/FeII_AGN/}\label{footnote:K10P19}} \citep[see also][]{Shapovalova+2012,Popovic+2013}, who
produced a multicomponent semi-empirical template covering $4000-5500$ \AA. Starting with a list of the 50 strongest \ion{Fe}{2} lines in this wavelength range, they divided these lines into four  groups based on atomic properties of the transitions, with an additional group of \FeII\ lines taken from I Zw 1 data of \citet{Veron-Cetty2004}. For most of the lines, the relative line intensities within each line group are determined based on atomic data and the assumed excitation temperature, while for some lines, the relative strengths were determined based on a fit to the spectrum of I Zw 1. The individual line profiles in their template are assumed to be Gaussian, with a uniform width for all lines. Their five-component template thus allows for different relative line intensities among these five groups. This approach provides more flexibility to fit diverse AGN spectra more accurately than can generally be done with monolithic empirical templates having a single overall intensity.

In the UV spectral range,
\citet{VestergaardWilkes2001} constructed empirical templates covering $1075-3090$ \AA\ for \ion{Fe}{2} and
\ion{Fe}{3} emission based on Hubble Space Telescope (HST) spectra of I Zw 1 \citep[previously studied by][]{Laor+1997}.
\citet{Tsuzuki+2006}\footnote{available at \url{http://www.ioa.s.u-tokyo.ac.jp/~kkawara/quasars/}. 
(its optical template is excluded in this work due to much poorer quality than others.)}  
extended the \FeII\ template redward up to $3500$ \AA\ using the same HST UV data and additional archival ground-based spectra of I Zw 1
and partially recovered the \FeII\ flux level underneath the \MgII\ $\lambda2798$ line using photoionization modeling results. 
Recently, for the limited region of $2653-3049$ \AA, 
\citet{Popovic+2019}\textsuperscript{\ref{footnote:K10P19}}
\citep[see also][]{KovacevicPopovic2015} provided a semi-empirical multicomponent template similarly as done by \citet{Kovacevic+2010}.

Iron emission is also important at near-infrared (near-IR) wavelengths, and a semi-empirical near-IR template was constructed by
\citet{Garcia-Rissmann2012} (see also \citealt{Marinello+2016} for its application) using a ground-based spectrum of I Zw 1 and the theoretical models of \citet{SigutPradhan2003}.

\begin{figure*}[!ht]
	\centering
	\includegraphics[width=0.98\textwidth]{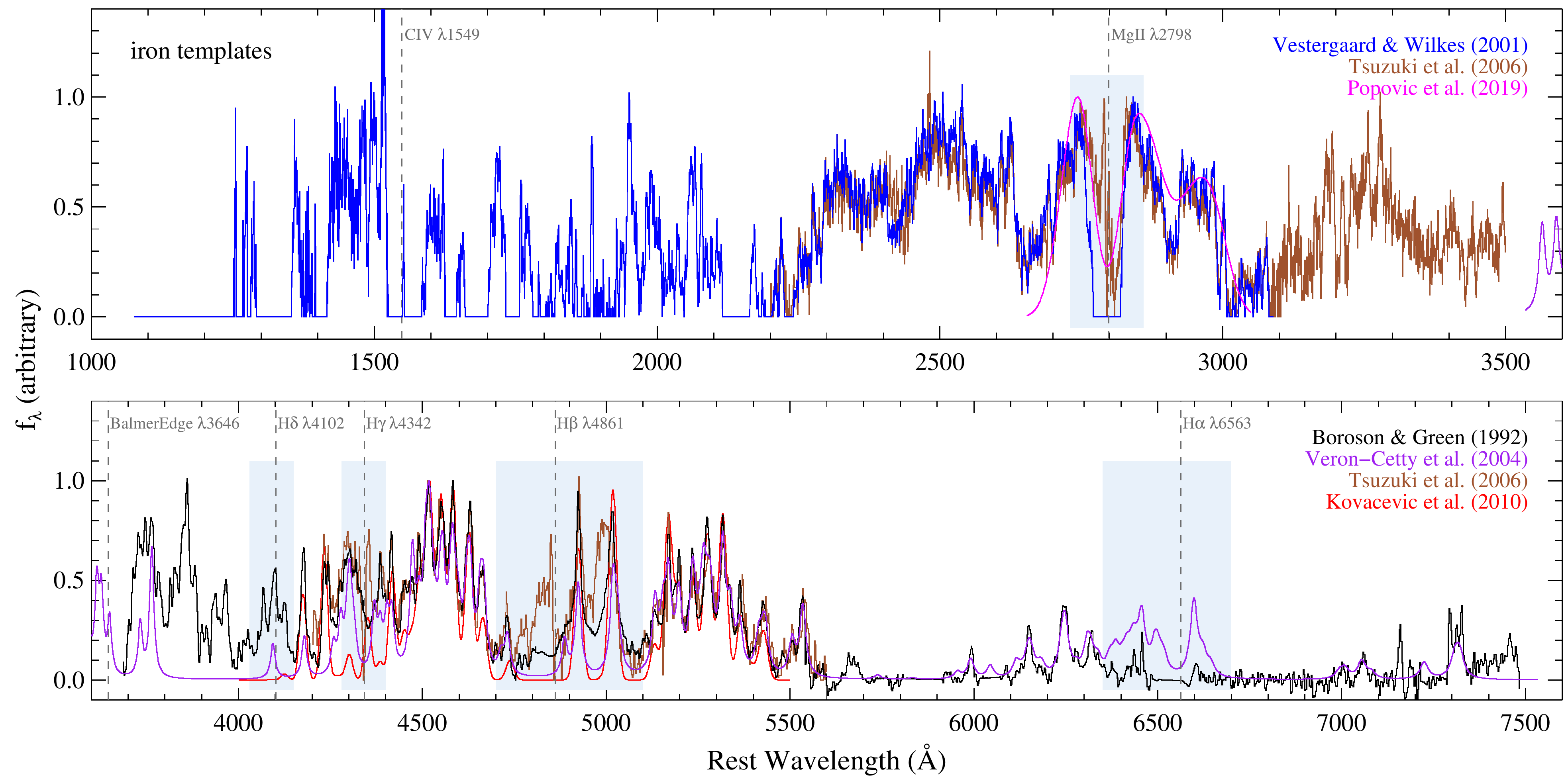} 
	\caption{
		Comparison of past UV (upper panel) and optical (lower panel) iron templates derived from HST and ground-based observations of I Zw 1.
		Blue shaded regions denote the locations of broad Balmer and \MgII\ emission lines where different templates have strongly differing structure.
		In the far-UV region surrounding \CIV, the \citet{VestergaardWilkes2001} template has low S/N and many gaps where the contribution of \FeII\ could not be determined. In the near-UV region, the S/N is also relatively poor and the \FeII\ flux underlying the \MgII\ line is very uncertain.
		In the optical region, the different templates make strongly divergent predictions for the strength of individual \FeII\ features, and there are particularly large discrepancies at wavelengths coincident with the Balmer lines.
		The existing templates also suffer from a gap between the UV and optical near the Balmer edge region, which consequently prevents accurate modeling of the Balmer continuum across this wavelength gap.
	}
	\label{fig:temp_compare}
\end{figure*}

\autoref{fig:temp_compare} shows a comparison between the most widely used UV and optical templates, giving an illustration of the wavelength ranges covered by the templates, their S/N, and the differences in their spectral properties. The comparison demonstrates that the relative strengths of \ion{Fe}{2} features in the strong optical blends between $\sim4000$ and 5600 \AA\ can vary strongly between templates. This is true even for templates that were derived purely empirically from observations of I Zw 1, indicating that the differences in template construction methodology lead to strong discrepancies in the inferred \ion{Fe}{2} emission spectrum.

\subsection{A Need for a New Template}

Despite the useful features of I Zw 1 as a very bright NLS1, it has some potential shortcomings as a template object: it exhibits an atypical narrow-line emission system \citep{Veron-Cetty2004}, 
an abnormal near-IR \FeII\ bump \citep{Garcia-Rissmann2012}, 
and a peculiar UV continuum shape, 
most likely due to substantial intrinsic reddening (see the overall continuum shape in \autoref{fig:spec_STIS_IZw1}
and also \citealt{Laor+1997,ConstantinShields2003}). 
I Zw 1 also exhibits atypical UV line ratios, 
specifically unusually weak \CIV\ emission and a high \ion{Si}{3} $\lambda1892$/\ion{C}{3}] $\lambda1909$
line ratio (Fig.~\ref{fig:spec_STIS_IZw1}), which are characteristic of extreme EV1 AGN \citep{Wills+1999}.
While templates derived from I Zw 1 have generally been shown to perform well in fitting optical \ion{Fe}{2} emission across a broad range of AGN properties \citep{Marziani2003}, it is still worthwhile to explore whether additional improvement could be obtained by creating alternative templates based on observations of different AGN.

Another concern regarding the use of empirical templates is that their quality is dependent on the S/N and wavelength coverage of the data used to derive them. Existing templates based on I Zw 1 are based on either ground-based data with less than ideal S/N, or HST data that has relatively low S/N in the UV and lacks simultaneous UV and optical observations (\autoref{fig:temp_compare}). For I Zw 1, the ground-based spectra used to construct optical templates were obtained with different observational apertures at different epochs than the HST UV data, making it difficult to produce a consistent model of \ion{Fe}{2} emission across the full UV-optical range. These issues consequently limit the accuracy of all the spectral measurements in AGN that are based on these templates. To date, there has not been a single empirical template based on data continuously covering the entire UV and optical wavelength range obtained at the same time. Having a complete and consistent UV-optical template would be particularly important for accurate modeling of the Balmer continuum and broad-band AGN continuum shapes,  and for 
simultaneous measurements of UV and optical emission lines together in the same object and comparison of BH masses derived from \ion{Mg}{2} and \Hb\ in the same object.

Moreover, the relative merits and disadvantages of the different iron templates have not been examined in detail. It would be extremely useful to carry out a systematic comparison between available templates in order to examine their relative precision for fitting AGN spectra, their relative accuracy in recovering line profiles and fluxes of non-iron lines in AGN including the Balmer lines and \ion{Mg}{2}, and to search for any systematic differences in line widths (and consequently in derived BH masses) when using different templates to fit AGN spectra.

To overcome the limitations of the previous empirical templates and data, we have obtained new spectroscopic data 
having high signal-to-noise ratio (S/N) and complete and quasi-simultaneous UV-optical coverage ($1150 - 10270$ \AA)
using the Space Telescope Imaging Spectrograph (STIS) aboard HST 
for a new iron template object, Mrk 493, which we have identified as superior in some respects to the canonical template object I Zw 1.
This new high-quality data enables us to generate a new empirical iron emission template, and will also provide an opportunity to test and refine theoretical \FeII\ emission models \citep[e.g.,][]{Sarkar+2021}.

The overall goal of our project is to create a new empirical iron template, based for the first time on high S/N, quasi-simultaneous spectroscopic observations across the full UV-optical wavelength range available to HST. Such a new template will enable better precision in AGN spectral modeling by more accurately constraining complex iron emission, which has been a long-standing source of systematic uncertainties in AGN spectral measurements.
This work is the first of a planned series of papers about the new Mrk 493-based iron emission template and its applications. As a first step, this paper presents a new empirical iron template from the STIS data for the optical wavelength range of $4000-5600$ \AA. Our template construction method is based in part on the work of \citet{Veron-Cetty2004}, but with several modifications and improvements. We also present extensive comparison tests between the templates available for this optical region to search for possible systematic biases in inferred AGN properties that might depend on the choice of \ion{Fe}{2} template used for spectral decompositions.

This paper is organized as follows.
In \autoref{sec:sampledata}, we describe our sample, observations, and data reduction.
We describe the method used to construct our new template in \autoref{sec:construct} and  present a 
qualitative comparison with other templates in \autoref{sec:comparison}.
 \autoref{sec:application} describes the results of comparison tests between our template and other widely used optical \ion{Fe}{2} templates, based on spectral fitting carried out on AGN spectra having diverse \ion{Fe}{2} emission properties.
We conclude with a summary and discussion in \autoref{sec:summary}.
All of the identified emission lines in Mrk 493 are listed in \autoref{app:linelist}, and spectral fitting plots for individual test objects are given in \autoref{app:allSDSSfits}.
The following standard cosmological parameters were adopted to calculate distances in this work:
$H_0 = 70$ km s$^{-1}$ Mpc$^{-1}$, $\Omega_{\rm m} = 0.30$, and $\Omega_{\Lambda} = 0.70$.

\section{Observations and Data Reduction} \label{sec:sampledata}

\subsection{Selection of Mrk 493}

We began by selecting a nearby NLS1 to use as the basis for constructing a new \ion{Fe}{2} template. An optimal template object would be defined by the following criteria: (1) extremely narrow BLR emission lines; (2) bright apparent magnitude, so that high S/N STIS spectra can be obtained in reasonable exposure times; (3) low foreground and intrinsic extinction. From an extensive search of catalogs and literature, we identified Mrk 493 at $z = 0.03102$ as an ideal target. Mrk 493 has long been known as a very narrow-lined NLS1 galaxy \citep{Osterbrock1985}, and its broad emission lines are substantially narrower than those of I Zw 1 (FWHM = $860$ versus $1240$ \kms; \citealt{Greene&Ho2007}),
enabling a more accurate decomposition between blended spectral features.  Mrk 493 also has very low Galactic foreground reddening of $E(B-V)=0.022$ mag \citep{Schlafly+2011} and insignificant intrinsic reddening \citep{Crenshaw+2002}, 
while it shows \FeII\ emission strength similar to that of I Zw 1 and its nucleus is very bright, with SDSS fiber magnitude $r\approx15.5$.
While I Zw 1 has an Eddington ratio of $\sim 2.5$ calculated from the BH mass of $10^{6.97}$ \msun\ and bolometric luminosity of $10^{45.47}$ \ergs\ \citep{Huang+2019}, Mrk 493 has a less extreme value of $\sim 0.5$, 
calculated from BH mass of $10^{6.18}$ \msun\ and bolometric luminosity of $10^{44.01}$ \ergs\ \citep{Wang+2014} using a bolometric correction factor of $9.26$ \citep{Shen+2008}.
Mrk 493 has been used previously as the basis of an empirical \ion{Fe}{2} template at optical wavelengths: \citet{Greene&Ho2007} used the SDSS spectrum of Mrk 493 to create a template used to fit other SDSS AGN spectra because they found that I Zw 1 was too broad for some of the low-mass AGNs in their sample. Existing archival HST spectra of Mrk 493 taken with the Faint Object Spectrograph (FOS) in 1997 have broad UV/optical wavelength coverage, but the S/N of the FOS data is much too low to be useful for creating an iron emission template.

\subsection{Observations}

We obtained UV and optical STIS spectra of Mrk 493 in HST program GO-14744 (PI: Park) during 2017 August 28 and 31. The data were obtained over three HST visits spanning this period using a total of $\sim10$ hours on-source exposure time. All observations were carried out with a consistent slit position angle (46\fdg0) and width (0\farcs2) across all wavelengths. These data have higher S/N and broader wavelength coverage than any previous HST observations of I Zw 1 or Mrk 493, resulting in what is likely to be the best-quality broad-band spectrum of an NLS1 ever obtained with HST.

We used the G140L, G230L, G430L, and G750L gratings to cover the full available wavelength range from $1150$ to $10270$ \AA, 
and the \texttt{52x0.2} slit, whose narrow width minimizes the host-galaxy and narrow-line region contributions to the data.
The spectral resolving power for the G140L grating data is $R=\lambda/\Delta\lambda\sim2400$, while it is $\sim1500$ for the G230L, G430L, and G750L grating data. As recommended in the STIS instrument handbook,
the D1 aperture position for the FUV-MAMA G140L grating was used to minimize FUV dark current.
The E1 aperture position was used for the CCD G430L grating to minimize losses due to the imperfect charge transfer efficiency.
The E2 aperture location for the CCD G750L grating was used for optimal fringe subtraction, 
along with CCDFLAT exposures observed immediately afterward.
Total integrations of 
14493 s for G140L, 
14493 s for G230L, 
3920 s for G430L,
and 3514 s for G750L were split into five or seven exposures, depending on the grating, 
and dithered along the slit for cleaning of cosmic-ray hits and bad pixels and for reduction of small-scale detector nonuniformity.
The overall observational setups and data reductions were done similarly to those described in \citet{Park+2017}, 
except for the G750 grating data and de-fringing process.

\subsection{Data Reduction}

For the UV grating data, we used the fully reduced data provided by the HST STIS pipeline.
However, for the optical grating data, we carried out additional steps to improve the cleaning of cosmic-ray charge transfer trails in raw images from the badly degraded STIS CCD and to remove fringe patterns in the G750 grating data at $\lambda>7000$ \AA. 
We performed a custom reduction for the optical grating data
based on the standard STIS reduction pipeline including trimming the overscan region, bias and dark subtraction, and flat-fielding
with supplementary steps for cosmic-ray and fringe pattern removals 
using the \texttt{LA\_COSMIC} \citep{vanDokkum2001} routine and \texttt{DEFRINGE} PyRAF \citep{PyRAF2012} task, respectively.
The multiple dithered exposures for each grating were then aligned and combined 
using the \texttt{IMSHIFT} and \texttt{IMCOMBINE} PyRAF tasks after performing wavelength calibration.
One-dimensional spectra from each grating were extracted with the \texttt{X1D} PyRAF task with the default extraction box heights of 11 pixels ($\sim0.27\arcsec$) for UV gratings and 7 pixels ($\sim0.36\arcsec$) for optical gratings 
and then joined together to produce a final single spectrum 
by taking into account the flux and noise levels in the overlap regions around $\sim1660$ \AA, $\sim3065$ \AA, and $\sim5610$ \AA.

\begin{figure*}[hbt!]
	\centering
	\includegraphics[width=0.90\textwidth]{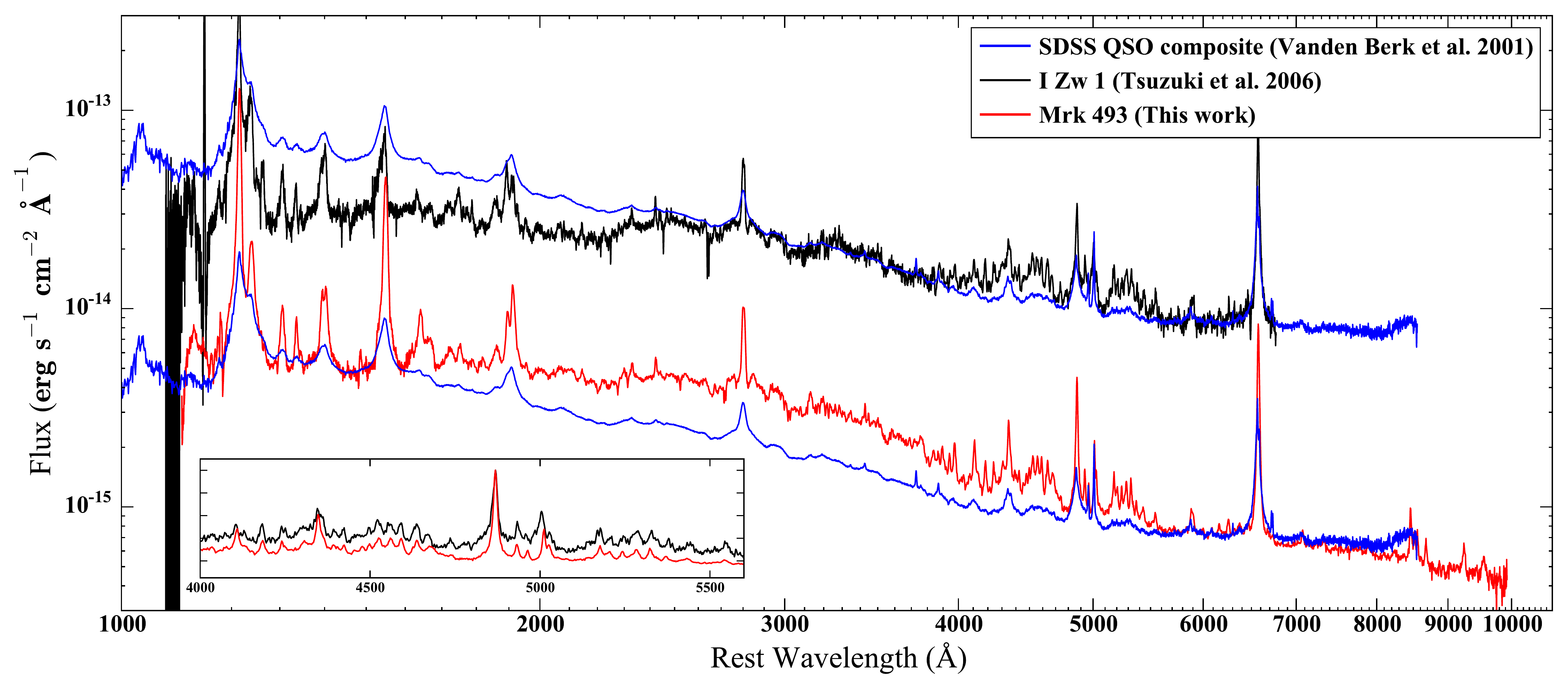} 
	\caption{
		Comparison of the Mrk 493 spectrum (red) from this work and the I Zw 1 spectrum (black) from \citet{Tsuzuki+2006}. The data are corrected for Galactic reddening. Overplotted in blue is the SDSS composite quasar spectrum  \citep{VandenBerk+2001}, which is arbitrarily normalized to the flux of each AGN at rest-frame 6300 \AA\  to facilitate comparison of the continuum shapes. 
		I Zw 1 has a much flatter UV continuum slope, likely due to significant intrinsic extinction within the host galaxy and/or the nuclear region, while Mrk 493 shows a similar far-UV to optical continuum slope to the SDSS quasar composite, while also exhibiting a stronger small blue bump feature in the near-UV.
		The inset shows a direct comparison of the I Zw 1 and Mrk 493 spectra, normalized by their maximum value at the \Hb\ line peak.
	}
	\label{fig:spec_STIS_IZw1}
\end{figure*}

The spectrum was corrected for Galactic extinction using the $E(B-V)$ value of $0.022$ mag from \citet{Schlafly+2011} listed in the
NASA/IPAC Extragalactic Database (NED) and the reddening curve of \citet{Fitzpatrick1999},
and then converted to the rest frame based on the known redshift from NED and the [\ion{O}{3}] $\lambda5007$ emission line on the data.
The final fully reduced and calibrated rest-frame spectrum has S/N per pixel of $44$ at 1350 \AA, 
$76$ at 2650 \AA, $78$ at 5100 \AA, and $89$ at 6200 \AA.  
\autoref{fig:spec_STIS_IZw1} displays the Mrk 493 STIS spectrum in comparison with the UV/optical spectrum of I Zw 1 and the SDSS composite quasar spectrum from \citet{VandenBerk+2001}.

\section{Construction of the Iron Template} \label{sec:construct}

To construct an empirical iron template, 
we follow an approach based on that used by \citet{Veron-Cetty2004} with several modifications and improvements.
Their template construction method is more complete and self-consistent than the earlier work of \citet{BorosonGreen1992}
in that they used an extensive line list (more than 400 lines; see references therein)
and performed fitting for all the model components simultaneously over the full spectral range of the observed data.
They built their template from the sum of all the fitted profiles to broad iron features in the I Zw 1 spectrum, resulting in a noise-free model template. In contrast, the \citet{BorosonGreen1992} template contains the observational noise of the data 
because the observed spectrum of I Zw 1 was adopted as a template after removing non-iron features from the data. The \citet{BorosonGreen1992} template may also be partly contaminated by residual non-iron lines and some amount of residual continuum emission 
due to the use of a more restricted list of lines and a somewhat simplified fitting strategy,
as well as the relatively lower quality and low spectral resolution of the observed I Zw 1 spectrum.

However, 
there are several limitations of the approach used by \citet{Veron-Cetty2004} and the resulting template.
One is that they adopted a Lorentzian profile for the broad emission line model.
This adopted model has extremely broad line wings that may not provide a good representation of broad-line profiles, and the line dispersion ($\sigma$) measurement for a Lorentzian profile is mathematically divergent.  
Also, they added an additional very broad Gaussian model to the Lorentzian profile for Balmer lines to obtain a good fit.
In fitting the data, no continuum model was specified even though there must be some amount of continuum emission from the AGN itself and 
host galaxy starlight due to the relatively large slit width of their ground-based observations. Consequently, the very broad wings of the Lorentzian line profiles could actually incorporate some AGN or host-galaxy continuum emission, which would then become part of the \ion{Fe}{2} template.
Finally, they constructed the final template using only iron emission lines identified as broad lines. 
However, it has later been empirically shown that incorporating both broad and narrow iron emission lines provides better spectral fitting results  (see, e.g., \citealt{Dong+2008,Dong+2010,Dong+2011}).

The three major modifications that we adopted in methodology for generating our template are as follows.
We used a Gauss-Hermite series function \citep{vanderMarelFranx1993,Cappellari+2002} to model the broad and narrow emission-line profiles. This model is more flexible than a symmetric simple Gaussian and has been commonly used in many AGN studies \citep[e.g.,][]{Woo+2007,Woo+2008,McGill+2008,Denney+2009,Denney+2013,Denney+2016,Wang+2009,Bennert+2011,Bennert+2015,Bennert+2018,Assef+2011,Barth+2011:mrk50,Barth+2013,Barth+2015,Park+2012,Park+2013,Park+2015,Park+2017,DeRosa+2014,Hu+2015,Hu+2021,Coatman+2016,Bahk+2019}. 
The continuum emission was modeled with a single power-law function. This continuum model is sufficient to describe the AGN continuum over the restricted wavelength range of our template (4000--5600 \AA), and the small STIS spectroscopic aperture ensures that any starlight contribution to the continuum is very small.
In modeling the STIS data, we include all the identified broad and narrow iron emission lines, with the addition of \ion{Ti}{2}, \ion{Ni}{2}, and \ion{Cr}{2} lines  
which are associated with \FeII\ line emission \citep{Veron-Cetty2004}, 
based on the approach used by \citet{Dong+2008,Dong+2010,Dong+2011}.  
Including these low-ionization metal lines in the iron template is beneficial as a major practical application of the iron template is to isolate the Balmer emission lines (primarily H$\beta$) and measure their fluxes and profiles accurately by deblending and cleanly subtracting off the thick forest of iron and related lines from AGN spectra.

\begin{figure*}[!htb]
	\centering
	\includegraphics[width=0.95\textwidth]{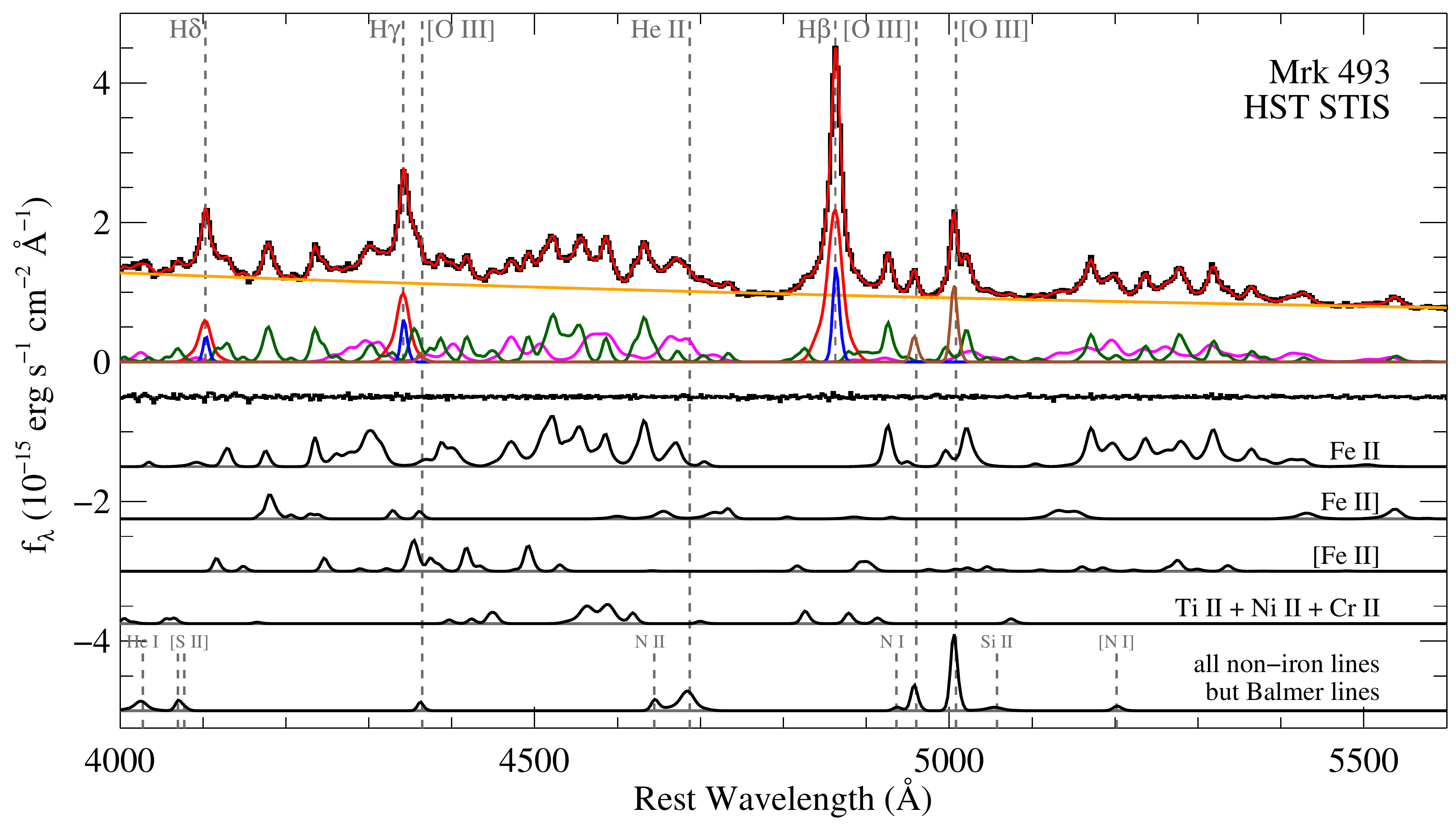} 
	\caption{
		Multicomponent spectral decomposition of the optical region ($4000-5600$ \AA) of Mrk 493.
		The observed HST STIS spectrum (black) is shown along with the best-fit model (red).
		A few strong emission lines (Balmer and [\ion{O}{3}] lines) are labeled with vertical dashed lines.
		The best-fit 
		power-law continuum (orange), 
		broad Balmer lines (red), 
		narrow Balmer lines (blue),
		all other broad emission lines (magenta),
		all other narrow emission lines (dark green), and
		[\ion{O}{3}] lines (brown) are also shown respectively.
		The residuals (black), representing the difference between the observed data and sum of all the best-fit model components, are arbitrarily shifted downward ($-0.5$ on the y-axis) for clarity.	
		The lower portion of the plot displays the models for the permitted, semi-forbidden, and forbidden iron lines (\ion{Fe}{2}, \ion{Fe}{2}], [\ion{Fe}{2}]), the additional low-ionization metal lines associated with \ion{Fe}{2} (\ion{Ti}{2}, \ion{Ni}{2}, and \ion{Cr}{2}), and all non-iron lines but the Balmer lines.
	}
	\label{fig:decomp_STIS}
\end{figure*}

\subsection{Spectral Decomposition Analysis} \label{subsec:decomposition}

In this work, we focus on the rest-frame wavelength range of $4000-5600$ \AA\ because this region includes the strongest optical iron emission complexes as well as the \Hb\ $\lambda4861$ emission line, which is of primary importance for AGN black hole mass estimation, reverberation mapping, and other studies of BLR physical properties.
Also, this range overlaps with the most widely used existing optical \ion{Fe}{2} templates, enabling direct comparison of fitting results.
The full UV and optical range will be examined in forthcoming papers, with the goal of producing a complete UV/optical template based on the high-quality Mrk 493 STIS spectrum. 

As a starting point for spectral decomposition of the STIS data, we adopted the line lists compiled by \citet{Veron-Cetty2004} (their tables 3, 4, A.1, and A.2). In their decomposition of the I Zw 1 spectrum, they identified lines belonging to a broad-line system, a high-excitation narrow-line systems, and a low-excitation narrow-line system. We include all of these features in our decomposition of Mrk 493. Based on the line identification and wavelength information taken from the tables,
we first converted air wavelengths to vacuum wavelengths for consistency with our STIS data products.
We added three emission lines to the line list that were not included in the \citet{Veron-Cetty2004} tables since these features were not identified in their I Zw 1 data:
\ion{He}{2} $\lambda4686$, [\ion{O}{3}] $\lambda4363$, and [\ion{N}{1}] $\lambda5199$ \citep[cf.][]{VandenBerk+2001,Calderone+2013}.
The total number of lines in our model includes 67 broad and 176 narrow lines. In our final best-fitting model, the number of lines having non-zero intensities is 46 broad and 123 narrow lines.

Following \citet{Veron-Cetty2004}, we categorized emission lines into a few systems sharing the same line profile.
Our model is constructed as a combination of the following components:   
(1) a power-law function for continuum emission,
(2) a single 4th-order Gauss-Hermite series function for the broad Balmer lines (\Hb, \Hg, and \Hd), 
(3) a single 4th-order Gauss-Hermite series function for all other broad lines, 
(4) a single 4th-order Gauss-Hermite series function for the [\ion{O}{3}] $\lambda\lambda4959,5007$ narrow lines, 
and (5) a single 4th-order Gauss-Hermite series function for all other narrow lines.

We experimented with the use of a double-Gaussian function representing possible core and wing components of the [\ion{O}{3}] doublet lines, but we did not clearly detect evidence of a distinct blue-shifted wing component in the STIS data. Thus, for our final model we opted to use a single Gauss-Hermite series function of 4th order to represent the slightly asymmetric profile of the line.
The flux ratio of [\ion{O}{3}] $\lambda5007$/[\ion{O}{3}] $\lambda4959$ was fixed to be the canonical value of $2.98$ \citep{StoreyZeippen2000}.
The same [\ion{O}{3}] $\lambda5007$ line profile (same velocity) was used for all narrow Balmer lines 
(\Hb, \Hg, \Hd) with amplitudes and velocity shifts as free parameters.
The [\ion{O}{3}] $\lambda4363$ narrow line was also modeled using the [\ion{O}{3}] $\lambda5007$ line profile as a template
based on the discussion by \citet{BaskinLaor2005}.
Also, all Balmer series lines (\Hb, \Hg, \Hd) were assigned the same flux ratio between broad and narrow components with this flux ratio allowed to vary as a free parameter, 
meaning that all the Balmer lines were forced to have the same line profile (same velocity and shift) with different overall strengths.
The Hermite coefficients of the 4th-order Gauss-Hermite series function were forced to share the same values among  
all broad lines and among all narrow lines, respectively. 
This indicates that the asymmetry of the model line profile is the same for broad line groups and for narrow line groups, respectively.
All of the individual line intensities were treated as free parameters for all of the line systems.
The total number of free parameters in the fit is thus 260. 
To optimize the model parameters, we used the Levenberg-Marquardt least-squares minimization technique, 
implemented in the \texttt{mpfit} \citep{Markwardt2009} package in IDL. All of the model components were fitted simultaneously.

The final spectral decomposition is displayed in \autoref{fig:decomp_STIS}, which illustrates the precise fit to the highly complex spectrum, leaving very small residuals.
The fitting results for all the broad (46) and narrow emission (123) lines fitted with non-zero flux 
are listed in \autoref{tab:BLRlinelist} and \autoref{tab:NLRlinelist}. Based on this model fit, we define our template to be the sum of all model components for \ion{Fe}{2} lines (including permitted, semi-forbidden, and forbidden lines) plus the related \ion{Ti}{2}, \ion{Ni}{2}, and \ion{Cr}{2} lines.

Although the model provides an excellent overall fit to the data, we caution that the fit results are not unique, considering the limited spectral resolution of the data, the large number of blended lines, possible incompleteness in the line list, and the simplifying assumptions made in the model fit such as assuming a common velocity broadening for numerous lines. Nevertheless, our spectral decomposition should result in a reliable separation and removal of the Balmer lines, [\ion{O}{3}], and other non-iron features. Thus, despite the possible fitting degeneracy between multiple blended iron features in the data, the sum of the model components for all iron lines (plus \ion{Ti}{2}, \ion{Ni}{2}, and \ion{Cr}{2}) can provide a robust empirical template.

\subsection{Measurement Uncertainty Estimates} \label{subsec:error}
We also estimated the uncertainty of the spectral decomposition
by using the Monte Carlo flux randomization method adopted by \citet{Park+2013,Park+2017} (see also \citealt{Shen+2011}).
First, we generated 100 mock spectra having resampled flux values
by adding Gaussian random noise to the original spectrum based on the value of the error spectrum at each spectral pixel.
We then applied our spectral decomposition method to the mock data, 
producing 100 realizations of the template.
The measurement uncertainties were then estimated by calculating the 68\% semi-interquantile range of the resulting distribution at each pixel.

\begin{figure*}[ht!]
	\centering
	\includegraphics[width=0.85\textwidth]{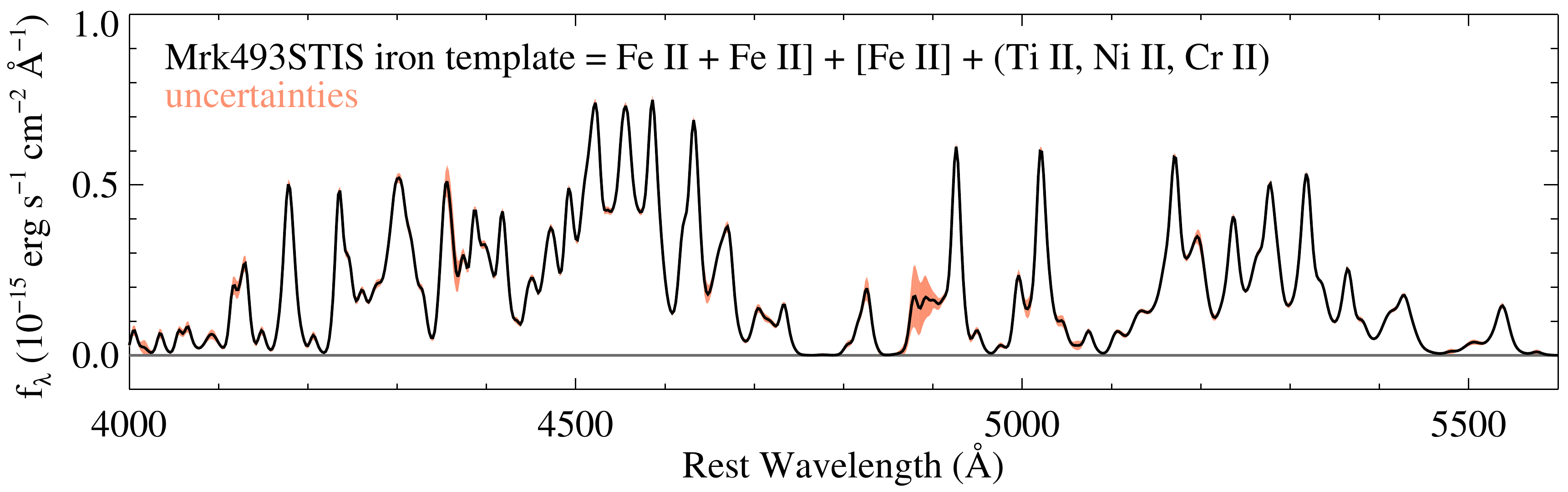} 
	\caption{
		The final iron template (Mrk493STIS) containing the broad and narrow iron lines and the \ion{Ti}{2}, \ion{Ni}{2}, \ion{Cr}{2} lines with measurement uncertainties estimated from Monte Carlo simulation.
	}
	\label{fig:decompError_STIS}
\end{figure*}

\autoref{fig:decompError_STIS} shows the final iron template (which we refer to as the Mrk493STIS template) with the measurement uncertainty level displayed as an error band around the template spectrum.
The largest uncertainties occurs around the wing areas of the \Hb\ line.
This region is particularly complex, and the model fits are subject to substantial degeneracy arising from many emission line components having flexible model profiles that are not uniquely distinguishable given the severe line blending. This blending is partly intrinsic to the AGN but also partly due to the limited spectral resolution of the STIS data.
The final iron template spectrum including its uncertainty is given in \autoref{tab:template}. For convenience in usage of the template for spectral fitting, it was rebinned from the best-fit model onto a uniform wavelength scale of 2 \AA\ per bin, which is slightly finer than the original scale of the observed STIS data ($\sim2.7$ \AA).

\begin{deluxetable}{ccc}
	\tablecolumns{3}
	\tablewidth{0pt}
	\tablecaption{Iron Template Spectrum}  
	\tablehead{
		\colhead{Wavelength} &
		\colhead{Flux} &
		\colhead{Uncertainty} \\
		\colhead{(\AA)} &
		\multicolumn{2}{c}{($\rm 10^{-15}\ erg\ s^{-1}\ cm^{-2}$ \AA$^{-1}$)} 
	}
	\startdata
	$4000$  &  $0.02977$  &  $0.00744$  \\
	$4002$  &  $0.05271$  &  $0.01186$  \\
	$4004$  &  $0.07097$  &  $0.01471$  \\
	$4006$  &  $0.07227$  &  $0.01531$  \\
	$4008$  &  $0.05746$  &  $0.01307$  \\
	$4010$  &  $0.03941$  &  $0.01097$  \\
	$4012$  &  $0.02834$  &  $0.01134$  \\
	$4014$  &  $0.02479$  &  $0.01614$  \\
	$4016$  &  $0.02387$  &  $0.02097$  \\
	$4018$  &  $0.02116$  &  $0.02188$  \\
	$4020$  &  $0.01596$  &  $0.01729$  \\
	$4022$  &  $0.01073$  &  $0.01132$  \\
	$4024$  &  $0.00791$  &  $0.00740$  \\
	$4026$  &  $0.00886$  &  $0.00536$  \\
	$4028$  &  $0.01569$  &  $0.00622$  \\
	$\cdots$ & $\cdots$ & $\cdots$ 
	\enddata
	\label{tab:template}
	\tablecomments{The full content of this table is given in the electronic version of the Journal.
		A portion is shown here for guidance regarding its form and content.}
\end{deluxetable}

\section{Comparison with Other Templates} \label{sec:comparison}

Here we present a qualitative comparison of our new template with other widely used templates covering this spectral region,
before describing quantitative comparison tests in the following section.

\begin{figure*}[!htb]
	\centering
	\includegraphics[width=0.70\textwidth]{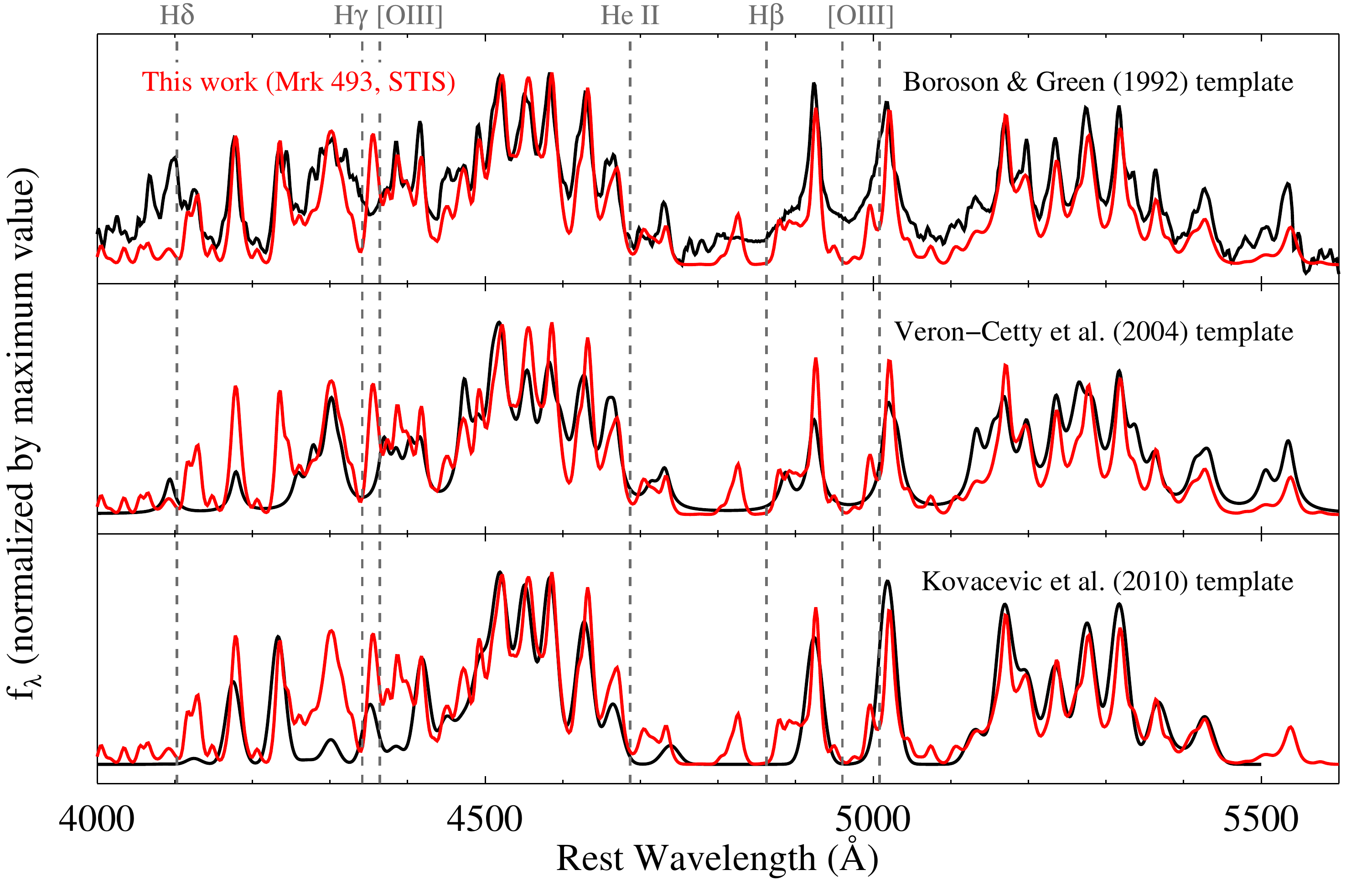} 
	\caption{
		Comparison of the new Mrk493STIS template ($4000-5600$ \AA) to the templates from \citet{BorosonGreen1992}, \citet{Veron-Cetty2004}, and \citet{Kovacevic+2010}.
		Locations of the Balmer lines, [\ion{O}{3}] $\lambda4363$, [\ion{O}{3}] $\lambda\lambda4959,5007$, and \ion{He}{2} $\lambda4686$ are depicted with vertical dotted lines.
		Major differences between templates mostly occur at around the Balmer line regions.
	}
	\label{fig:compare_templatesSTIS}
\end{figure*}

In \autoref{fig:compare_templatesSTIS}, we show direct comparisons between the templates to illustrate their differences. 
We selected three representative optical iron templates from
\citet[BG92 hereafter]{BorosonGreen1992}, \citet[VC04 hereafter]{Veron-Cetty2004}, and \citet[K10 hereafter]{Kovacevic+2010},
that have been widely and commonly used.
It is worth noting that the five \ion{Fe}{2} line groups in the K10 template can have different relative strengths (see Fig.~13 in  \citealt{Shapovalova+2012}) but for purposes of this illustration we display the template with the original relative scaling between the line groups and a broadening velocity of 900 \kms\ provided by K10.
As can be seen, the Balmer line regions show significant differences over the templates.
Those differences in template structure would introduce systematic differences in inferred emission-line properties including the flux and profiles of the Balmer lines, depending on the choice of template used to decompose an AGN spectrum.
Such differences would stem from the complex combined effects of 
intrinsic differences in the emission-line spectra of the template basis objects (I Zw 1 versus Mrk 493), 
differences in adopted modeling details (e.g., line profile model, continuum model, line list, and fitting technique), 
and data quality resulting from different slit widths (and consequently different host galaxy and NLR contributions), spectral resolution, and S/N for the spectra used as the basis of template construction.

An advantage of the Mrk493STIS template is that it should be relatively more free of possible contaminants than other empirical templates, as a result of (1) the intrinsically narrower lines of Mrk 493 compared with I Zw 1; (2) the high-quality, narrow-slit STIS data used to construct the template, and (3) our more robust model fitting approach relative to the earlier empirical templates of BG92 and VC04.
Overall, the VC04 template shows the largest difference against our template, which is somewhat expected 
because their template was constructed using only the lines from their broad \ion{Fe}{2} line list and excluding the narrow-line complexes, as well as their adoption of a Lorentzian broad-line profile.

The Mrk493STIS template and BG92 template are particularly strongly discrepant over the wavelength range of the \Hb\ and [\ion{O}{3}] lines. In this region, the BG92 template appears to contain some residual continuum emission and non-iron line contributions, 
which would likely lead to oversubtraction of iron features when applied to AGN spectra. This behavior is somewhat expected because their template was constructed by using non-simultaneous manual fitting of the I Zw 1 spectrum using a more simplified fitting approach, and included only a limited number of strong emission lines. Our Mrk493STIS template includes a larger number of identified iron features than other templates in this spectral region, 
including \ion{Cr}{2} 30 (4825 \AA, 4878 \AA) and \ion{Ti}{2} 114 (4913 \AA) lines as well as several Fe lines.
It is worth noting again that the \ion{Ti}{2}, \ion{Ni}{2}, and \ion{Cr}{2} lines identified in our fit to the Mrk 493 spectrum were included in the new template as they appear to contribute to the forest of blended lines commonly considered to be `iron' emission blends. Thus, our template construction assumes that the strengths of these metal lines in typical AGN spectra are closely tied to the iron line strengths.

The K10 template was constructed semi-empirically by using
a total of 65 \FeII\ lines identified as the strongest in the range of $4000-5500$ \AA. In the \Hg\ line region, the K10 template shows a large difference compared with other templates including the new Mrk493STIS template. The lower strengths of the iron features in the K10 template could cause under-subtraction of iron emission over the \Hg\ line region when applied, thus resulting in biased (over-estimated) \Hg\ line fluxes. This discrepancy might be due to their use of a less extensive line list and the fact that narrow iron lines were not included in their template. 

Overall, except for the regions immediately surrounding the Balmer lines, the K10 template appears more consistent with our Mrk 493-based template than others, 
which is notable given the substantial difference in construction methodology 
between the K10 template and the other empirical templates.

In the \Hd\ line region,  the BG92 template includes elevated flux attributed to \ion{Fe}{2} emission that is not present in any of the other templates. This may have resulted from imperfectly subtracting \Hd\ and other nearby narrow emission lines
when constructing the template. If so, use of the BG92 template in AGN spectral decomposition would tend to cause oversubtraction in the \Hd\ line region and would bias measurements of the \Hd\ flux and profile.

\section{Spectral Fitting Tests} \label{sec:application}

To quantitatively assess and compare the performance of the new Mrk493STIS template with the three other templates discussed above (BG92, VC04, and K10),  
we have carried out spectral decompositions of quasar spectra selected from the SDSS DR7 quasar catalog 
\citep{Shen+2011}\footnote{\url{http://quasar.astro.illinois.edu/BH_mass/data/catalogs/dr7_bh_Nov19_2013.fits.gz}}.
In this section, we present detailed results comparing the templates in terms of statistical performance, decomposition differences, measurements of Balmer line profiles, and other physical parameter estimates derived from spectral measurements.

\begin{figure}[ht!]
	\centering
	\includegraphics[width=0.47\textwidth]{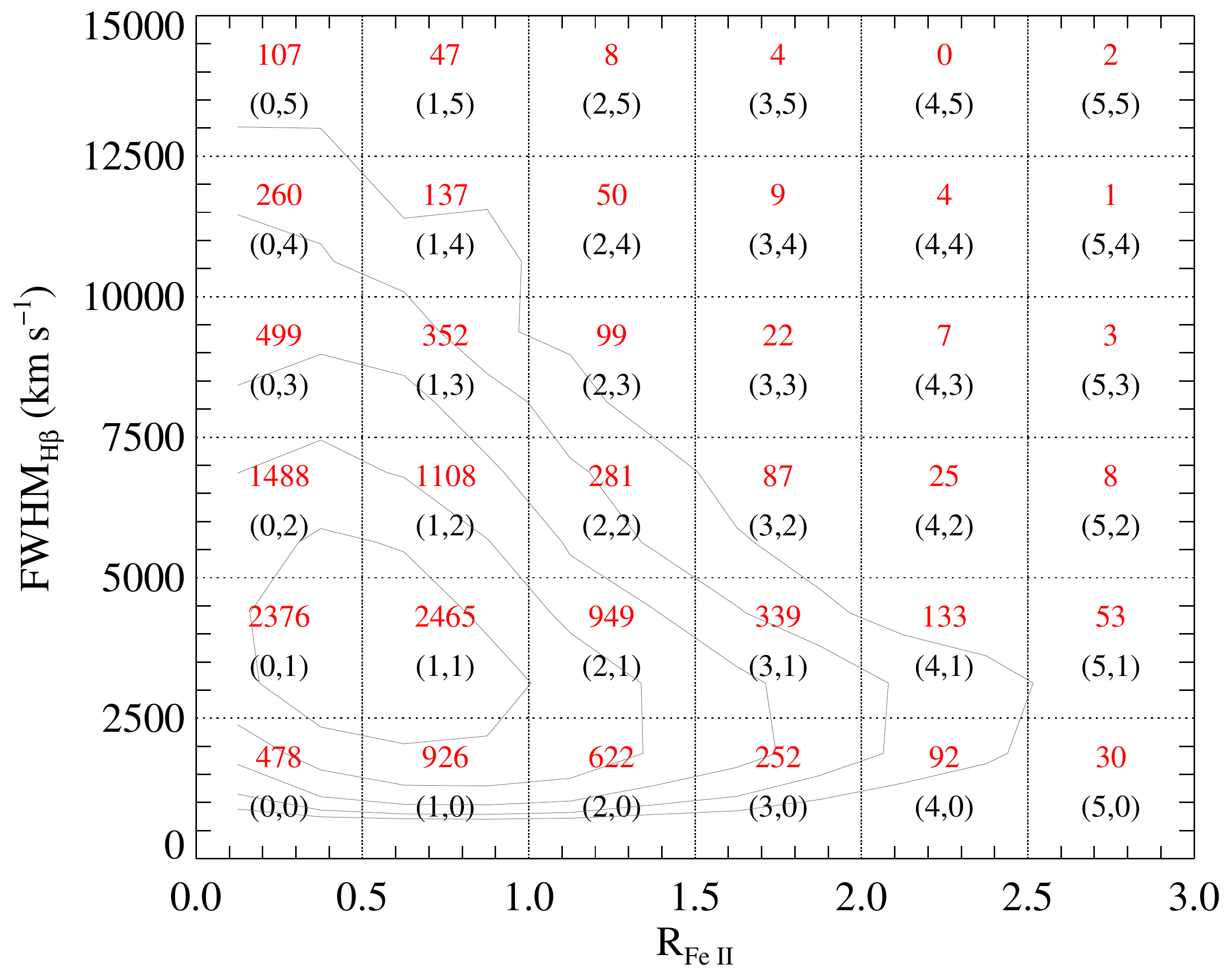} 
	\caption{
		Sampling grid in the EV1 plane (FWHM$_{\rm H\beta}$-R$_{\rm Fe\ II}$) to uniformly select test samples from the SDSS DR7 quasar catalog.
		The grid numbers used to define the object ID are given in parentheses within each grid cell.
		The total count of objects belonging to each grid cell is given in red, and dark gray contours indicate the distribution of DR7 AGN over the EV1 plane.
	}
	\label{fig:SDSSsampling}
\end{figure}

\subsection{Comparison Test Sample Selection} \label{subsec:sdss-sampling}

Our goal for these comparison tests was to carry out fits to AGN spectra spanning the full range of \ion{Fe}{2} strengths and velocity widths in order to assess the differences in performance between templates. Starting from the quasar distribution on the EV1 plane as described by \citet{ShenHo2014},
we selected a sample of AGN covering full parameter space of the EV1 
(FWHM$_{\rm H\beta}$-$R_{\rm Fe\ II}$, i.e., broad \Hb\ FWHM - relative \FeII\ strength) domain. We divided the EV1 space into a uniformly spaced $6\times6$ grid over the parameter values $R_\mathrm{Fe~II}$ from 0 to 3, and FWHM$_{\rm H\beta}$ from 0 to 15000 \kms. The adopted sampling grid is depicted in \autoref{fig:SDSSsampling}.

To select AGN for our comparison tests, we first picked the object having the highest-S/N spectrum from within each grid cell, 
giving a total of 35 objects (one grid cell did not contain any AGN in the DR7 quasar catalog). The selected objects are designated as A00, ... , A55, where
the initial letter A denotes the highest-S/N object in the grid cell (the object with the second-highest S/N is denoted by the label B),  and the two digits represent the horizontal ($R_{\rm Fe\ II}$) and vertical (FWHM$_{\rm H\beta}$) grid locations. 
From those, seven spectra with S/N$< 20$  at 5100 \AA\ were discarded since such low S/N does not yield sufficiently reliable fitting results for our tests. 
Two spectra showing virtually no \Hb\ emission were also removed. 
Additionally, seven objects were discarded from the sample for having extremely weak and very broad iron features, since their \ion{Fe}{2} emission was insufficient to provide any useful test to distinguish between templates. 
After removing these objects, the remaining sample of A objects (selected as having the highest S/N in a grid cell) contained 19 AGN.
To increase the sample size for our tests, we picked the second highest S/N object on each grid cell (designated as B00, ... , B55), for the 34 grid cells containing more than two AGN.
Among those 34 objects, we discarded eight AGN having S/N$<20$ at 5100 \AA, one AGN for which the spectral region $5050-5350$ was missing from the data, and three objects having extremely weak \ion{Fe}{2} features.
The final test sample selected from the above process then includes $41$ ($=19+22$) AGN 
and is listed in \autoref{tab:testSampleList}.
Galactic extinction corrections and conversion to the AGN rest frame applied to each object as described in \autoref{sec:sampledata}.

\startlongtable
\begin{deluxetable*}{clrrrccc}
	\tablecolumns{8}
	\tablewidth{0pt}
	\tablecaption{Test Sample from the SDSS DR7 quasar catalog}
	\tablehead{
		\colhead{Object ID} &
		\colhead{SDSS name} &
		\colhead{Plate} &
		\colhead{MJD} &
		\colhead{Fiber} &
		\colhead{$z$} &
		\colhead{FWHM$_{\rm H\beta}$} &
		\colhead{R$_{\rm Fe II}$} \\
		\colhead{(1)} &
		\colhead{(2)} &
		\colhead{(3)} &
		\colhead{(4)} &
		\colhead{(5)} &
		\colhead{(6)} &
		\colhead{(7)} &
		\colhead{(8)}
	}
	\startdata
	A00  &   SDSS J032213.89+005513.4  &      414  &    51901  &      341  &      0.1849  &        2440  &      0.41  \\
	A01  &   SDSS J172026.70+554024.2  &      367  &    51997  &      472  &      0.3592  &        2959  &      0.49  \\
	A02  &   SDSS J094715.56+631716.4  &      487  &    51943  &      391  &      0.4873  &        5842  &      0.13  \\
	A10  &   SDSS J085334.23+434902.2  &      897  &    52605  &      242  &      0.5142  &        2482  &      0.60  \\
	A11  &   SDSS J084302.97+030218.9  &      564  &    52224  &      471  &      0.5110  &        2759  &      0.51  \\
	A12  &   SDSS J131204.70+064107.5  &     1795  &    54507  &      106  &      0.2419  &        7460  &      0.63  \\
	A13  &   SDSS J105237.24+240627.3  &     2481  &    54086  &      532  &      0.3970  &        7615  &      0.51  \\
	A20  &   SDSS J154732.17+102451.2  &     2520  &    54584  &      249  &      0.1381  &        1448  &      1.18  \\
	A21  &   SDSS J115117.75+382221.5  &     1997  &    53442  &      639  &      0.3345  &        4345  &      1.34  \\
	A22  &   SDSS J094755.99+535000.3  &      769  &    52282  &      225  &      0.4875  &        5630  &      1.24  \\
	\multicolumn{8}{c}{$\cdots$}
	\enddata
	\label{tab:testSampleList}
	\tablecomments{
		Column 1: Object ID.
		Column 2: SDSS name.
		Column 3: Spectroscopic plate number.
		Column 4: MJD of spectroscopic observation.
		Column 5: Spectroscopic fiber number.
		Column 6: Redshifts.
		Column 7: FWHM of broad \Hb\ line (\kms). 
		Column 8: Ratio of EW of Fe within $4434-4684$ \AA\ to EW of broad \Hb. 
		All but Object ID are taken from the SDSS DR7 quasar catalog. 
		The full content of this table is given in the electronic version of the Journal.
		A portion is shown here for guidance regarding its form and content.}
\end{deluxetable*}

\subsection{Fitting} \label{subsec:sdss-fitting}

To perform a systematic spectral fitting analysis for the test sample,
we use a modified and improved version of the multicomponent spectral decomposition code developed by \citet{Park+2015,Park+2017}.
The original code consists of separate procedures for modeling the continuum and the emission lines. 
The continuum emission over the \Hb\ line region is first fitted  
using a linear combination of the following three pseudo-continuum model components 
($F_{\lambda }^{\rm PL} + F_{\lambda }^{\rm iron} + F_{\lambda }^{\rm host}$):
\begin{itemize}
	\item[(1)] AGN featureless power-law (PL) continuum
	$$F_{\lambda }^{\rm PL} (a,\beta) = a\ \lambda ^\beta$$
	\item[(2)] AGN iron emission blends
	$$F_{\lambda }^{\rm iron} (c_i,v_s, \sigma_w ) = \sum_{ i=1 }^{ d }{ c_i\ T_{\lambda,i}^{\rm iron}  \otimes G_\lambda (v_s, \sigma_w) }$$
	\item[(3)] Host galaxy starlight 
	$$F_{\lambda }^{\rm host} (k_j,v_s^*, \sigma_w^* ) = \sum_{ j=1 }^{ 7 }{  k_j\ T_{\lambda,j} ^{\rm star}  \otimes G_\lambda (v_s^*, \sigma_w^*) }$$
\end{itemize}
where $T_{\lambda,i}^{\rm iron}$, $T_{\lambda,j} ^{\rm star}$, $G_\lambda$, and $\otimes$ denote 
an iron template from one of those four choices (BG92, VC04, K10, Mrk493STIS), 
a host galaxy template composed of seven stellar spectra from the Indo-US spectral library \citep{Valdes+2004},
a Gaussian broadening kernel with free velocity shift and width parameters, and a convolution process, respectively. For the iron template, $d=1$ for monolithic templates (Mrk493STIS, BG92, and VC04), and $d=5$ for the K10 template that consists of five components that can vary independently.
The model parameters are optimized based on the $\chi^2$ statistic in the continuum regions of 
$4170-4260$ \AA, $4430-4720$ \AA, and $5080-5500$ \AA\ where the \ion{He}{2} $\lambda4686$ and several weak AGN narrow emission lines are masked out during the fitting.

For the K10 iron template, we use the multicomponent template data, as provided on their website,
which is calculated by adopting a fixed excitation temperature of 9900 K.
Although the temperature parameter allows additional variations for relative strengths between iron lines within each line group,
we opt not to change the temperature value 
since varying this parameter does not substantially alter the spectral fits, and the temperature parameter is subject to large uncertainties 
due to the very approximate calculation formula as discussed by K10.
The host galaxy template is added to the model 
only if host galaxy features are clearly seen in the observed data and the fit converges reliably.

After subtracting off the best-fit pseudo-continuum model from the observed data, 
the remaining \Hb\ emission region is then fitted with a linear combination of
a 6th-order Gauss-Hermite series function for the \Hb\ broad component,
a Gaussian function (tied to the [\ion{O}{3}] line profile) for the \Hb\ narrow component,
two Gaussian functions representing narrow core and broad blueshifted components for the [\ion{O}{3}] $\lambda\lambda4959,5007$ doublet lines respectively, 
and two Gaussian functions for the \ion{He}{2} $\lambda4686$ broad and narrow components.
From examination of each object's spectrum, the Gauss-Hermite function order used for the \Hb\ broad-line profile was determined individually for each AGN (either 6th order, 4th order, or a simple Gaussian function) in order to fit the line profile adequately while avoiding over-fitting the noise or obtaining negative flux values in the line wings.
The region surrounding the \ion{He}{2} line  was masked for some objects during the fits if it was too weak to be clearly identified in the data.
The \Hg\ and \Hd\ line regions ($4260-4430$ and $4000-4170$ \AA, respectively) were excluded during fitting to reduce the model complexity, since including these lines in the model (with broad and narrow components) would add several additional free parameters.

For a further refinement of the fits and results, 
we modify the above two-step (continuum first and then emission) fitting code 
by combining the two separate model fits into a single simultaneous (one-step) fit that optimizes all the model parameters at once,
enabling a consistent and straightforward comparison between the fits with four different templates.
The number of all free parameters during the one-step fitting ranges from 21 to 36 (with the number of degrees of freedom ranging from 888 to 966) depending on the adopted setup for each object. 
The best-fit results from the prior two-step fitting are used as the
initial parameter values for the final one-step fitting.
The fitting windows are same as those from combination of the separate continuum and emission fitting windows, or slightly wider 
to continuously cover the spectral ranges over the \Hb\ region complex 
if there are small gaps between the previous separate continuum and emission region fitting windows.
The \texttt{mpfit} \citep{Markwardt2009} routine is used to optimize the one-step fits,  while for the two-step fitting procedure the \texttt{bvls} routine \citep[see][]{Park+2015} is also used internally to constrain intensities of the iron and stellar templates. 

All the fits are performed uniformly and consistently  
using the same code with consistent setup for each of the 41 objects, except for the iron template difference (4 choices). 
Measurement uncertainties for all quantities are estimated using the same Monte Carlo method described in \autoref{subsec:error}, by creating and re-fitting 100 noise-added realizations of each spectrum. Plots of the model fits illustrating the comparison of results for the different iron templates are shown in  \autoref{app:allSDSSfits}.

\subsection{Comparison Results} \label{subsec:sdss-results}

\begin{figure*}[htb!]
	\centering
	\includegraphics[width=0.45\textwidth]{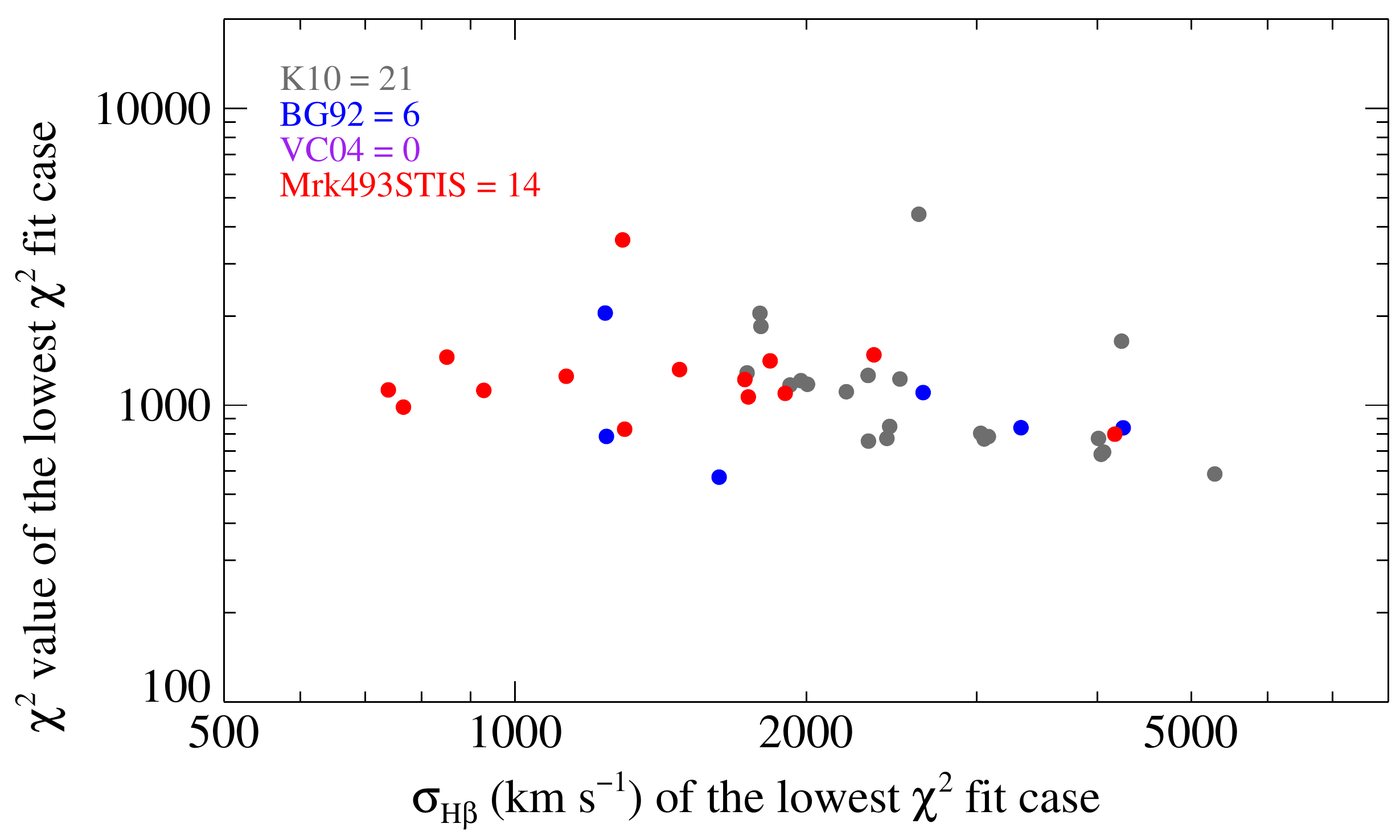} 
	\includegraphics[width=0.45\textwidth]{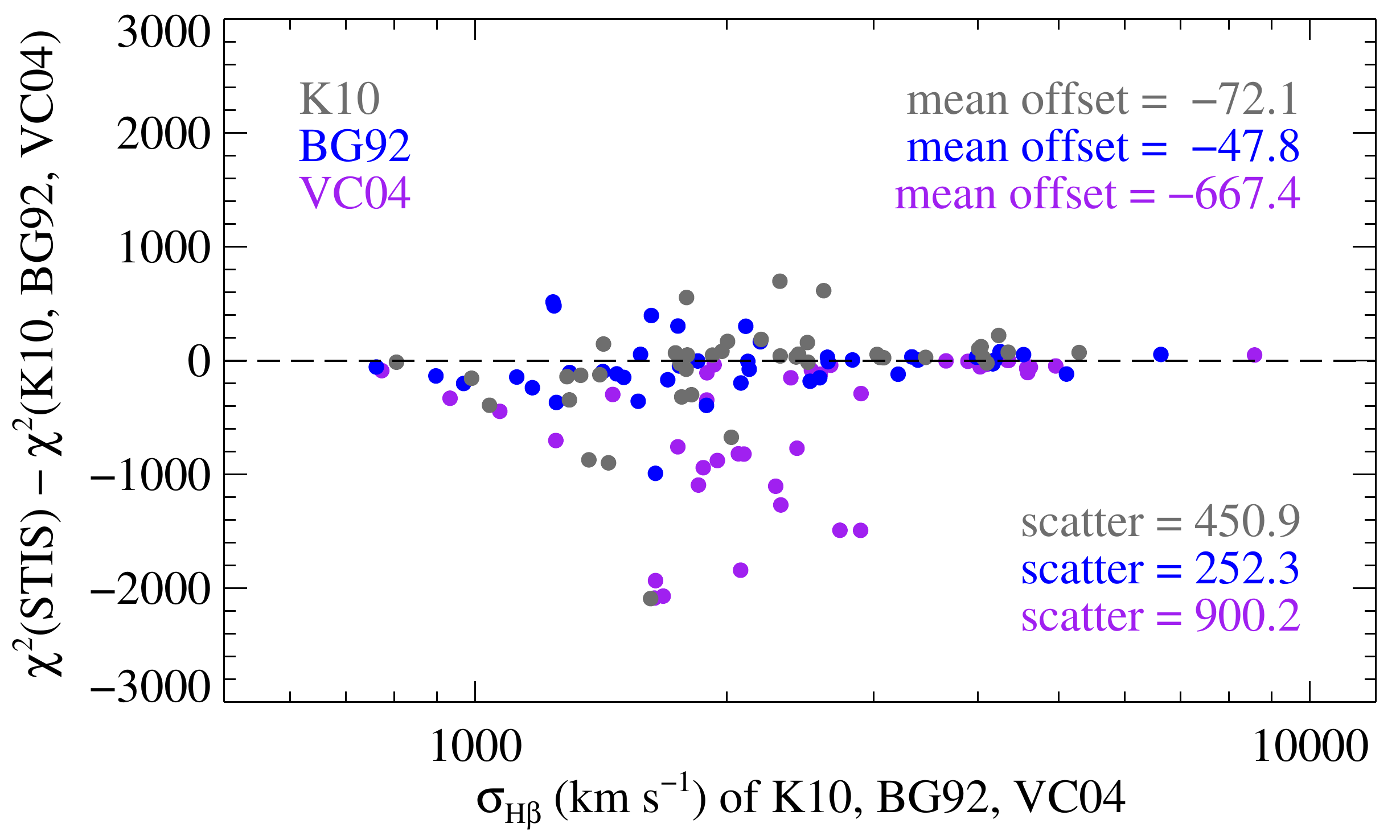}\\
	\includegraphics[width=0.45\textwidth]{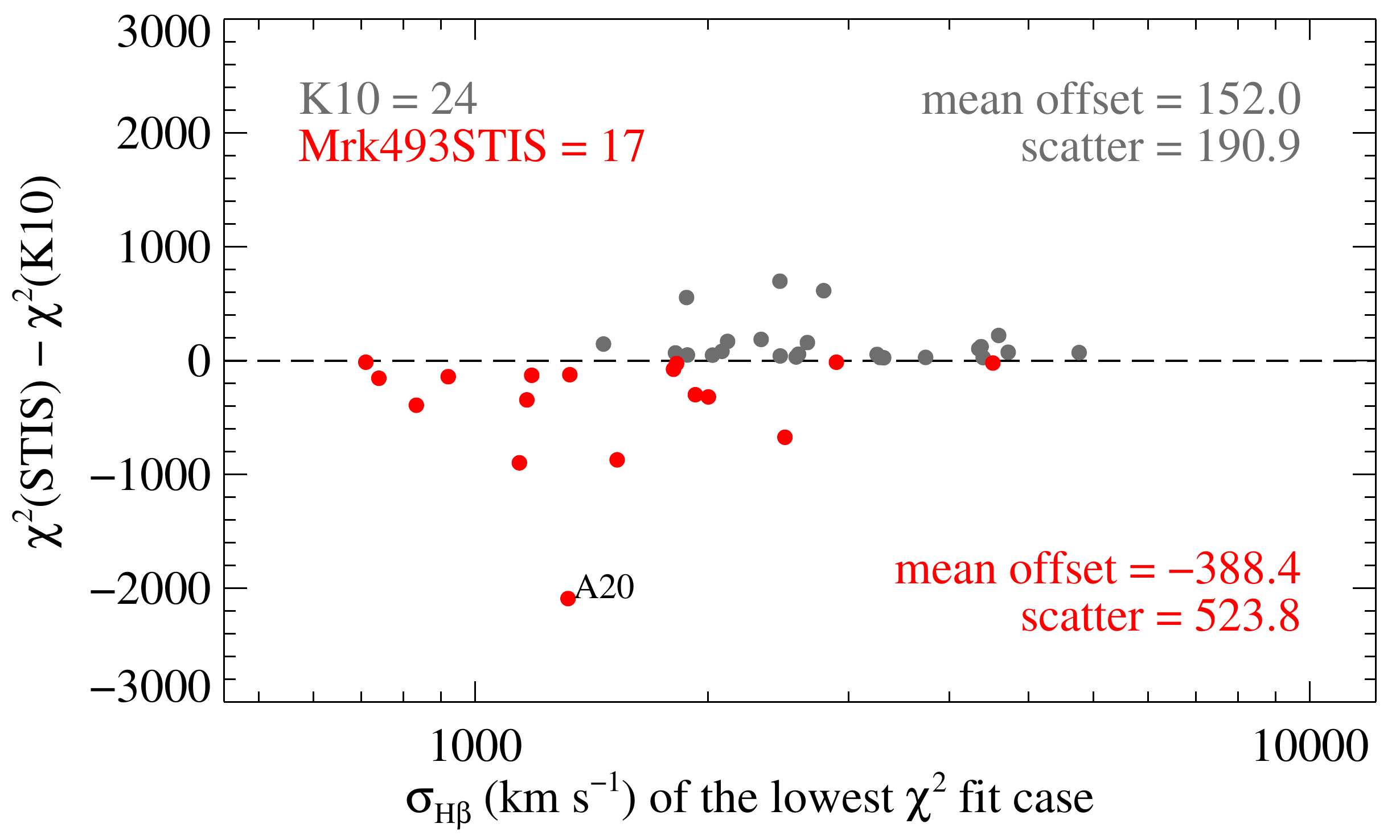}
	\caption{
		Iron template fitting tests using the sample of 41 SDSS quasars described in the text.
		Top left panel: distribution of $\chi^2$ (i.e., sum of weighted residuals) as a function of \Hb\ broad line widths ($\sigma_{\rm H\beta}$, line dispersion) of the lowest $\chi^2$ fit case among results using four different iron templates (dark gray for K10, blue for BG92, purple for VC04, red for Mrk493STIS) for each object. The symbol color of each object shows which template works best for that object. The number of objects having the lowest $\chi^2$ among the four different template fit cases is given in the upper left corner. 
		A segregation between the Mrk493STIS and K10 fit results occurs at $\sim2000$ \kms in \Hb\ broad-line dispersion.
		Top right panel: distribution of $\chi^2$ differences of the K10, BG92, VC04 fit cases against the Mrk493STIS fit as a function of $\sigma_{\rm H\beta}$. A negative value of the $\chi^2$ differences indicates that our Mrk493STIS template works statistically better in the fits, achieving smaller residuals than others. Mean offsets and $1\sigma$ scatter values of the $\chi^2$ differences are given as well.
		Bottom panel: distribution of $\chi^2$ differences between the Mrk493STIS and K10 fit cases as a function of $\sigma_{\rm H\beta}$ of the lowest $\chi^2$ fit case. Mean offset and $1\sigma$ scatter values for the positive difference (i.e., the K10 template does better) and negative difference (i.e., the Mrk493STIS template does better) objects respectively are also given in the panel.
	}
	\label{fig:SDSS_fit_chi2DistAll}
\end{figure*}

\subsubsection{Statistical performance}
In \autoref{fig:SDSS_fit_chi2DistAll}, we provide comparisons of statistical performance between the four iron templates 
based on $\chi^2$ values from the test sample fits described above.
The lowest $\chi^2$ case is selected for each object from the four different iron template fit results.
According to the numbers of objects having the lowest $\chi^2$ among the four different template fit cases 
as shown in the top left panel of \autoref{fig:SDSS_fit_chi2DistAll},
the K10 multicomponent template shows on average better performance than any other monolithic templates, achieving the lowest $\chi^2$ for 21 objects in the sample versus 6, 0, and 14 for the BG92, VC04, and Mrk493STIS templates, respectively.
Our Mrk493STIS template works best among the monolithic templates in terms of the number of objects for which it achieves the lowest $\chi^2$ (14 versus 6, 0).
We also observe a somewhat clear segregation at $\sim2000$ \kms\ between monolithic and multicomponent template results,
which indicates that the fit quality with different templates depends on the line width of the object at work.
The multicomponent semi-empirical K10 template generally performs best for objects having broader line widths, 
while the monolithic templates perform relatively better when fitting objects with narrower lines. For AGN having $\sigma$(\Hb)$\lesssim1740$ \kms, monolithic templates always perform better than the K10 template, but for larger line widths there are several objects for which the best fit is achieved by monolithic templates even though the K10 template achieves the best-quality fit for most objects in this range.

This trend can be understood as follows.
The K10 multicomponent template, having more free parameters to adjust the line fluxes across five sub-groups, naturally provides more flexibility to fit AGN exhibiting a diverse range of iron line group flux ratios. 
The monolithic templates lack this internal freedom and cannot compete with the flexibility of the K10 template.
However, the K10 semi-empirical template was constructed by assuming that all the identified iron lines (the strongest $65$ only) originate from a single system in terms of velocity broadening, and are all described by a single Gaussian line profile. 
Thus, the K10 template may be missing some fine details representing some narrow-line region origin iron lines and possibly more complex line profile shapes, which may be important to obtain a good fit for NLS1 galaxies with strong iron emission. 
Such fine details, however, would not be as important in fitting objects with broader lines, 
since they will be smeared out by convolution with a broad kernel. 
For objects with broader lines, the additional degrees of freedom associated with the K10 template are a major advantage, 
which makes the K10 template best overall for most broader-lined objects.

In the top right panel of \autoref{fig:SDSS_fit_chi2DistAll},
we show the distribution of $\chi^2$ differences between our Mrk493STIS fits and those done with other templates.
The negative mean offset values indicate that our template on average works better than the others (i.e., it results in smaller $\chi^2$ values).
The much larger mean offset size in $\chi^2$ for the VC04 template compared with the Mrk493STIS template ($\Delta \chi^2 = -667.4$) indicates that the VC04 template overall provides the least precise fits to the spectra in the SDSS sample.
The results with the BG92 template are on average most similar to those of our template, having the smallest absolute $\chi^2$ offset (47.8) and scatter (252.3). This similarity is unsurprising since both the BG92 and Mrk493STIS template include narrow-line contributions, and both are monolithic templates.

The lower panel of \autoref{fig:SDSS_fit_chi2DistAll} presents a one-to-one comparison between the Mrk493STIS and K10 templates, illustrating which of the two templates achieves a lower $\chi^2$ for each object in the sample.
The absolute size of the mean offset (388.4) for those having better fits with our Mrk493STIS template (red symbols)
is larger than that (152.0) for those with the K10 template (dark gray symbols), which means that in objects for which the Mrk493STIS template provides a better fit, it outperforms the K10 template by a somewhat larger margin than the degree to which the K10 template does better for the objects where it provides the better fit.
Although the K10 template works slightly better than ours on average in terms of the overall number of best-fitted objects (24 versus 17),
our template shows on average a much larger fit improvement (i.e., smaller fit residuals) for AGN having narrower line widths,
compared with the fit improvement achieved by the K10 template in the broader line width range.

\begin{figure*}[ht!]
	\centering
	\includegraphics[width=0.47\textwidth]{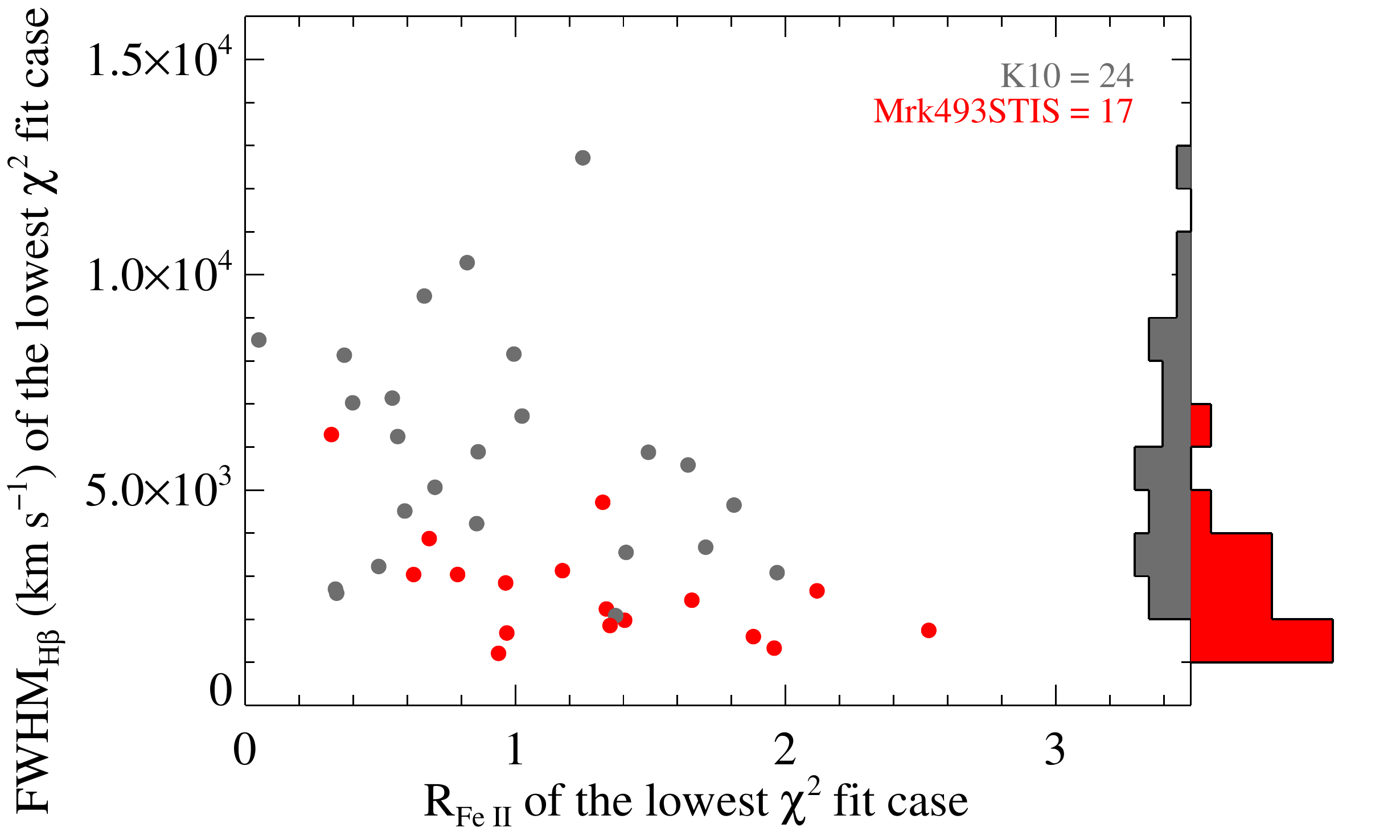}
	\includegraphics[width=0.47\textwidth]{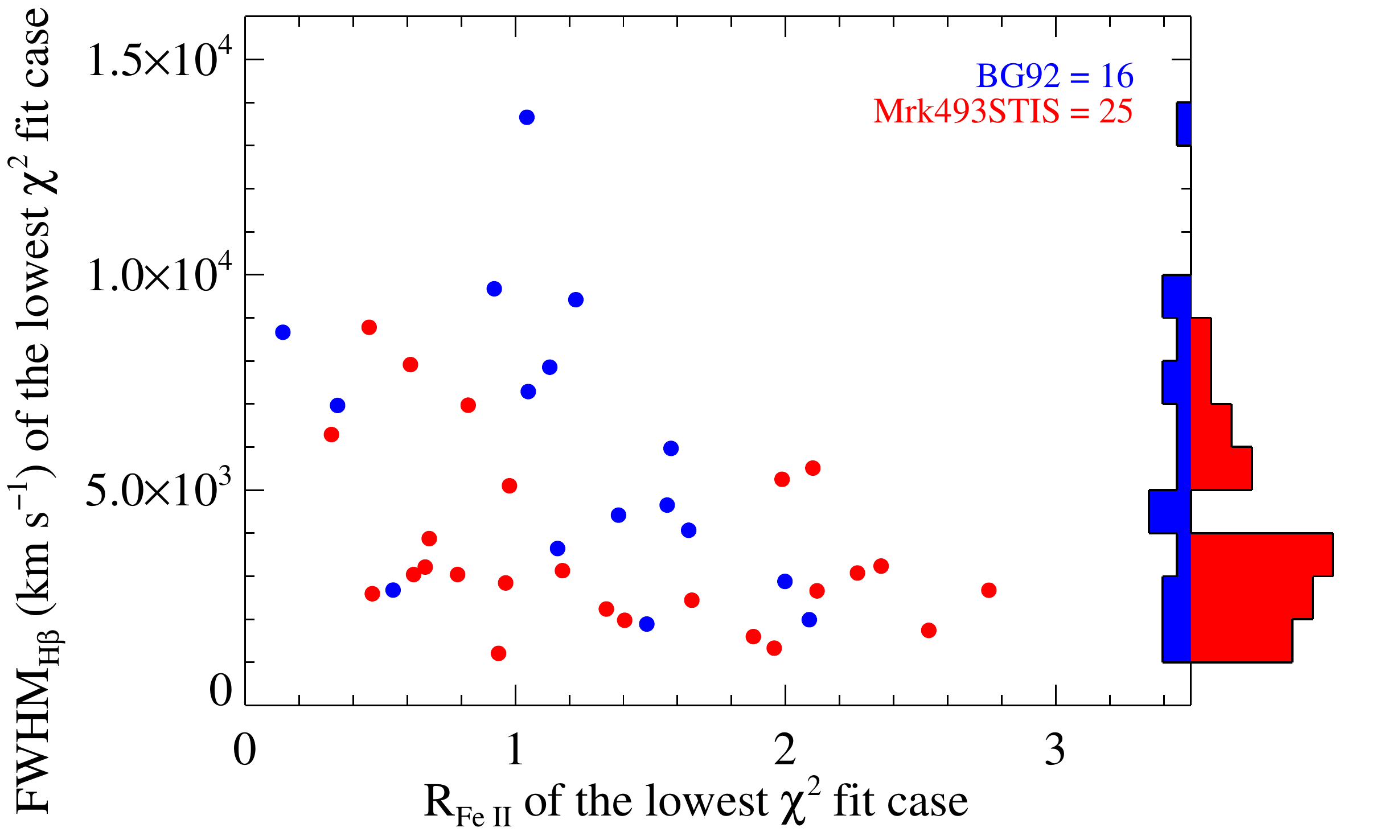}\\
	\includegraphics[width=0.47\textwidth]{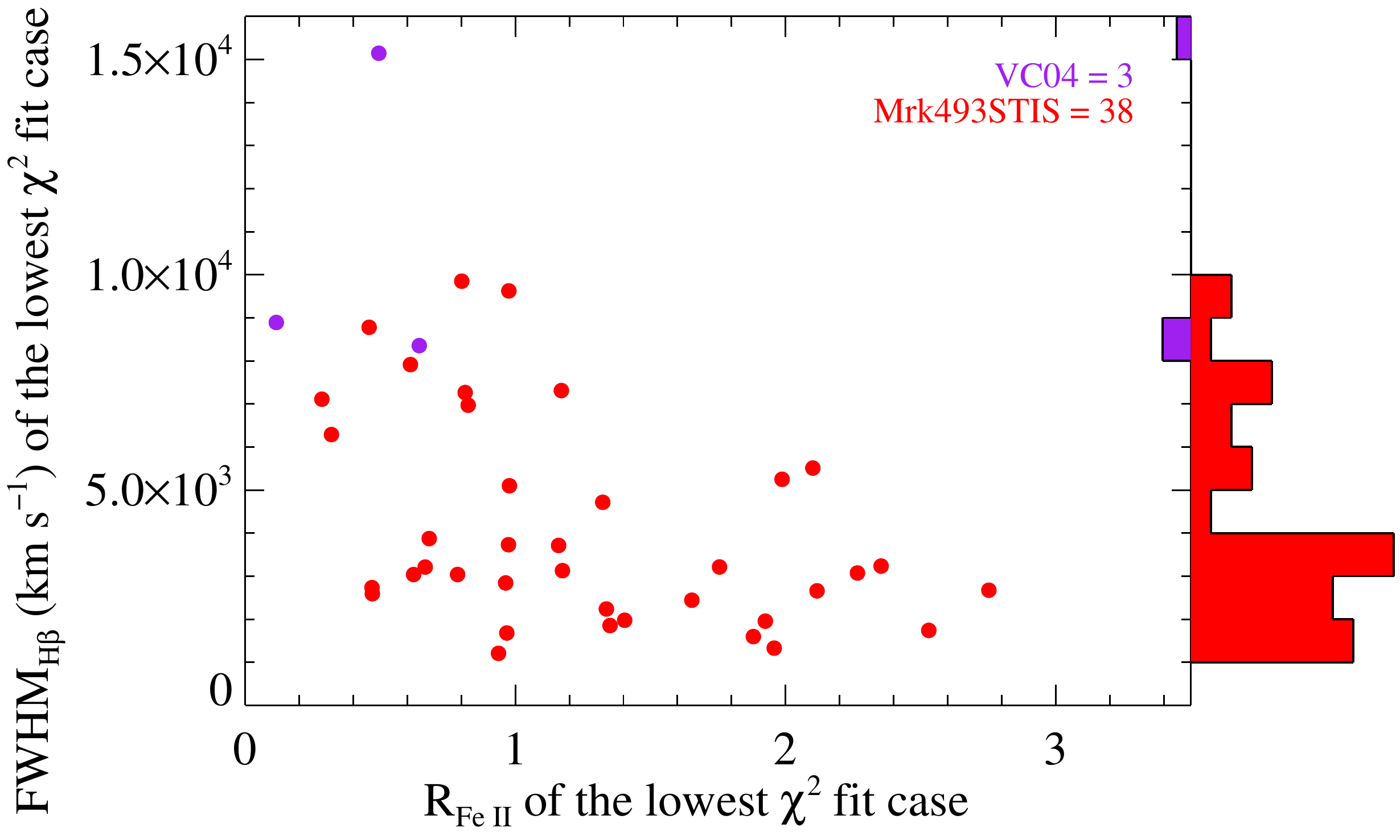}
	\caption{
		Distribution of resulting properties in the EV1 plane between each two-case template combination of	Mrk493STIS vs. K10 (top left), Mrk493STIS vs. BG92 (top right), and Mrk493STIS vs. VC04 (bottom).
		In each panel, the color indicates the lowest $\chi^2$ fit template case between the two results for each object, 
		with the corresponding total number of objects and histogram plots in the right side.
		For FWHM(broad \Hb) below $\sim4000$ \kms, our Mrk493STIS template works statistically better than any other templates. Values of $R_\mathrm{Fe~II}$ and FWHM in these plots are from the best-fitting model in each comparison, and differ from the DR7 catalog values.
	}
	\label{fig:SDSS_fit_chi2DistEV1plane}
\end{figure*}

\autoref{fig:SDSS_fit_chi2DistEV1plane} compares the distribution of resulting properties of the objects in the EV1 plane 
for the four different template fit cases. Each panel presents a one-on-one comparison of the Mrk493STIS template with each of the other templates, illustrating which template provides the better fit.
Comparing Mrk493STIS against the K10 template, we find that a similar segregation occurs at FWHM(\Hb)$\sim4000$ \kms\
in the distribution between our template and the K10 template fits,
indicating that the Mrk493STIS template performs better on average for narrower-lined objects.
In comparison with the other monolithic templates (BG92 and VC04), 
our template outperforms the others over most of the EV1 parameter space.

It is worth noting that the results of all of these comparison tests are specific to the sample used for the tests, and any statistical inferences derived from these tests would be subject to sample selection effects for our test sample. Our goal was to select the sample uniformly over the EV1 plane. As a result, our sample spans essentially the full range of \ion{Fe}{2} emission properties of AGN, but it does not represent the statistical distribution of different \ion{Fe}{2} properties across the AGN population, since our selection only chooses two objects from each cell in the EV1 grid space. Additional biases could result from our selection of the spectra with the highest S/N in each grid cell.

\begin{figure*}[ht!]
	\centering
	\includegraphics[width=0.24\textwidth]{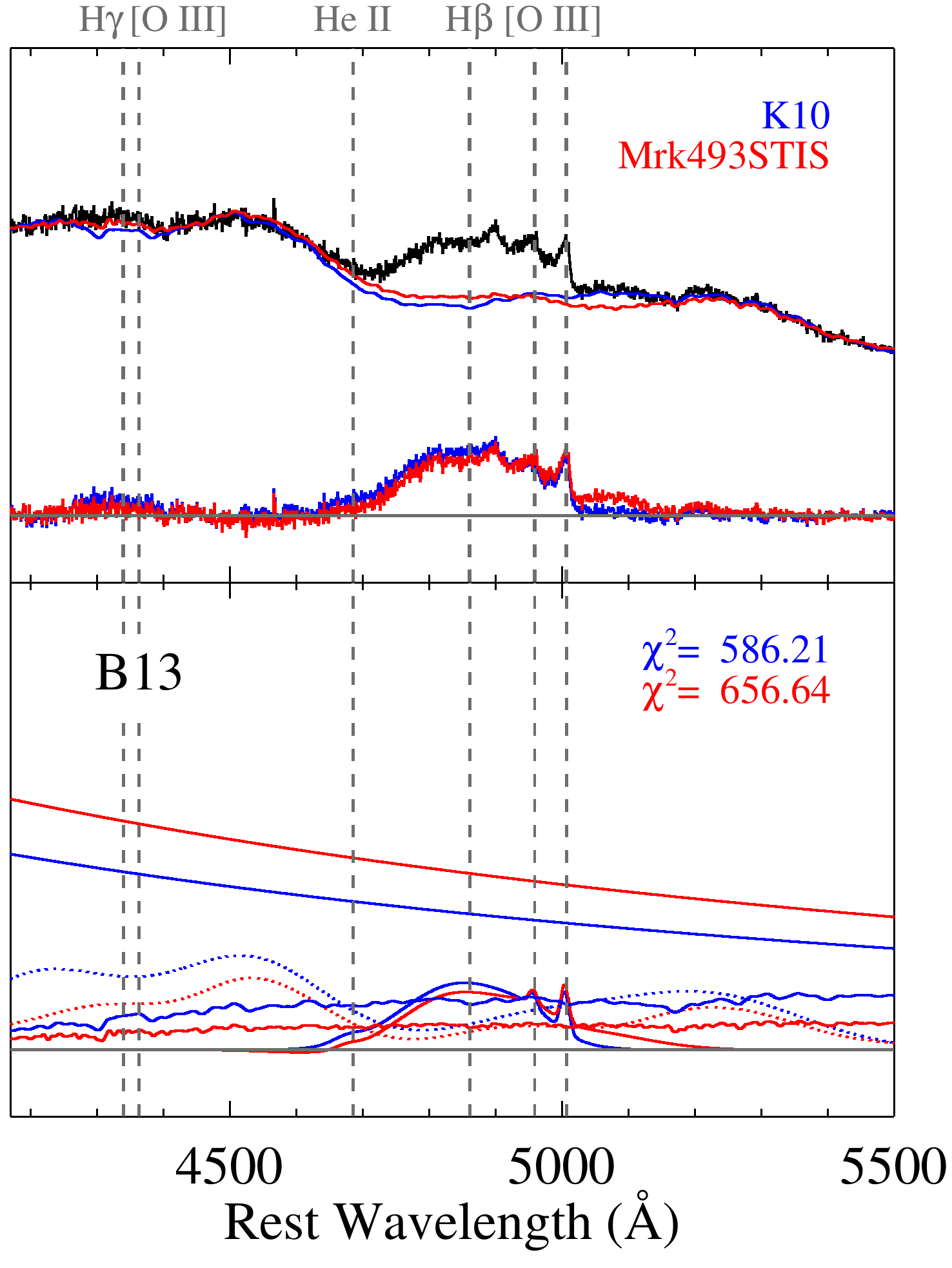}
	\includegraphics[width=0.24\textwidth]{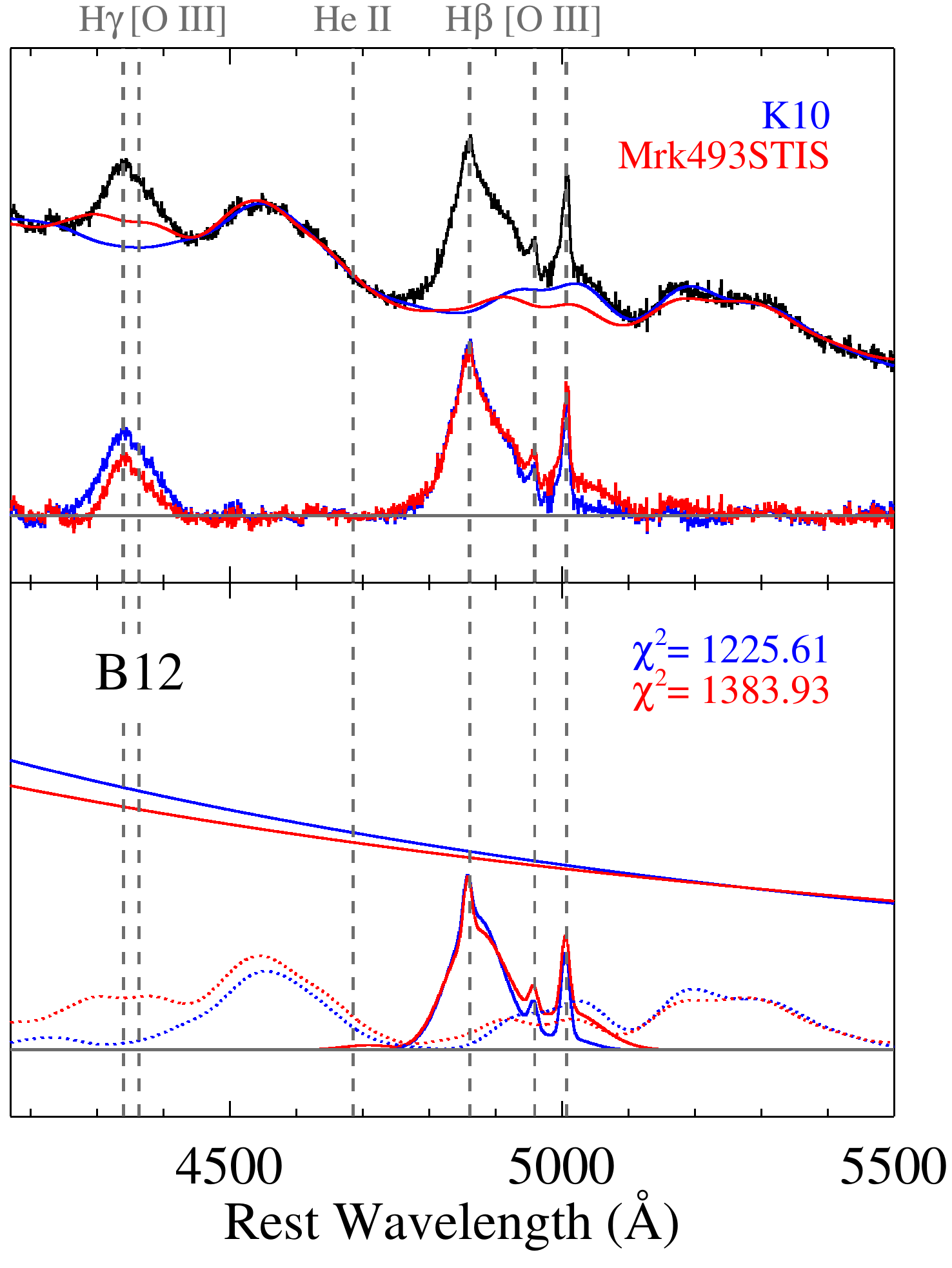}
	\includegraphics[width=0.24\textwidth]{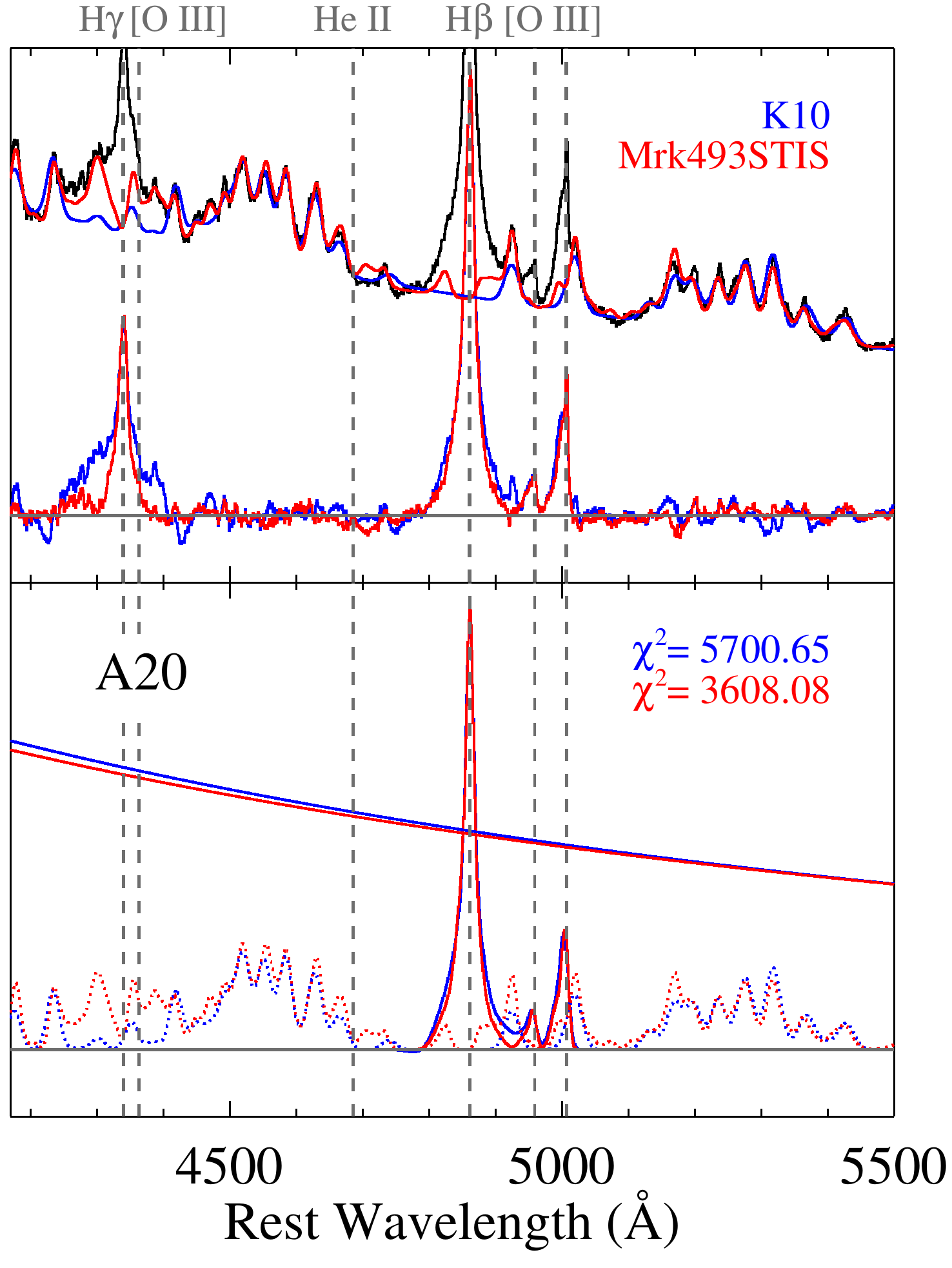}
	\includegraphics[width=0.24\textwidth]{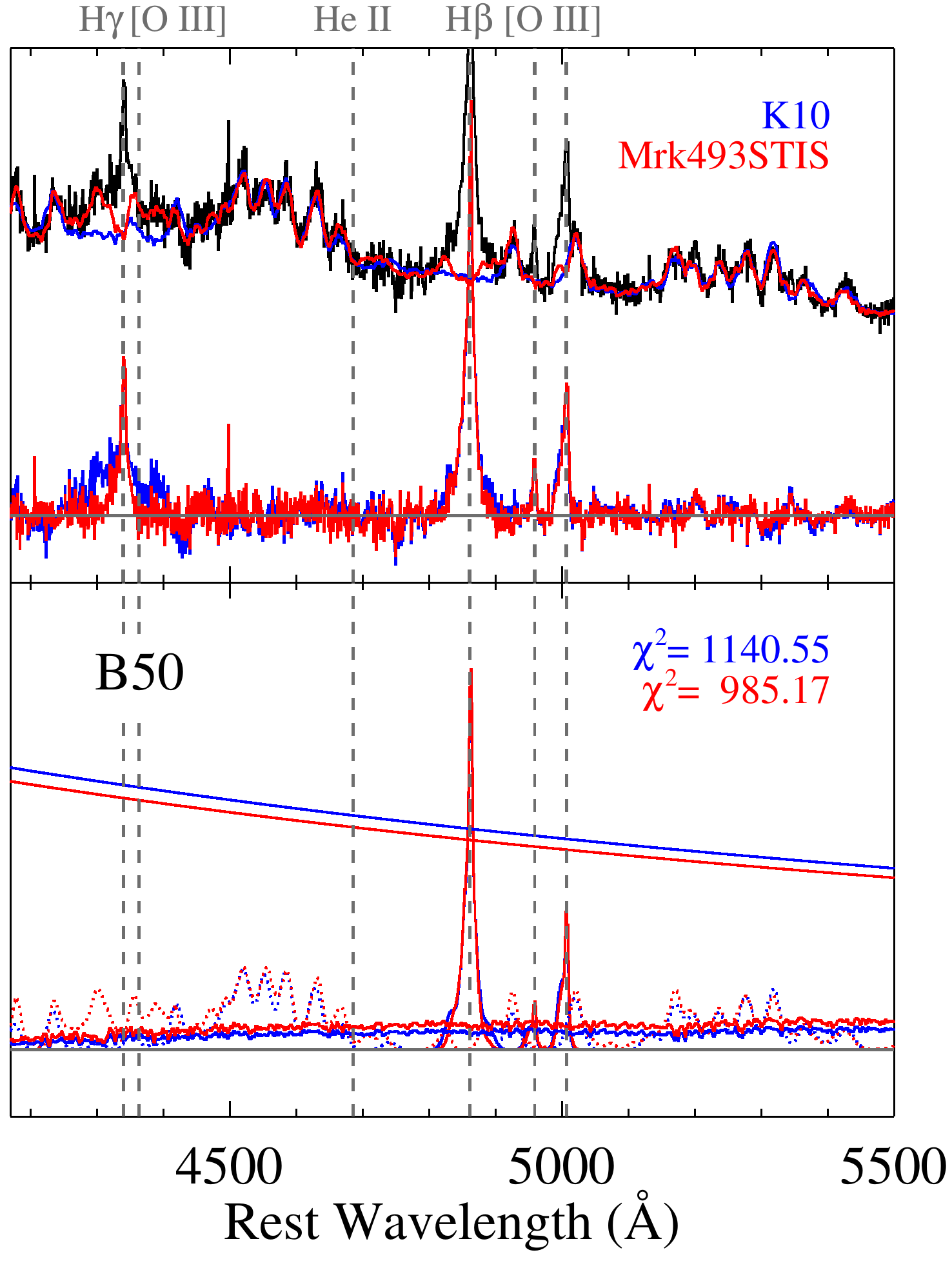}
	\caption{
		Multicomponent spectral fitting results using the two iron templates (Mrk493STIS and K10) respectively for four representative test objects with velocity widths ranging from broad to narrow (left to right panels).
		The observed spectrum (black) and best-fit models (blue for K10 and red for Mrk493STIS) with the resulting $\chi^2$ values and object IDs are shown in each panel.
		Upper panels show
		the best-fit pseudocontinuum model, consisting of power-law continuum + iron emission + host galaxy starlight (if any), overplotted on the data. 
		Below this is the residual emission-line spectrum after subtraction of the pseudocontinuum model from the data.
		Lower panels show each best-fit model component, including solid lines for the power-law continuum, emission lines, and host galaxy templates, and dotted lines for the iron templates.
	}
	\label{fig:SDSS_fit_decomp}
\end{figure*}

\begin{figure*}[ht!]
	\centering
	\includegraphics[width=0.24\textwidth]{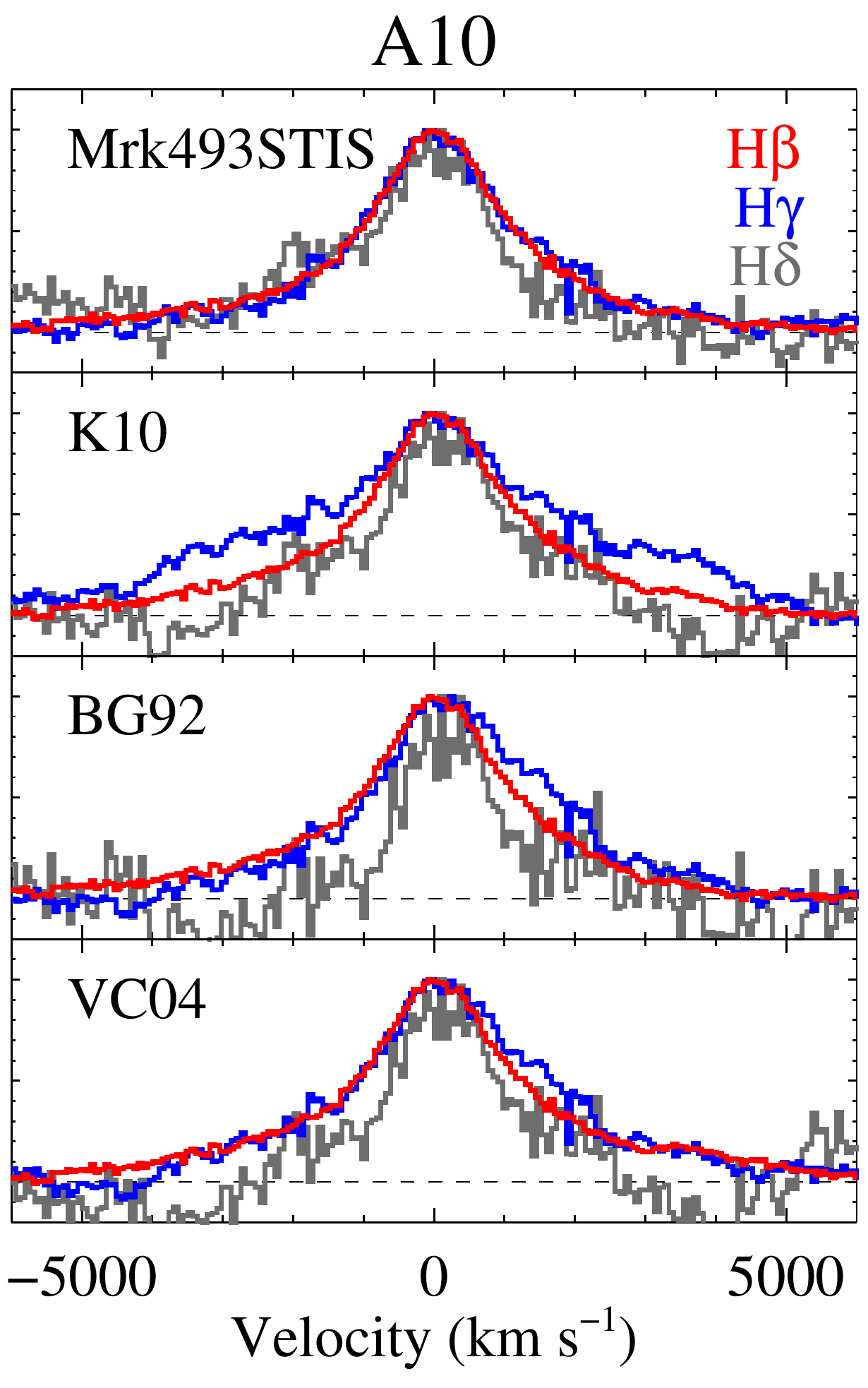}
	\includegraphics[width=0.24\textwidth]{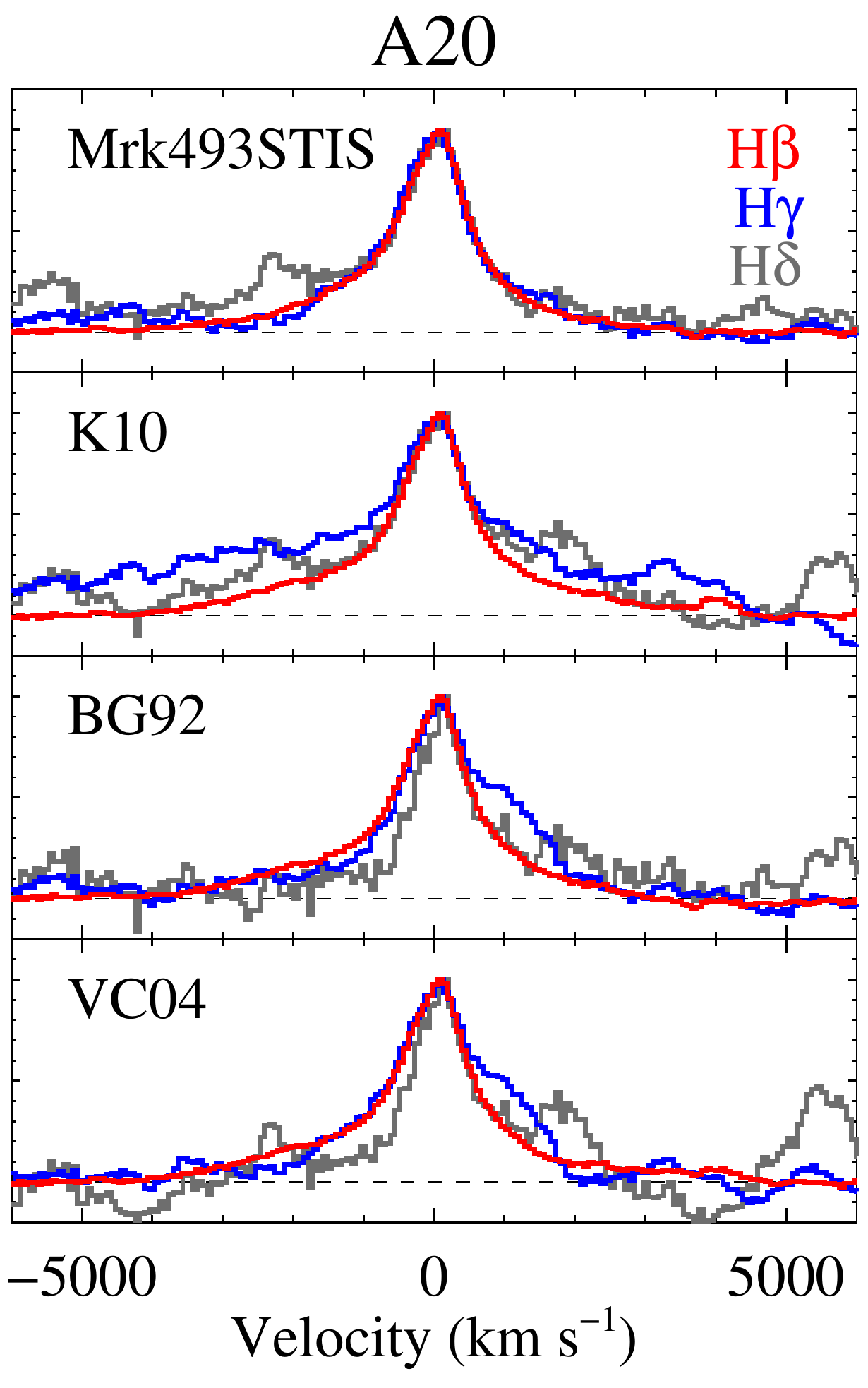}
	\includegraphics[width=0.24\textwidth]{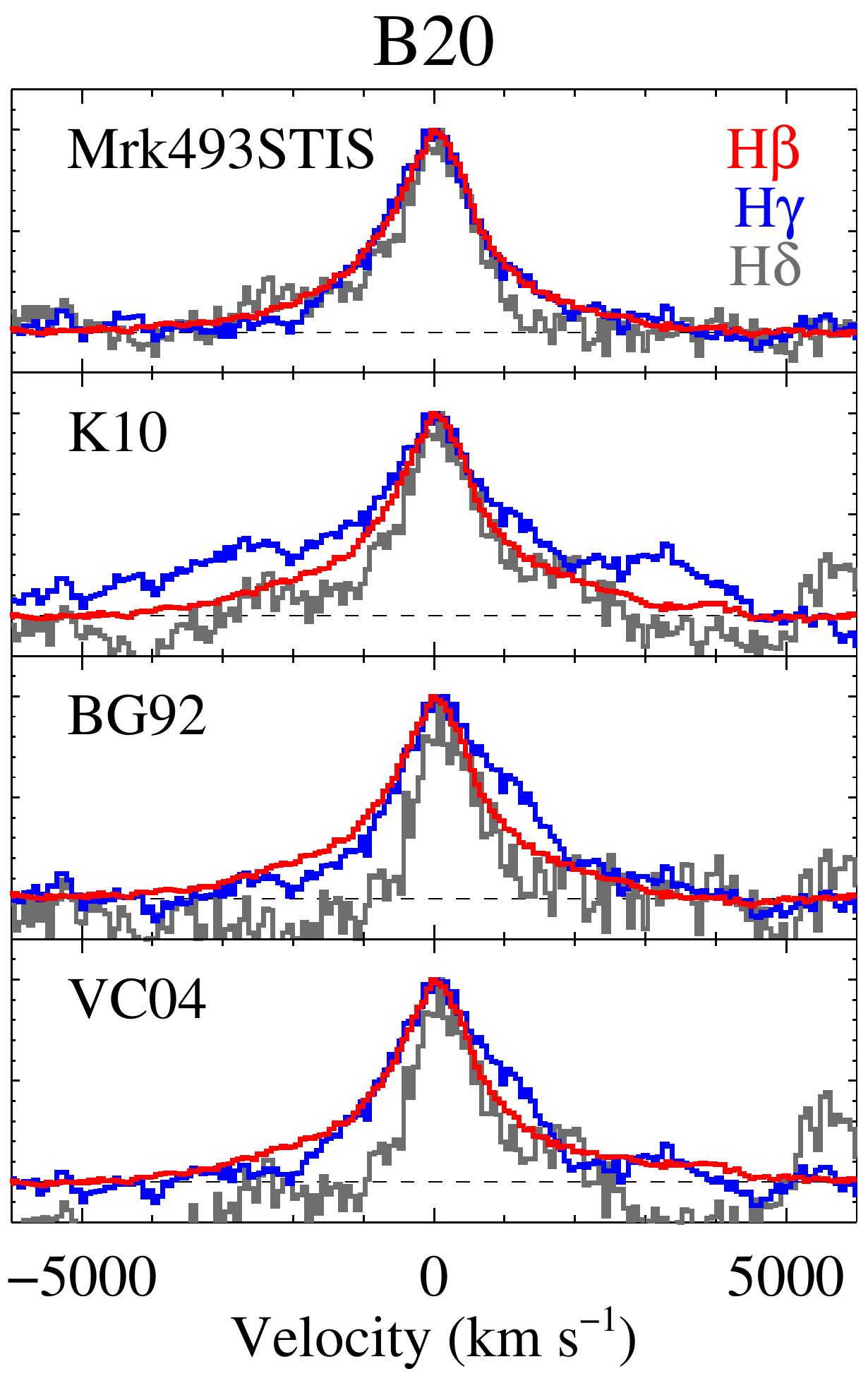}
	\includegraphics[width=0.24\textwidth]{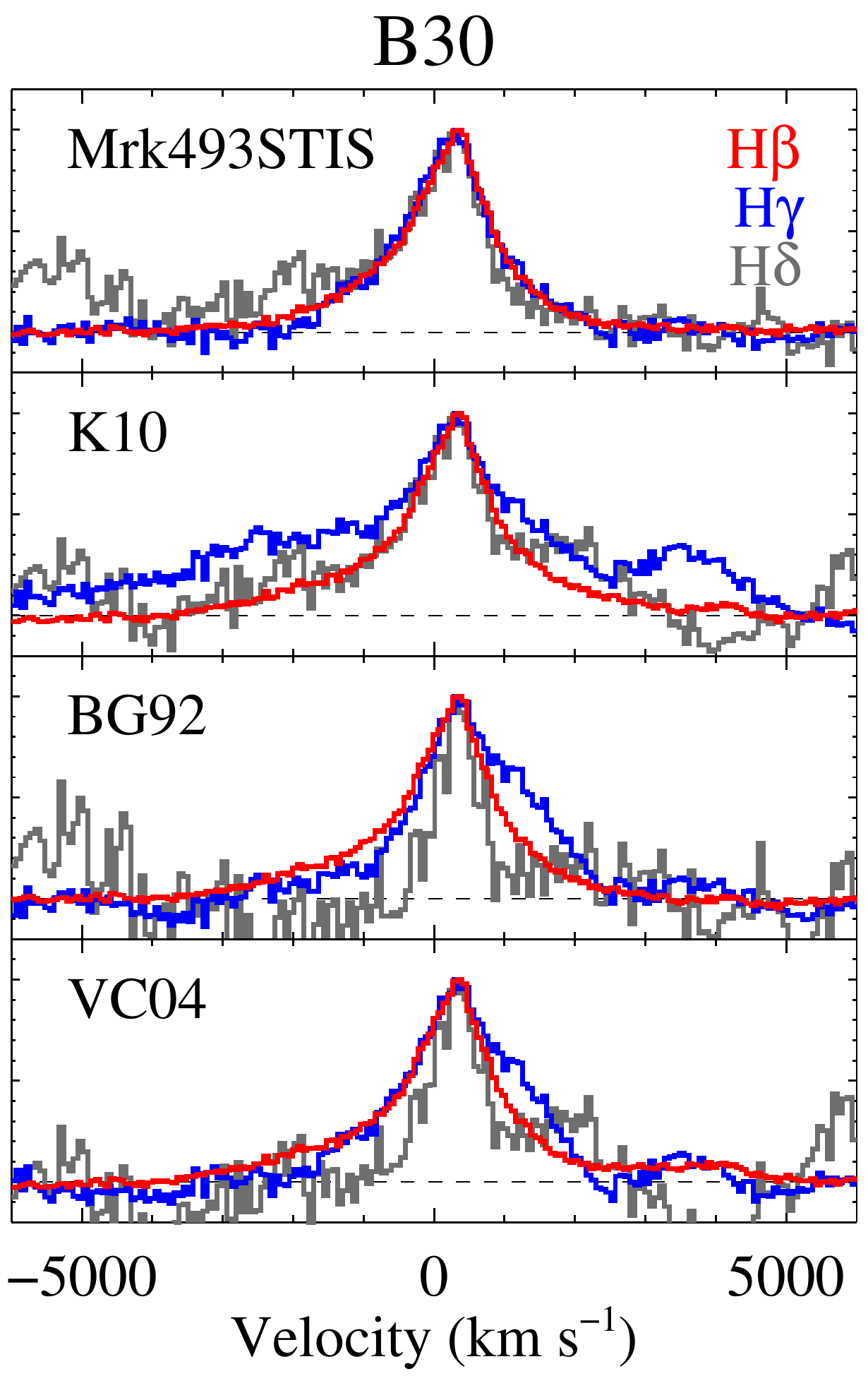}
	\caption{
		Comparison of the resulting Balmer line (\Hb, \Hg, \Hd) profiles as a function of line-of-sight velocity, after subtracting all the other model components from the data, for each template fit. The object ID is given in the top of each panel.
		Each Balmer line flux is normalized by its maximum value to facilitate comparison of profiles across the line series. 
		The \Hg\ profiles obtained with the K10 template show significant differences with other Balmer line profiles, 
		while the three Balmer line profiles from our Mrk493STIS template fits are most consistent with each other.
	}
	\label{fig:SDSS_fit_BalmerProfile}
\end{figure*}

\subsubsection{Spectral decomposition and Balmer line profiles}
\autoref{fig:SDSS_fit_decomp} shows direct comparisons of best-fit models using the two best overall templates (Mrk493STIS and K10) for several objects ranging from very broad to very narrow line widths.
As described above, the K10 multicomponent template provides lower fit residuals (i.e., smaller $\chi^2$) than the Mrk493STIS template for objects with broader lines,
while our Mrk493STIS template works better than the K10 template for the narrower-lined objects.

The flexibility of the multicomponent template is illustrated particularly well in the fit to object \texttt{B12}. This broad-lined AGN has a strong extended \Hb\ ``red shelf'' \citep{Veron+2002}, seen as the very strong line wing extending redward beyond the [\ion{O}{3}] $\lambda5007$ line. With the K10 template, this feature is well fitted by independently increasing the iron flux from one line group  (group S consisting of five lines from \FeII\ multiplets 41, 42 and 43; see K10 for a detailed description), while the Mrk493STIS template fit exhibits a lower flux in this region that fails to match the red shelf.
However, the origin of the flux in the red shelf region remains ambiguous, and this region could also be fitted with the addition of
broad \ion{He}{1} $\lambda\lambda4922,5016$ lines as suggested by \citet{Veron+2002} (see also \citealt{Barth+2015}).
With these data, we are unable to determine whether the enhanced red shelf flux originates from 
\FeII,  \ion{He}{1},  \Hb\ itself, or some combination of these components. We did not include the \ion{He}{1} lines in our fits in order to avoid intractable degeneracy between \ion{He}{1} and \ion{Fe}{2} lines over this region. When fitting models to this spectral region using monolithic \ion{Fe}{2} templates, adding the \ion{He}{1} lines provides additional modeling flexibility that can often significantly improve the fitting results, although the interpretation of the enhanced flux in this region remains unclear.

Relatively large residuals are apparent with the K10 template over almost all fitted areas in the (narrower) object \texttt{A20}. 
As noted above, this directly reflects the template shape differences, which are more pronounced in narrower-lined objects. 
There is a dramatic difference in the resulting \Hg\ line profiles as well due to the template difference between the K10 and Mrk493STIS
(as previously discussed in \autoref{sec:comparison}), where the K10 template fit leaves a (probably spurious) strong broad component residual on the \Hg\ line profile.

It is also worth noting that there are often non-negligible differences in overall flux level between the individual pseudo-continuum components for fits using different \ion{Fe}{2} templates, although the best-fit combined pseudo-continuum model is generally very similar between the Mrk493STIS and K10 model fits.
This is due to the degeneracy between the power-law continuum model, iron template, and host galaxy template over the limited wavelength range used in these fits.

\begin{figure*}[htb!]
	\centering
	\includegraphics[width=0.33\textwidth]{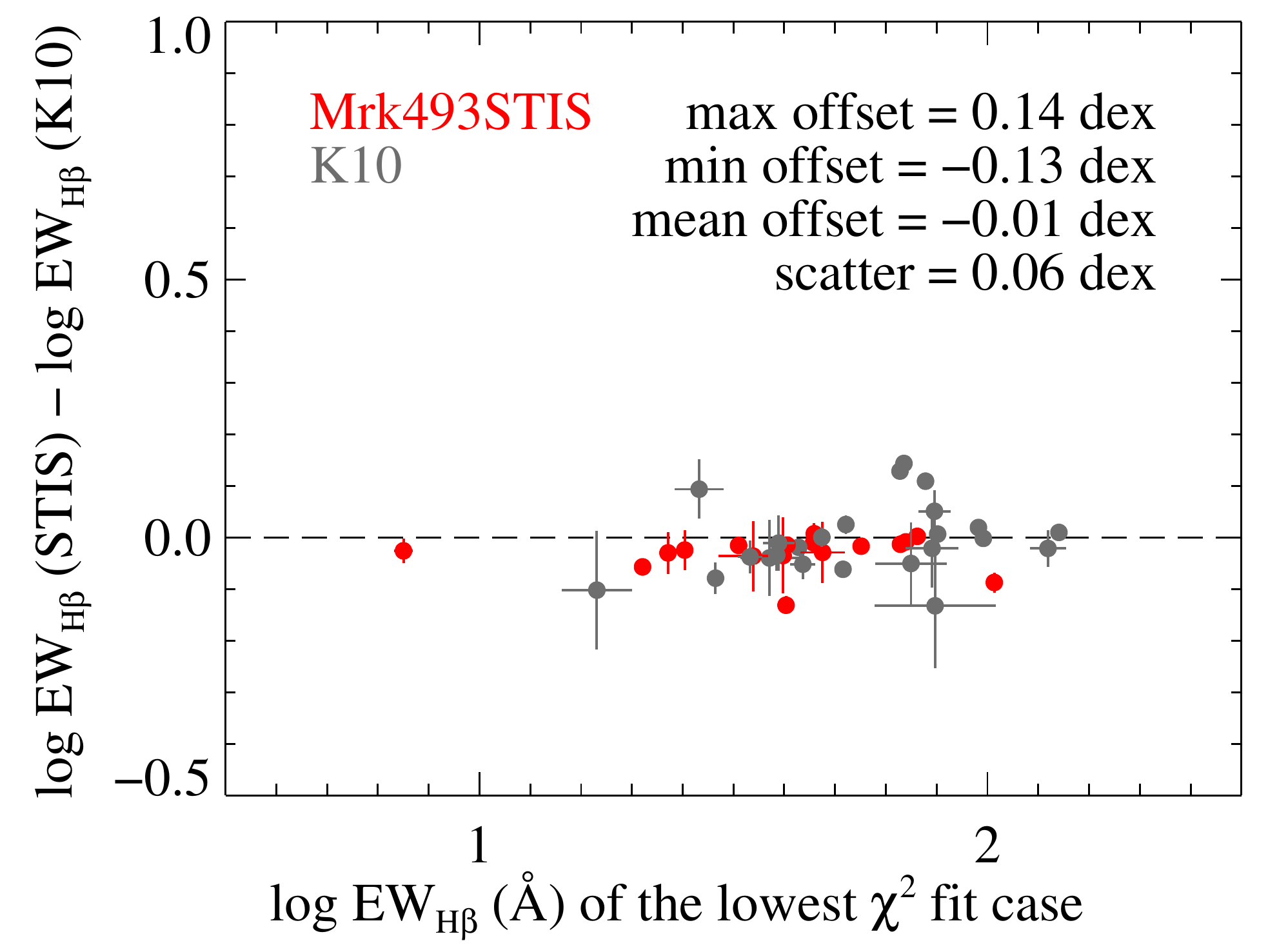}
	\includegraphics[width=0.33\textwidth]{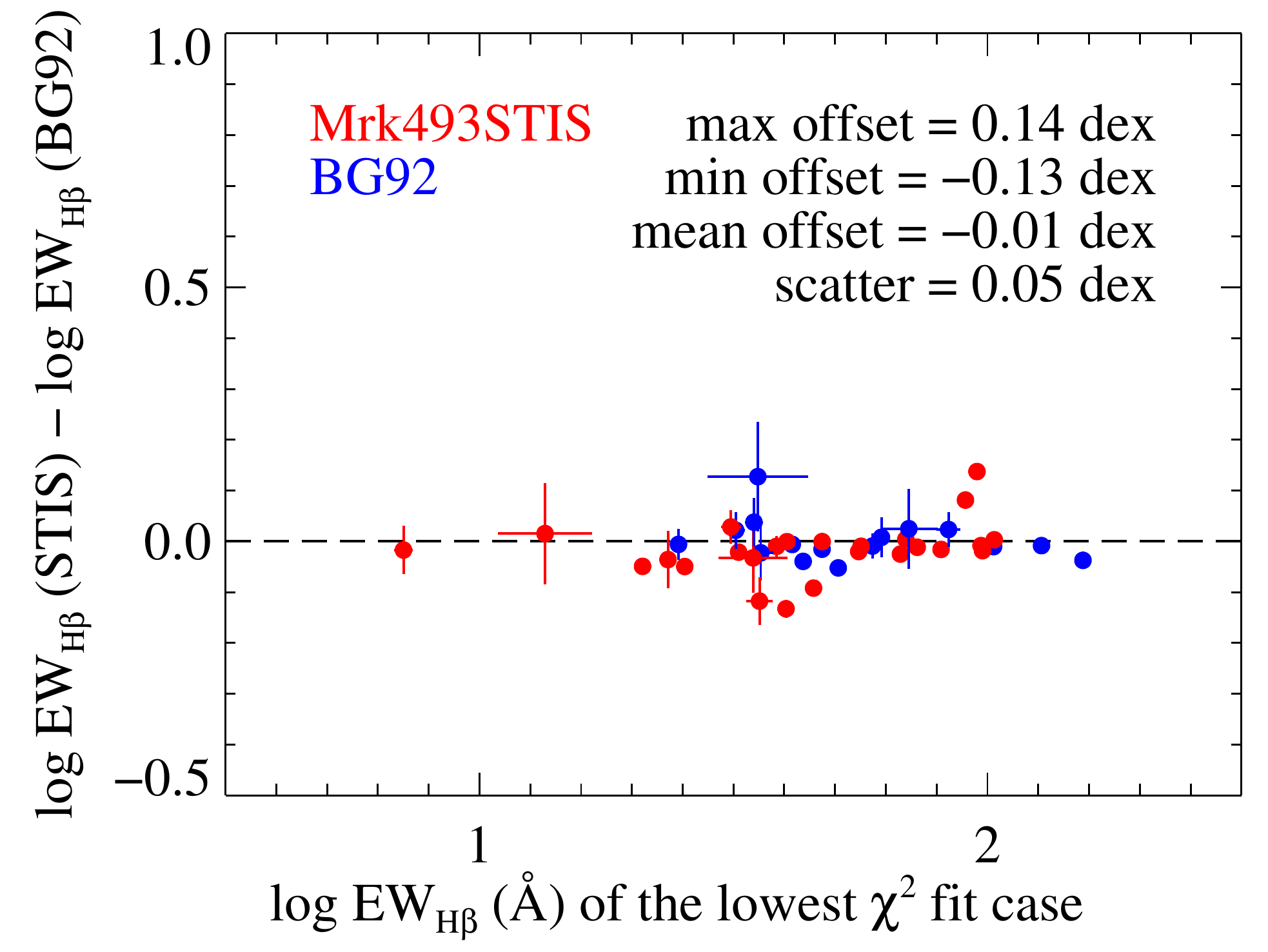}
	\includegraphics[width=0.33\textwidth]{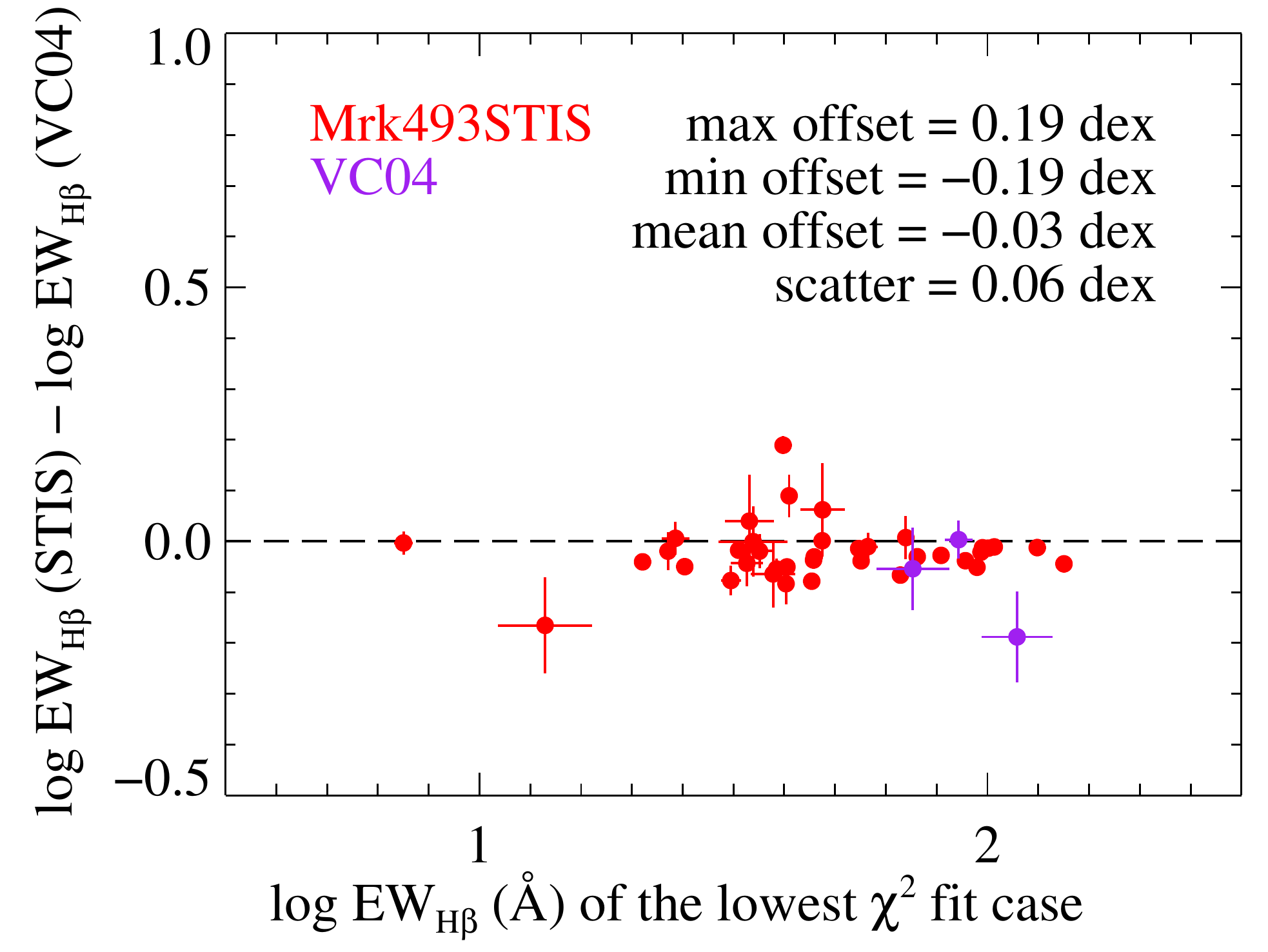} \\
	\includegraphics[width=0.33\textwidth]{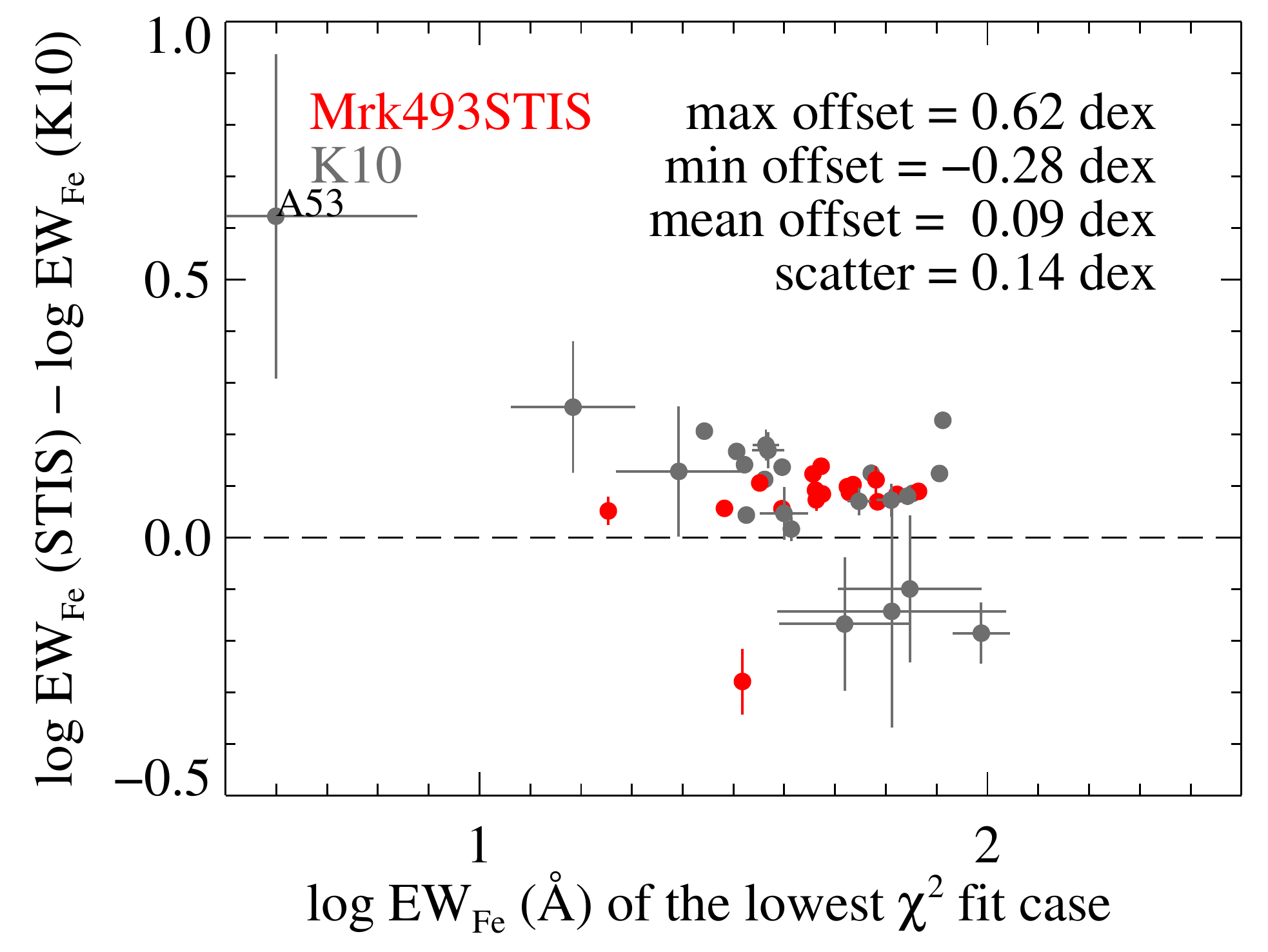}
	\includegraphics[width=0.33\textwidth]{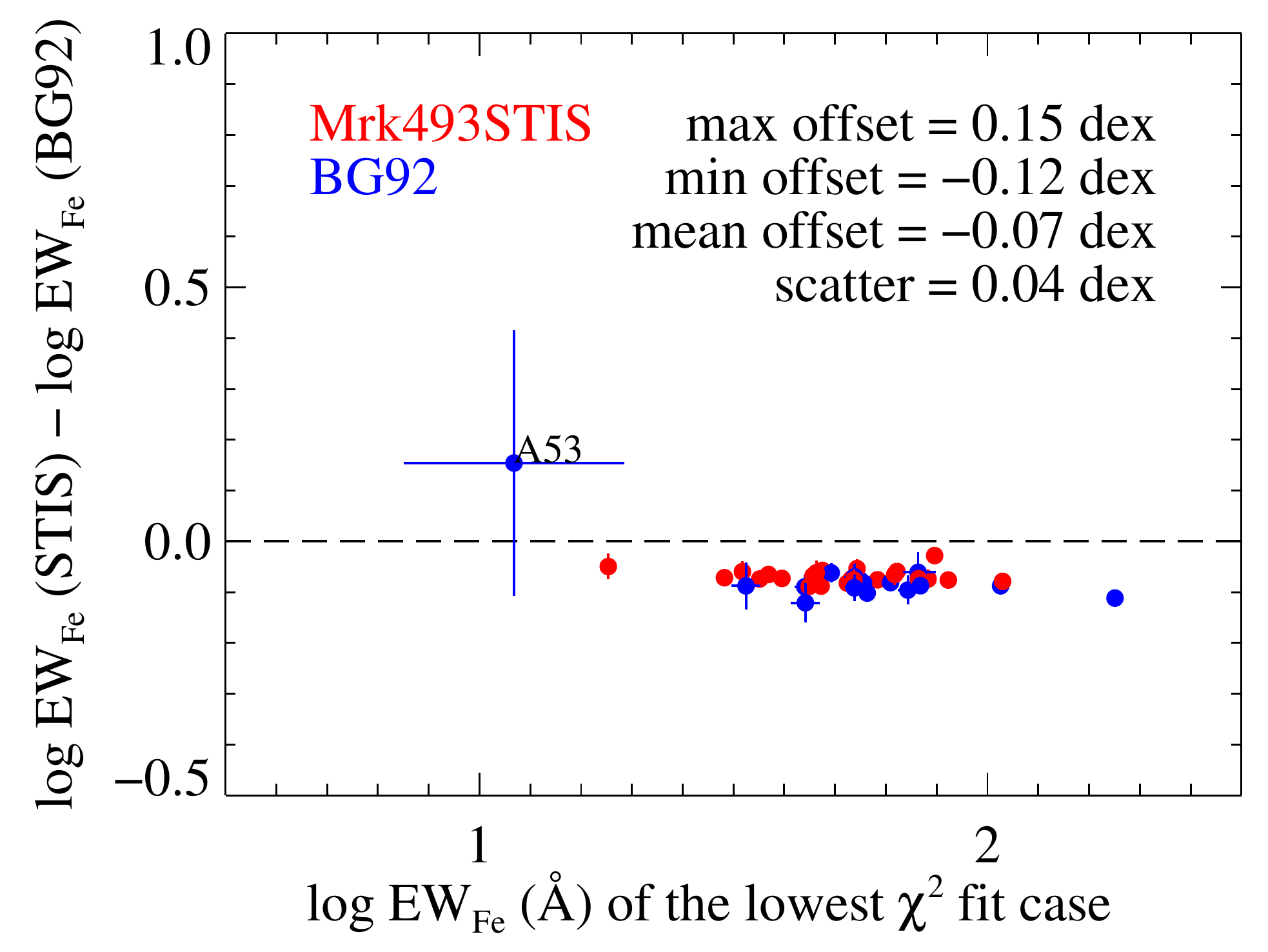}
	\includegraphics[width=0.33\textwidth]{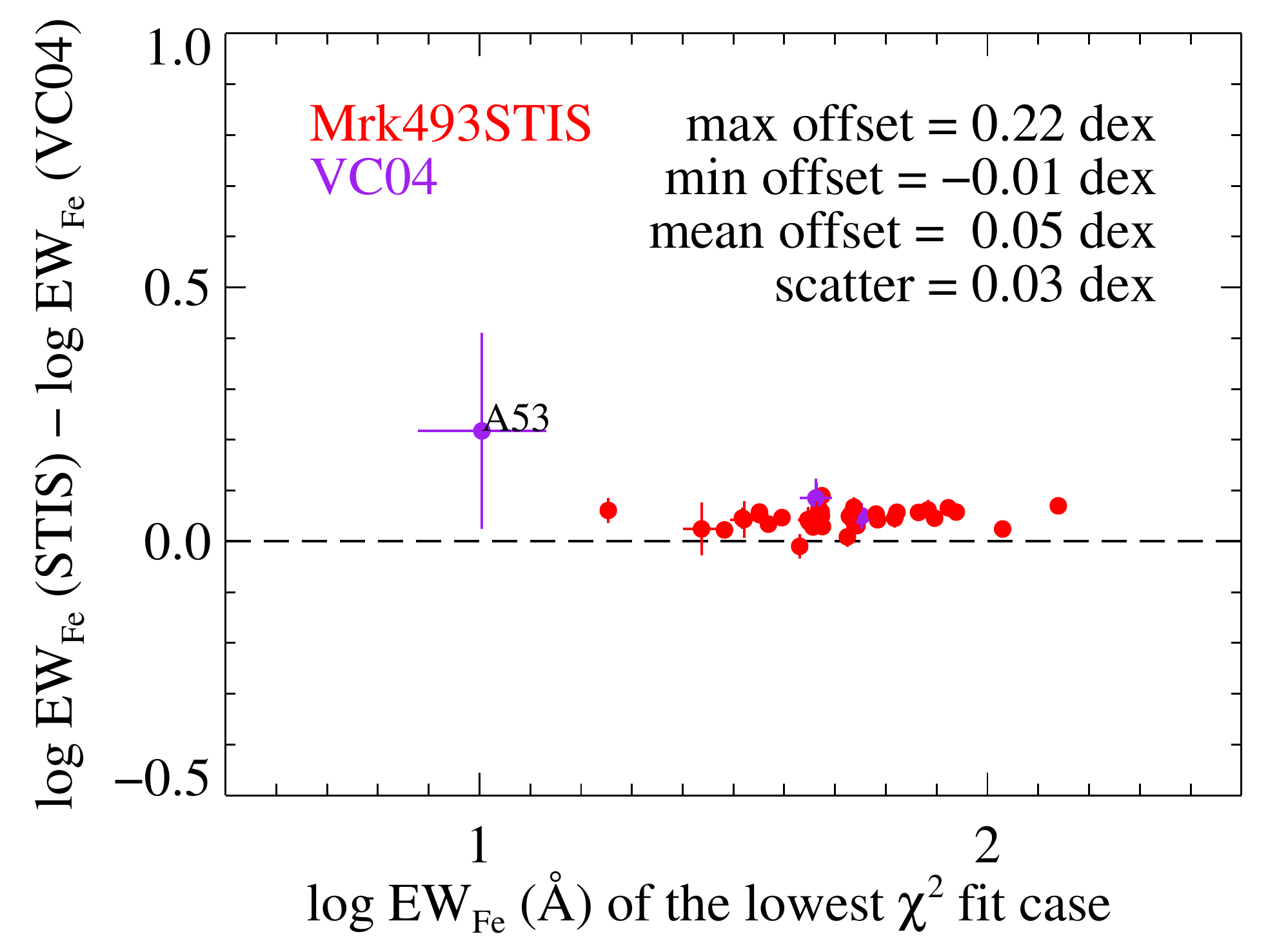} 
	\caption{
		Distribution of differences of rest-frame equivalent widths (EW) 
		for \Hb\ broad emission (upper row) and iron emission within 4434-4684 \AA\ (lower row) 
		between fits using two different templates.
		The maximum, minimum, average offsets, and $1\sigma$ scatter of differences of the two measurements are given in the upper right corner in each panel. The object IDs for significant outliers are labeled as well.
		For the y-axis labels, `STIS' is used to refer to the Mrk493STIS template.
		The adopted symbol color for each object indicates which template yielded the lower $\chi^2$ value between the two templates compared in each panel.
	}
	\label{fig:SDSS_fit_EWdiff}
\end{figure*}

\autoref{fig:SDSS_fit_BalmerProfile} compares the resulting Balmer line profiles for several objects 
highlighting large differences between the \Hg\ line profile relative to \Hb\ when fitted with the K10 template.
In the results with the K10 template, 
the \Hg\ line profile appears too broad in comparison with \Hb,
while the agreement between the Balmer line profiles appears significantly better in the fits with the Mrk493STIS template. 
This under-subtraction of flux surrounding the \Hg\ line with the K10 template is a direct reflection of the intrinsic template differences
as previously seen in Fig.~\ref{fig:compare_templatesSTIS} and discussed in \autoref{sec:comparison}.

It is also notable that the \Hd\ line profile is roughly consistent with \Hb\ in the results with the Mrk493STIS template, 
while the \Hd\ line region appears over-subtracted in the BG92 and VC04 template fits. Our fitting procedure did not model the \Hg\ and \Hd\ line profiles: these regions were just masked out during the fitting.
The over-subtracted \Hd\ line with the BG92 template is a natural outcome of the residual emission in the \Hd\ line regions in that template as illustrated in \autoref{sec:comparison}.
The unexpected over-subtraction in \Hd\ with the VC04 template is due to the much weaker iron features in the $4170-4260$ \AA\ range 
for that template, which consequently force the power-law continuum level in the fits to be higher than the actual continuum level
(see also \autoref{AppFig:allSDSS_fits1} for illustrations of this same tendency in many objects). 

This comparison provides good evidence that our new template performs better than others for recovering accurate profiles of the Balmer lines from multicomponent fits to AGN spectra.
These results highlight the crucial importance of the completeness and accuracy of a template construction method (in terms of line identification and removal of continuum components),   
because if some unnecessary spectral components remain in the template or some relevant components are missing, model fits using the template can significantly bias the inferred fluxes and profiles of various lines of interest when applied.

\begin{figure*}[!htb]
	\centering
	
	\includegraphics[width=0.33\textwidth]{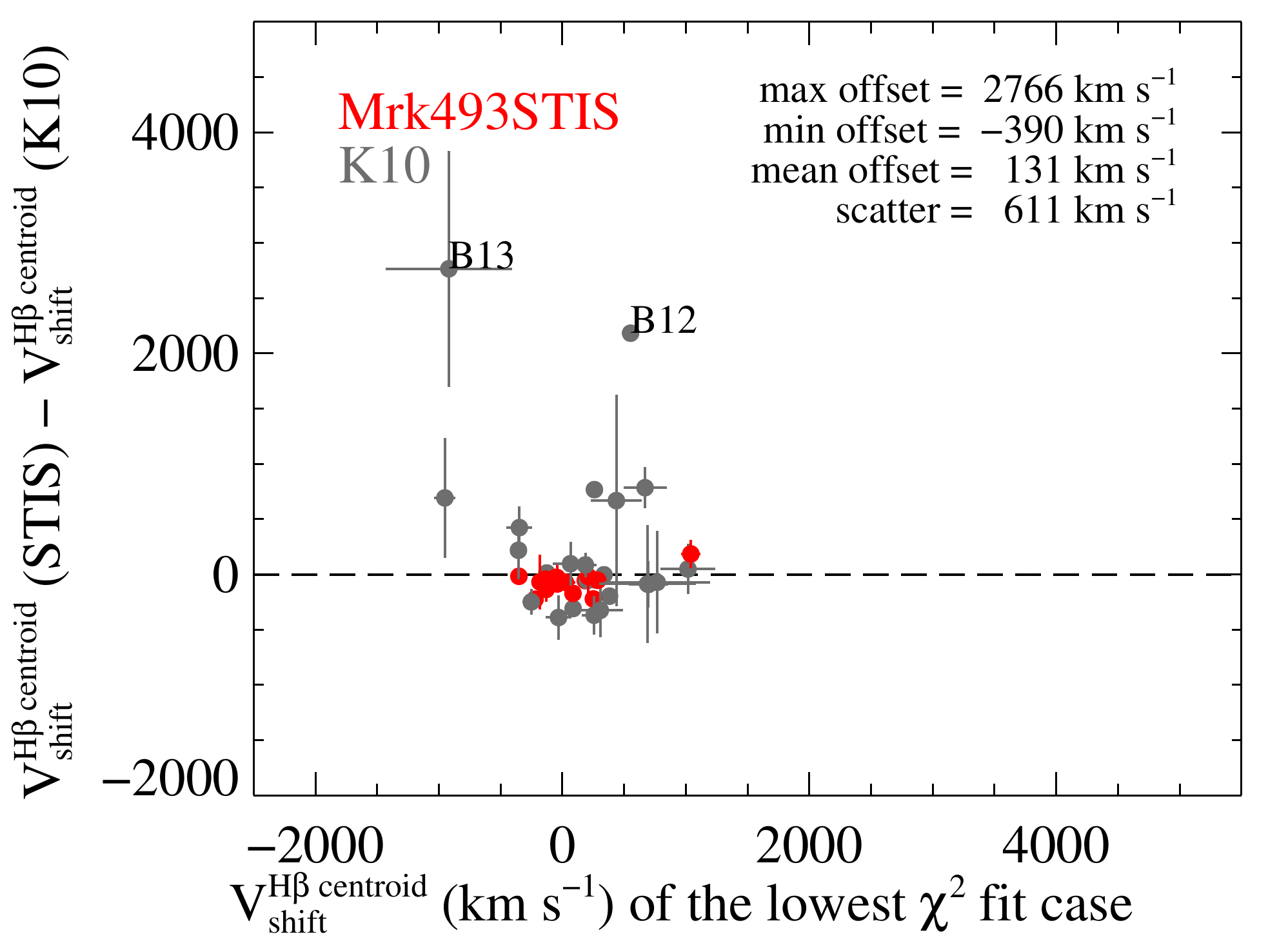}
	\includegraphics[width=0.33\textwidth]{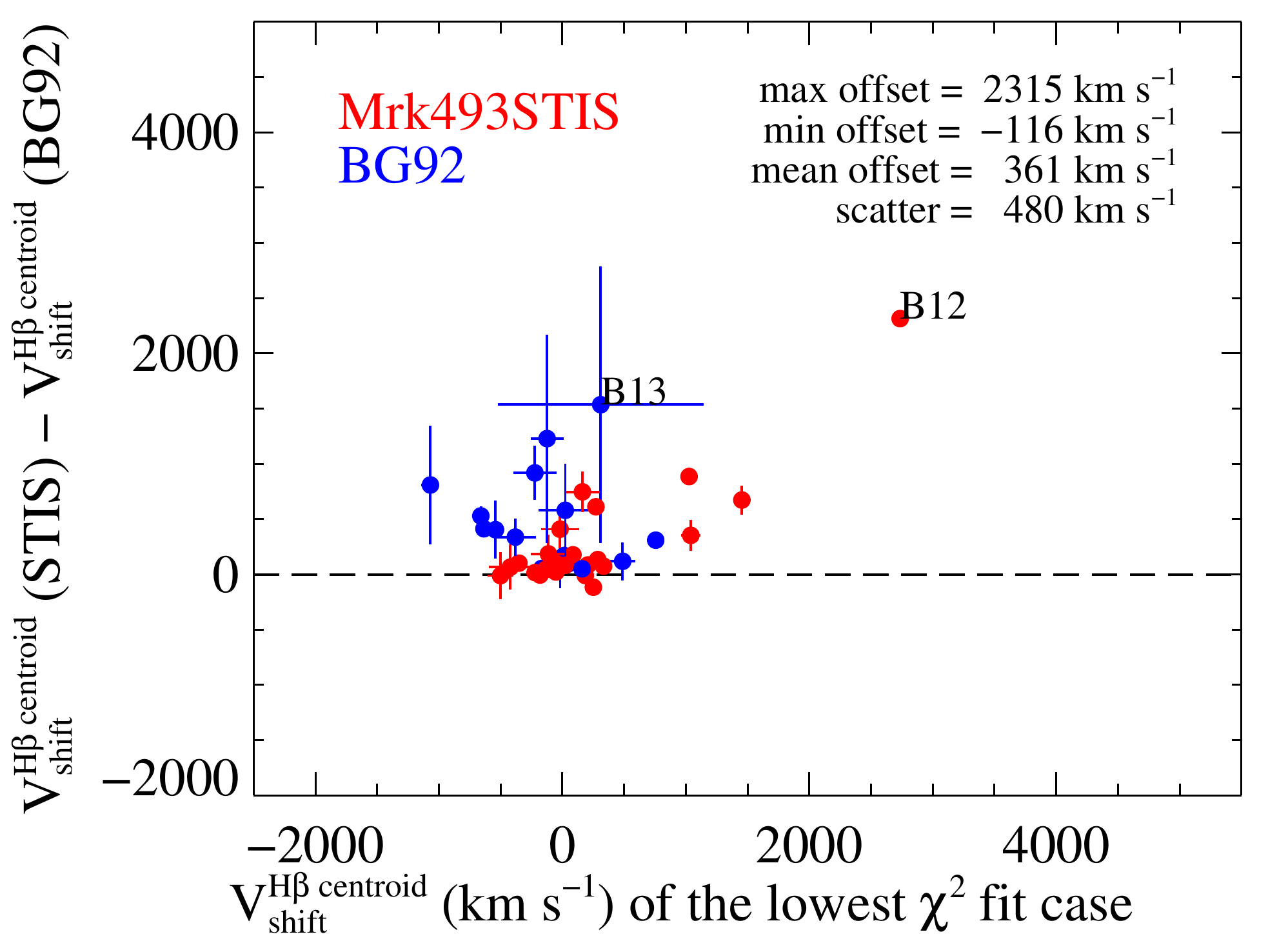}
	\includegraphics[width=0.33\textwidth]{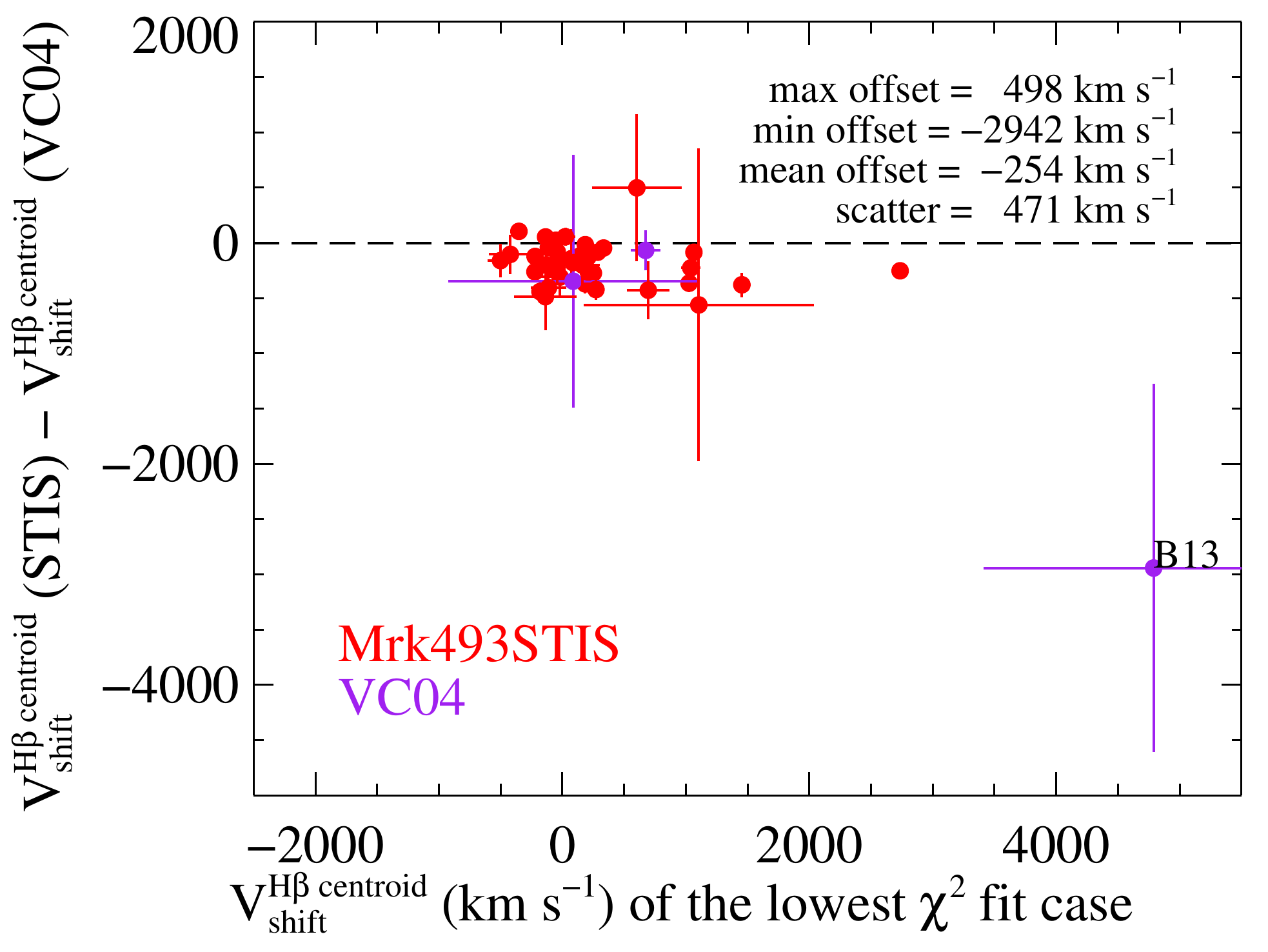} \\
	\includegraphics[width=0.33\textwidth]{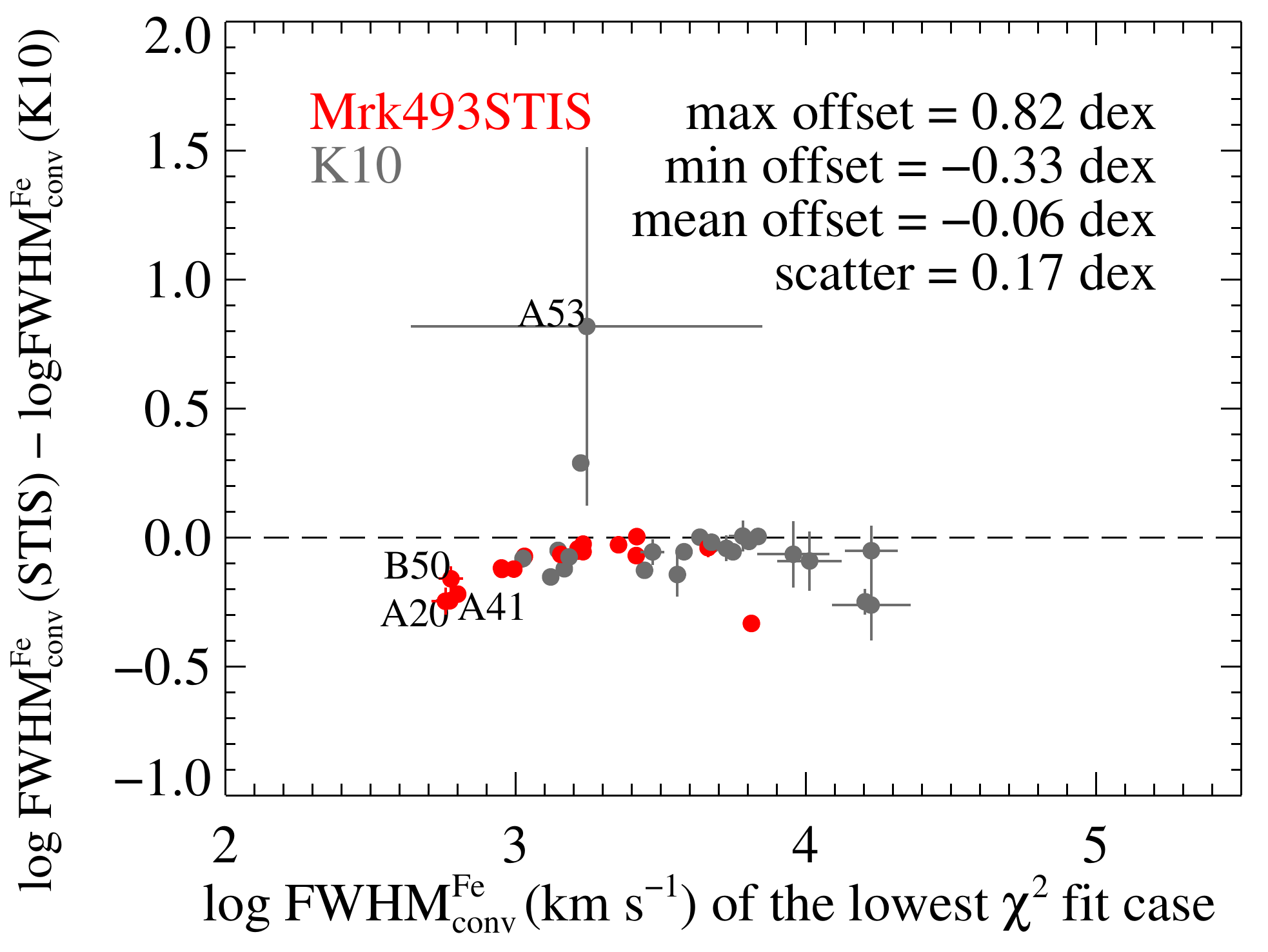}
	\includegraphics[width=0.33\textwidth]{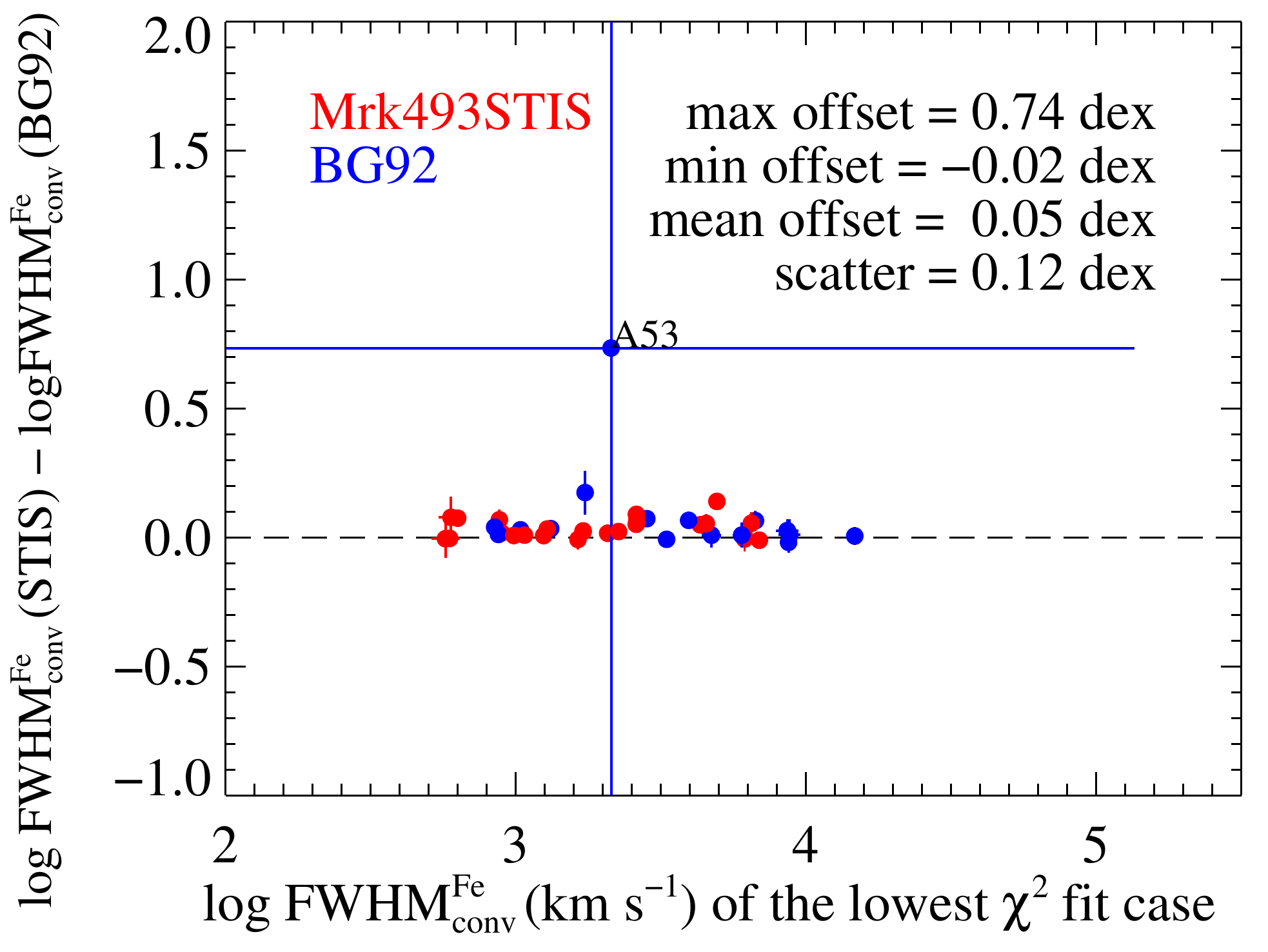}
	\includegraphics[width=0.33\textwidth]{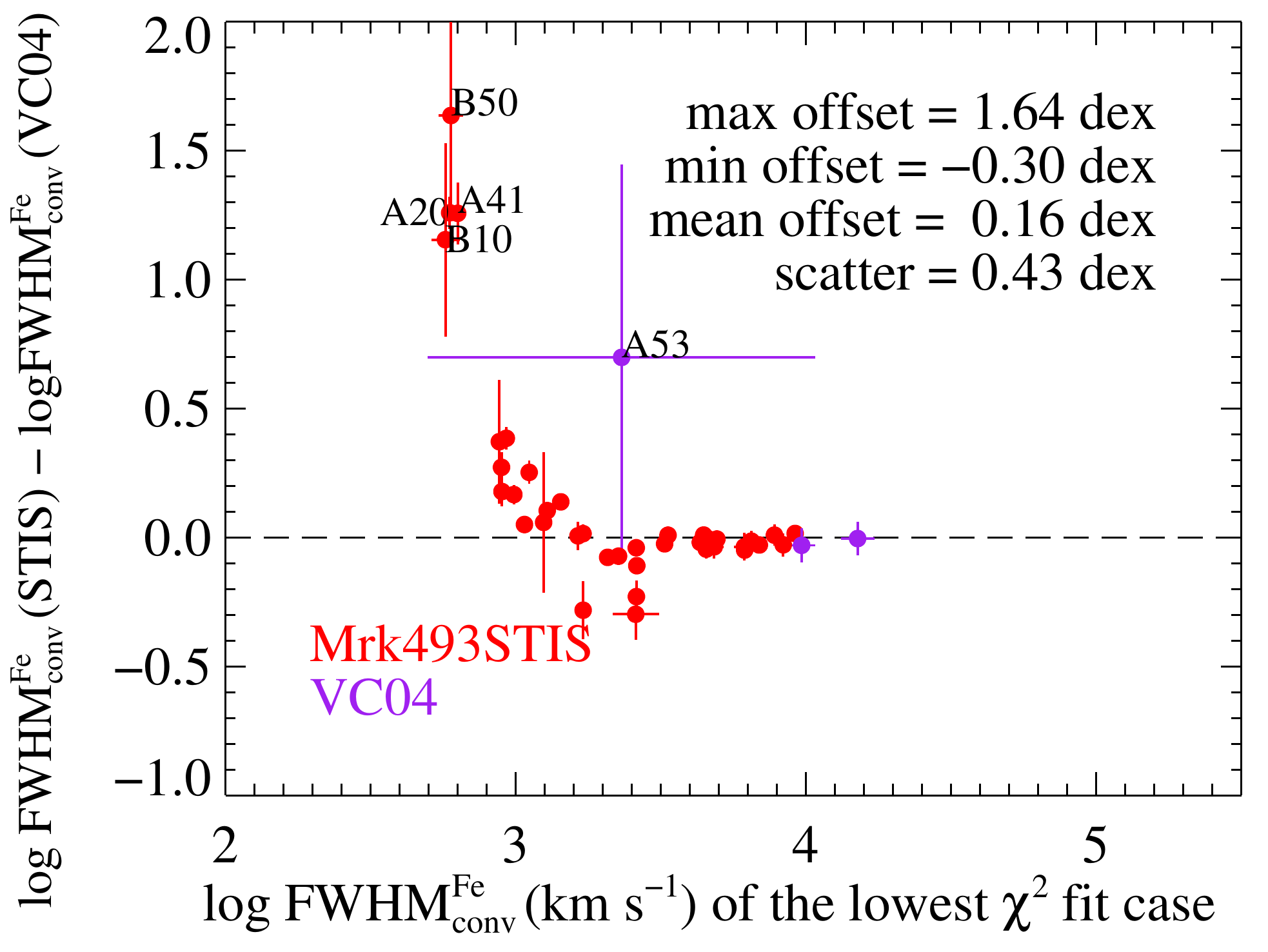}
	\caption{
		Same as Fig.~\ref{fig:SDSS_fit_EWdiff}, but for 
		\Hb\ broad emission line velocity shifts $V_{\rm shift}$ based on line centroid (upper row) and 
		broadening velocity widths FWHM$_{\rm conv}$ (lower row) from Gaussian convolution process of iron template fits. The adopted symbol color for each object indicates which template yielded the lower $\chi^2$ value between the two templates compared in each panel.
	}
	\label{fig:SDSS_fit_HbVelshift_ironFWHM}
\end{figure*}

\subsubsection{Measured and estimated physical properties}
\autoref{fig:SDSS_fit_EWdiff} shows differences of equivalent width (EW) measurements 
for the \Hb\ broad emission line and iron emission blends within the region $4434-4684$ \AA, respectively, depending on template choice.
When comparing the Mrk493STIS fit results against each of the other templates, the measured EWs of resulting \Hb\ broad lines are on average consistent with each other except for a few outliers, with relatively small offsets ($0.01-0.03$ dex) and scatter ($0.05-0.06$ dex).
This is because the EW as an integrated property of the \Hb\ line flux distribution is not significantly affected by relatively small changes in the line wing regions. The \Hb\ line wings can be substantially affected by the choice of iron templates as can be seen in Fig.~\ref{fig:compare_templatesSTIS}, but the EW of the line overall remains reasonably consistent independent of template choice.

On the contrary, there are noticeable average offsets ($0.05-0.09$ dex) on the distributions of EWs of the iron emission lines for different templates.
The direction and magnitude of such systematic offsets can be understood in terms of the intrinsic template differences originating from each template's construction method.
We find that on average, the EW of integrated \ion{Fe}{2} emission is inferred to be larger with the Mrk493STIS template than with the K10 template. This may result from the fact that the K10 template contains only broad \FeII\ lines identified as the strongest over these regions, which means that narrow iron lines and other iron-related lines are not included in the template.  Also, the large scatter ($0.14$ dex in \ion{Fe}{2} EW) between the K10 and Mrk493STIS results is likely due the increased fitting flexibility arising from the five independent line groups combined with degeneracy between \ion{Fe}{2} and other pseudo-continuum components (i.e., power-law model and host galaxy template) over this wavelength range.

Comparing \ion{Fe}{2} EWs between the Mrk493STIS fits and the other monolithic template fits, we find much lower scatter but distinct offsets in the inferred EWs.
The BG92 template contains several residual non-iron lines and some continuum emission due to the simplified template construction method, which would produce on average larger integrated iron fluxes (i.e., over-estimation) when using the BG92 template relative to the Mrk493STIS template.
For fits done with the VC04 template, we find lower EWs for \ion{Fe}{2} emission than for the Mrk493STIS template.  This can be attributed to the fact that the VC04 template contains only broad \FeII\ lines.

The significant outlier on the upper left corner is the object A53.
This object has weak iron features for which it is difficult to obtain a unique fit, given the degeneracy and data quality. 
Given the region of the EV1 parameter space from which this object was selected, it would be expected to have strong iron emission, 
but it actually does not, as can be seen in Figure \ref{AppFig:allSDSS_fits1}.\footnote{We attribute this discrepancy to a flaw in the iron flux measurement originally provided by the SDSS DR7 quasar catalog for this object (SDSS J125343.71+122721.5).
The DR7 catalog lists a very large value of iron EW for this object, 
although there is no iron component fitted (i.e., zero contribution) in the quality assessment plot provided along with the catalog. Thus, the catalog listing for the EW of iron emission over $4434-4684$ \AA\ appears to be incorrect for this object.}

\begin{figure*}[ht!]
	\centering
	\includegraphics[width=0.33\textwidth]{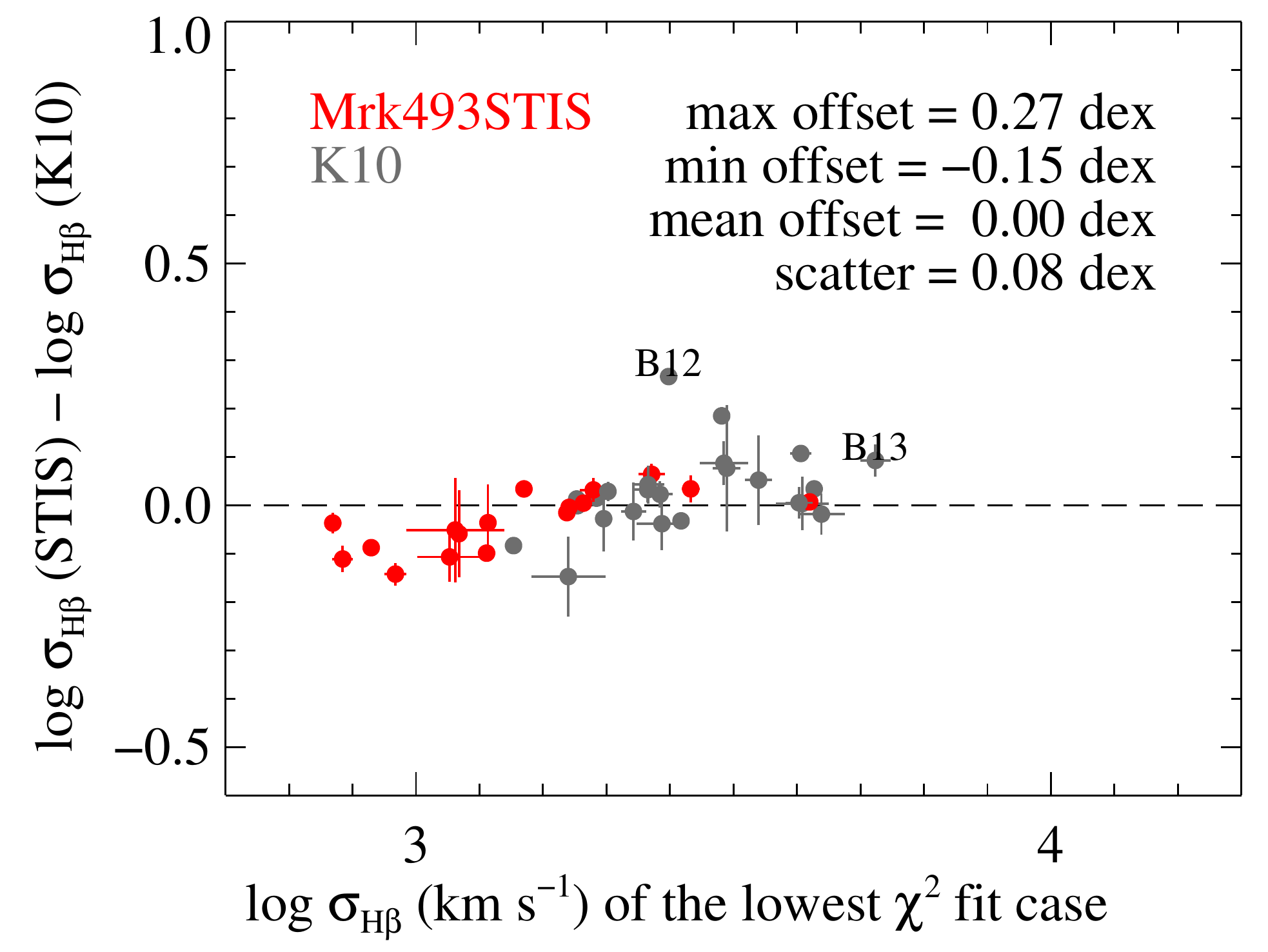}
	\includegraphics[width=0.33\textwidth]{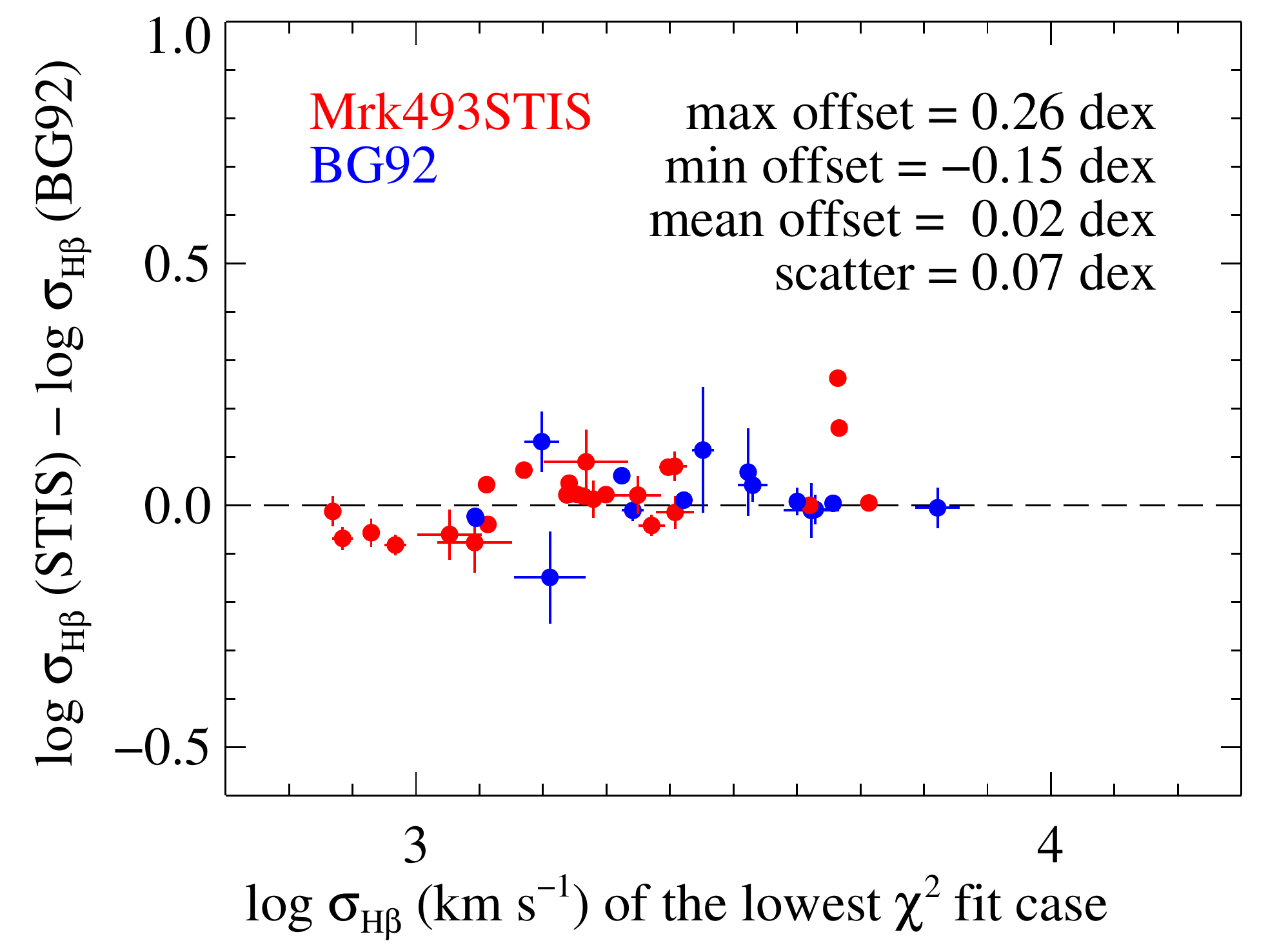}
	\includegraphics[width=0.33\textwidth]{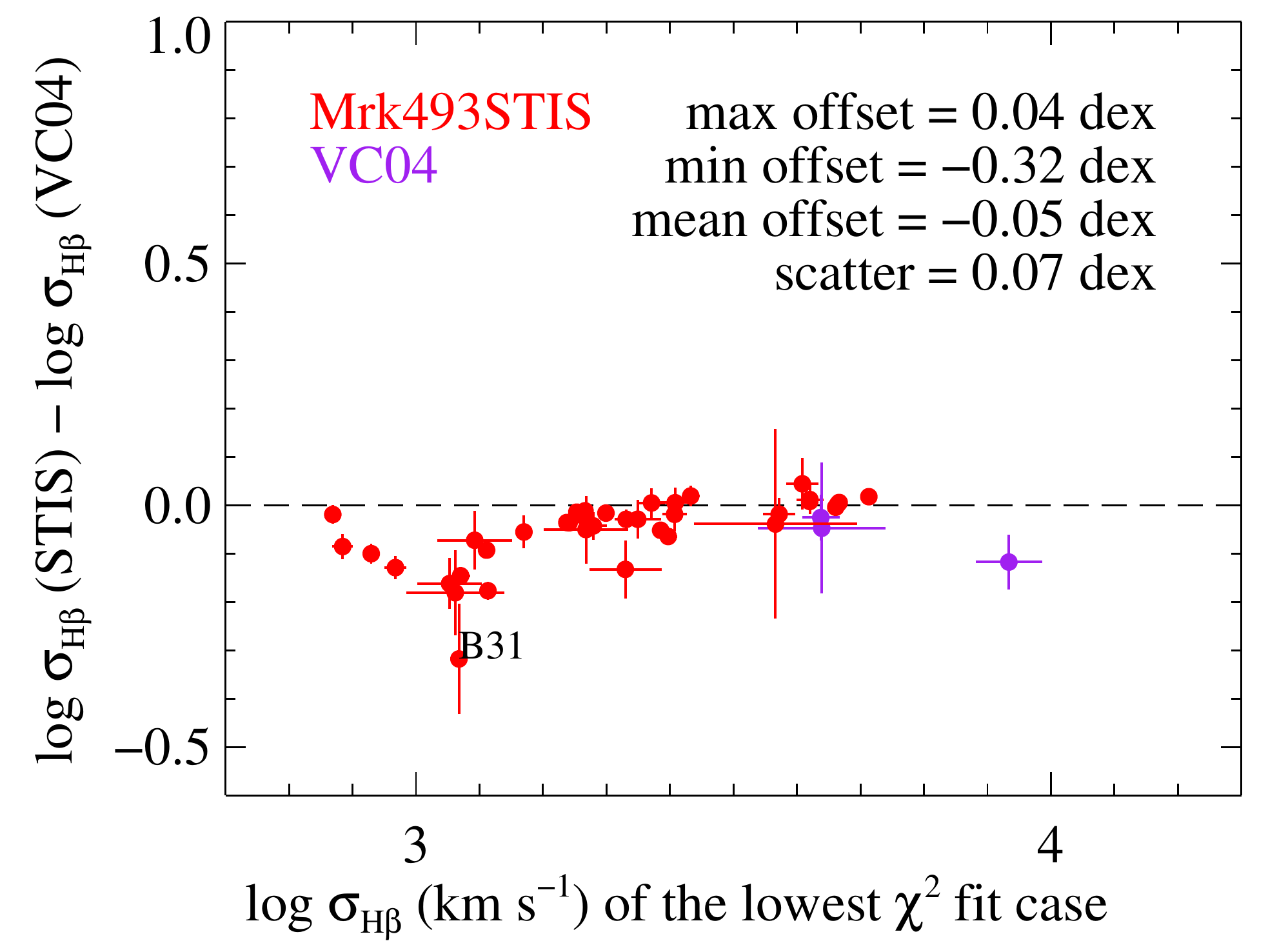} \\
	\includegraphics[width=0.33\textwidth]{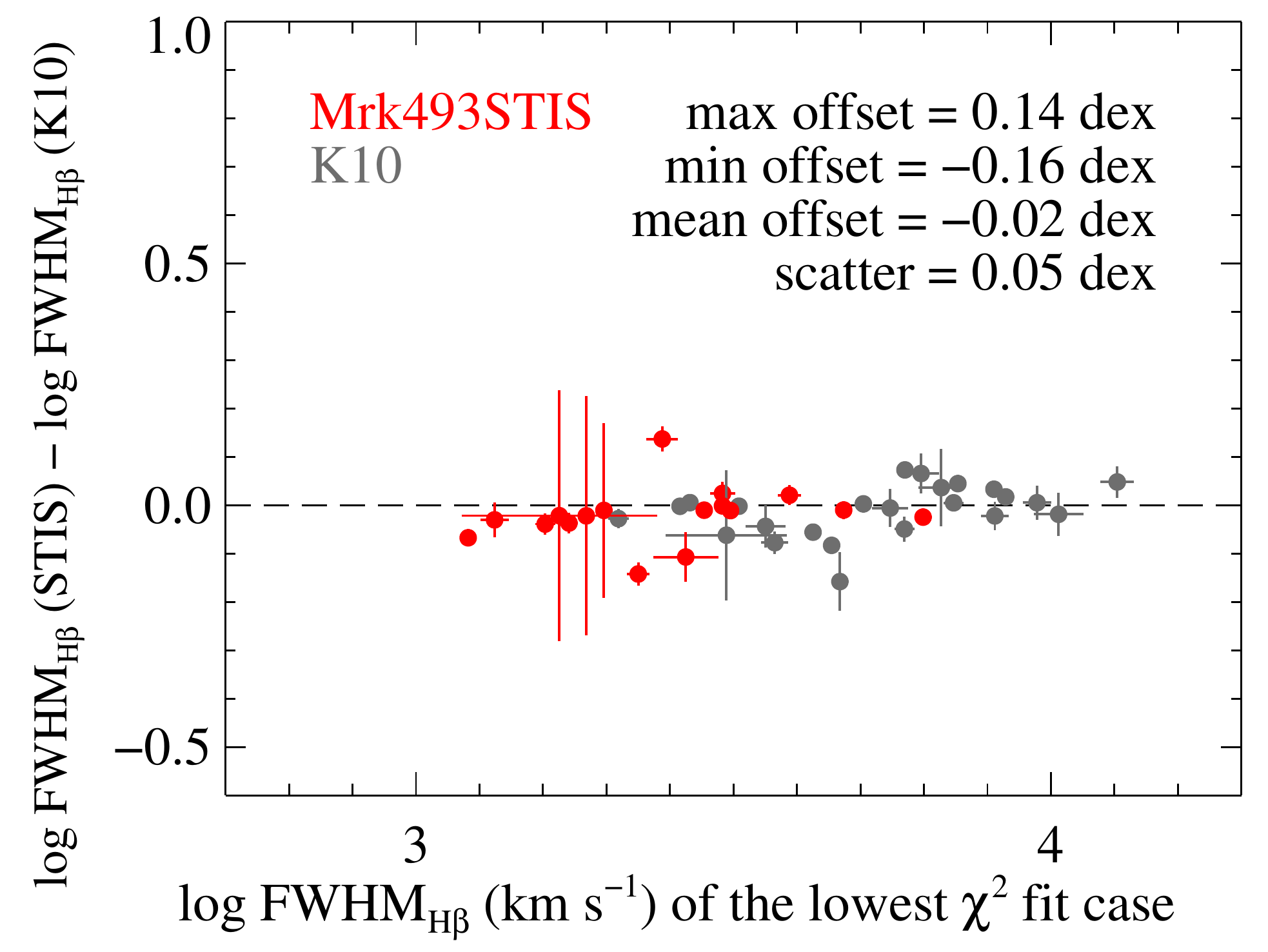}
	\includegraphics[width=0.33\textwidth]{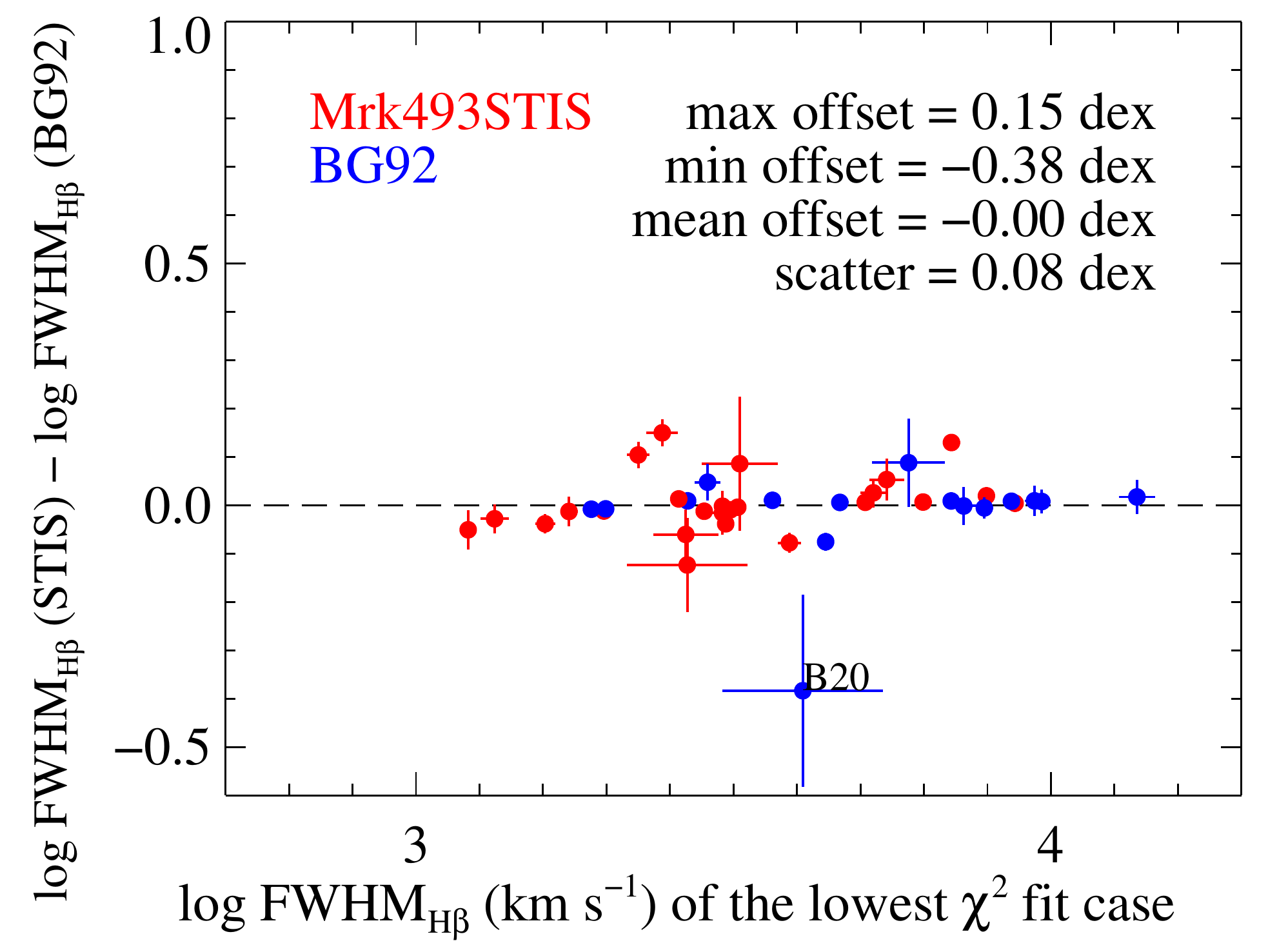}
	\includegraphics[width=0.33\textwidth]{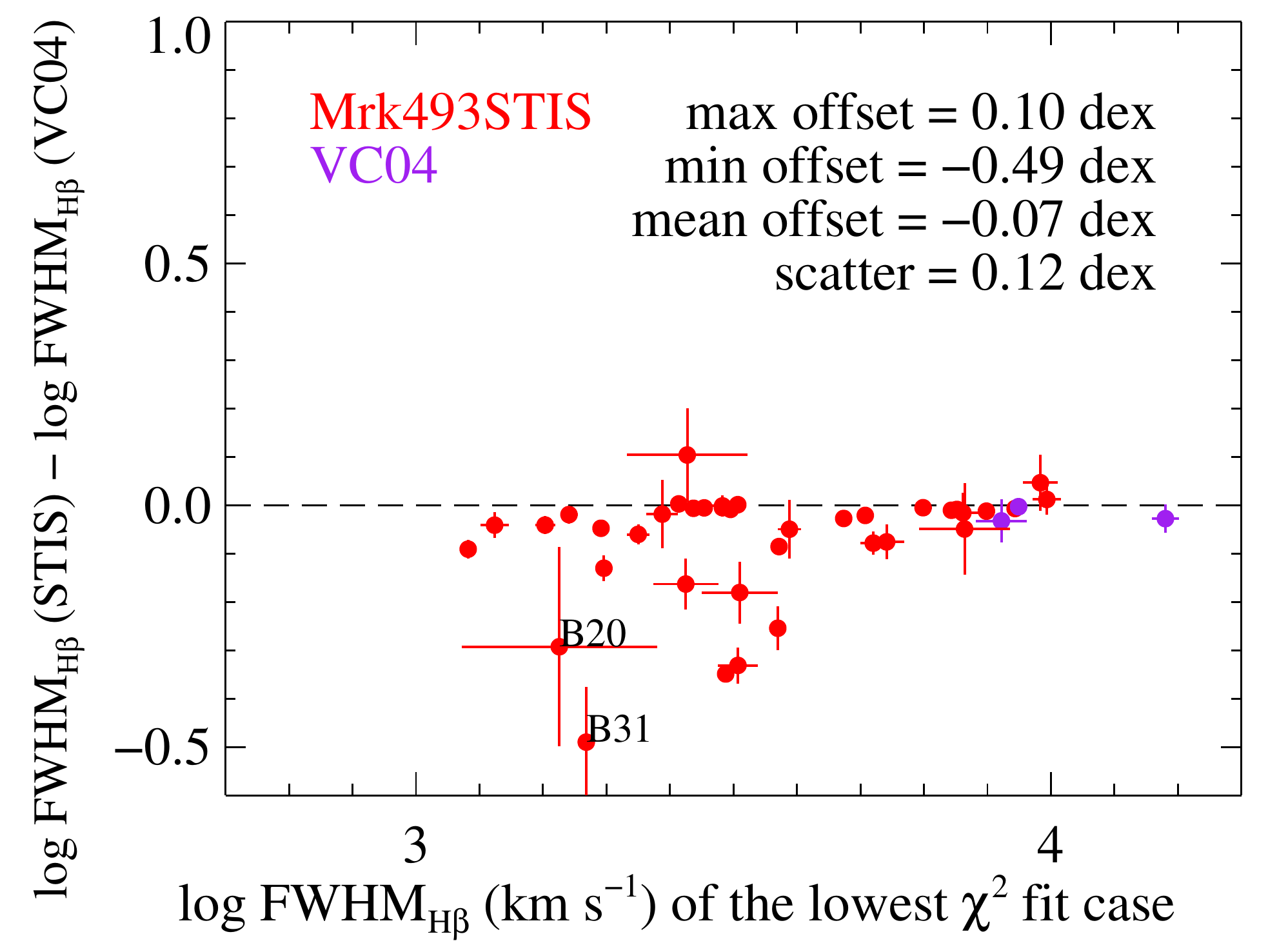} 
	\caption{
		Same as Fig.~\ref{fig:SDSS_fit_EWdiff}, but for \Hb\ broad emission line widths, 
		$\sigma_{\rm H\beta}$ (upper row) and FWHM$_{\rm H\beta}$ (lower row). The adopted symbol color for each object indicates which template yielded the lower $\chi^2$ value between the two templates compared in each panel.
	}
	\label{fig:SDSS_fit_Vdiff}
\end{figure*}

In the upper panels of \autoref{fig:SDSS_fit_HbVelshift_ironFWHM},
we compare differences of the measurements of velocity shift of the \Hb\ broad line based on its centroid, 
which incorporates information on the asymmetry of the full line profile, against the laboratory rest wavelength of \Hb. 
The measurements of \Hb\ centroid for the K10 and Mrk493STIS fits are mostly consistent, except for a few outliers.
The two strong outliers, B12 and B13, are both objects with extremely broad \Hb\ lines and very strong red-shelf emission, and the inferred velocity centroid differences can be attributed to the differences in how the two templates model the red shelf region. The K10 template fits the red shelf flux with iron emission, while the model fits done with the Mrk493STIS template ascribe the red shelf to \Hb\ emission, as can be seen in Fig.~\ref{fig:SDSS_fit_decomp}, resulting in a substantial \Hb\ velocity offset between the two fits.
For similar reasons, we find a positive systematic offset in \Hb\ velocity centroids from the Mrk493STIS fits in comparison with the BG92 fits, indicating
on average weaker \Hb\ red wing emission (i.e., over-subtraction) in the BG92 fits than those of our Mrk493STIS fits. 
This trend is naturally expected because the BG92 template includes stronger iron lines over the \Hb\ red wing region compared to the Mrk493STIS template (see Fig.~\ref{fig:compare_templatesSTIS}).
These differences are particularly apparent in the fits to the same outlier objects, B12 and B13, as described above for the K10 fits.
The relatively small negative systematic velocity offset on average in comparison with the VC04 fits is also expected to stem from the
slightly weaker iron features in the \Hb\ red wing region of the VC04 template (Fig.~\ref{fig:compare_templatesSTIS}).
The only large outlier, object B13, is discrepant due to strong under-subtraction of pseudo-continuum emission over the \Hb\ red wing region resulting from  severe degeneracy between the pseudo-continuum components with the VC04 template. As previously described, the fits with the VC04 template produce much poorer results (in terms of $\chi^2$) than the other templates for almost all objects.

In the lower panels in \autoref{fig:SDSS_fit_HbVelshift_ironFWHM},
we show differences between measurements of the Gaussian broadening (convolution) velocity widths of the iron template, FWHM$_{\rm conv}$, depending on template choice.
We find that the measurements using the Mrk493STIS template are consistent overall with those using the BG92 template, 
except for the one outlier.
This is probably because both templates are monolithic and constructed to contain narrow iron lines, 
which are especially  important for fitting narrower-lined objects, as well as the broad iron lines.
The strong outlier having the large error-bar on the upper region is again object A53, 
which has weak iron features rendering its fit very uncertain due to heavy degeneracy with other continuum components.
In contrast, there are systematic trends particularly in the low-velocity range ($\lesssim1000$ \kms) 
in comparisons with the K10 and VC04 template fits,
which struggle to obtain good fits for such narrower line objects.
This can be attributed to the combined effect of 
the absence or presence of narrow line contributions to the templates, the difference in intrinsic velocities depending on template objects and construction methods, and different iron line profiles adopted (a Gaussian for K10, a Lorentzian for VC04, a Gauss-Hermite series for Mrk493STIS).

\begin{figure*}[ht!]
	\centering
	\includegraphics[width=0.33\textwidth]{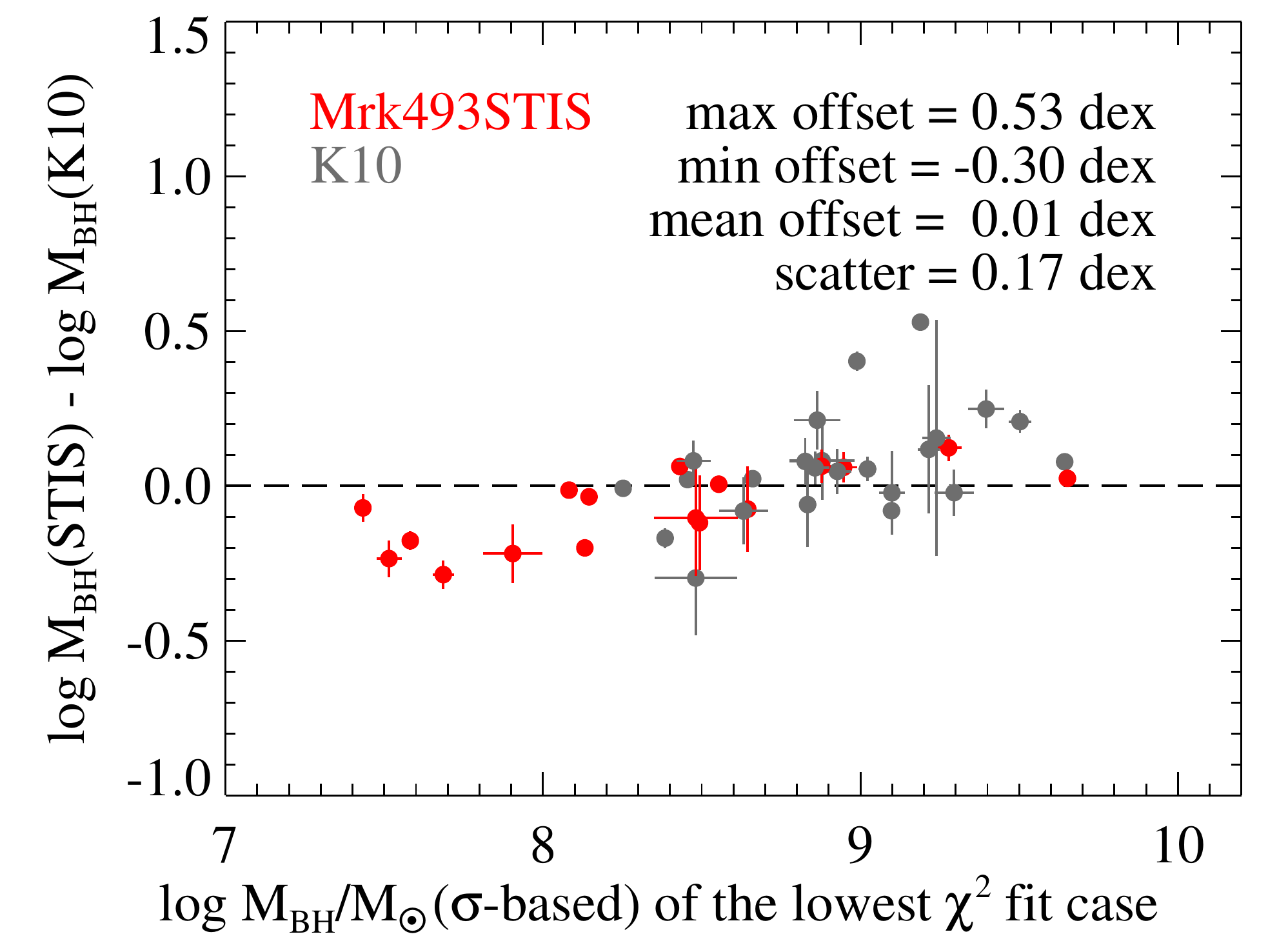}
	\includegraphics[width=0.33\textwidth]{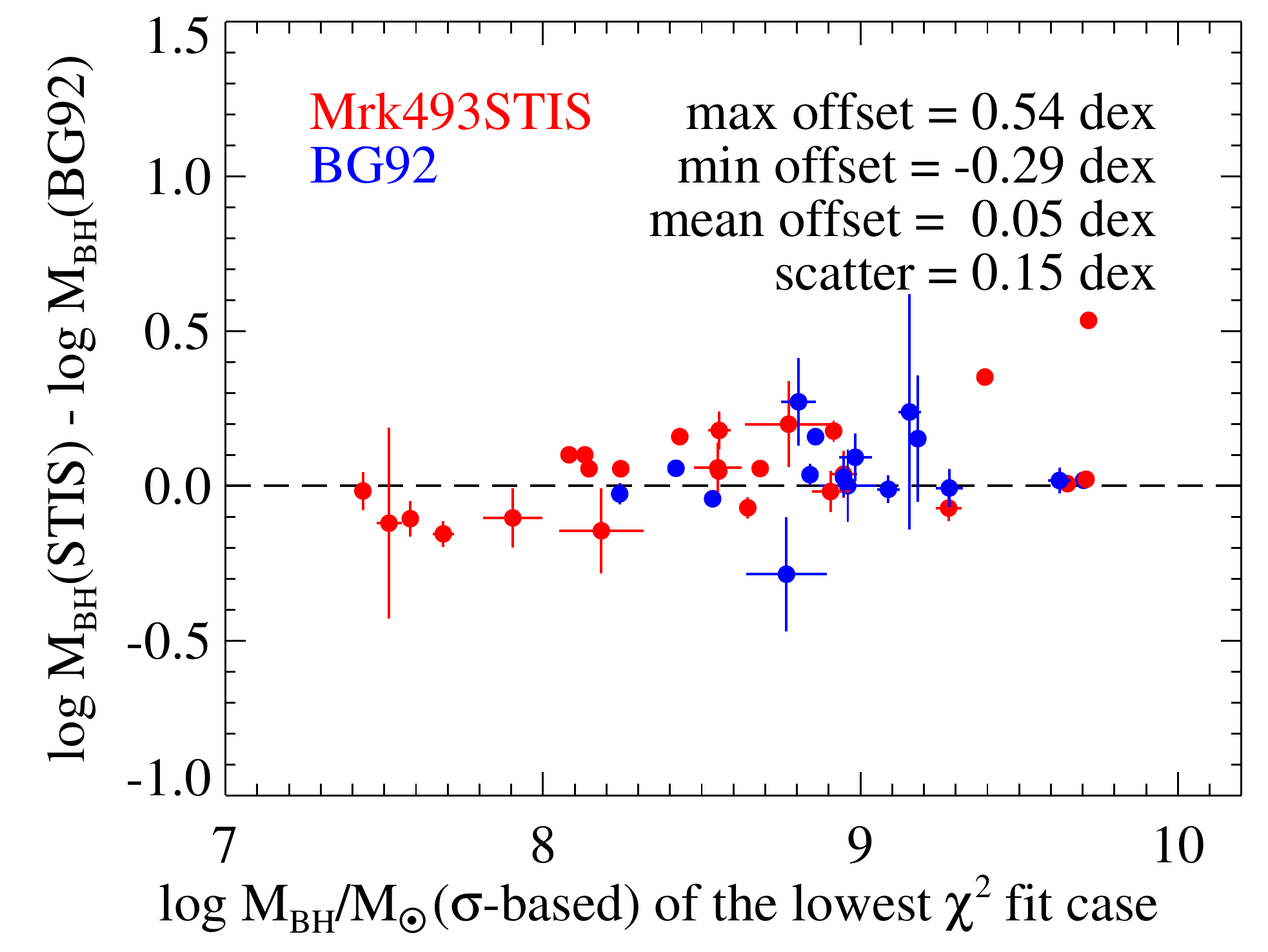}
	\includegraphics[width=0.33\textwidth]{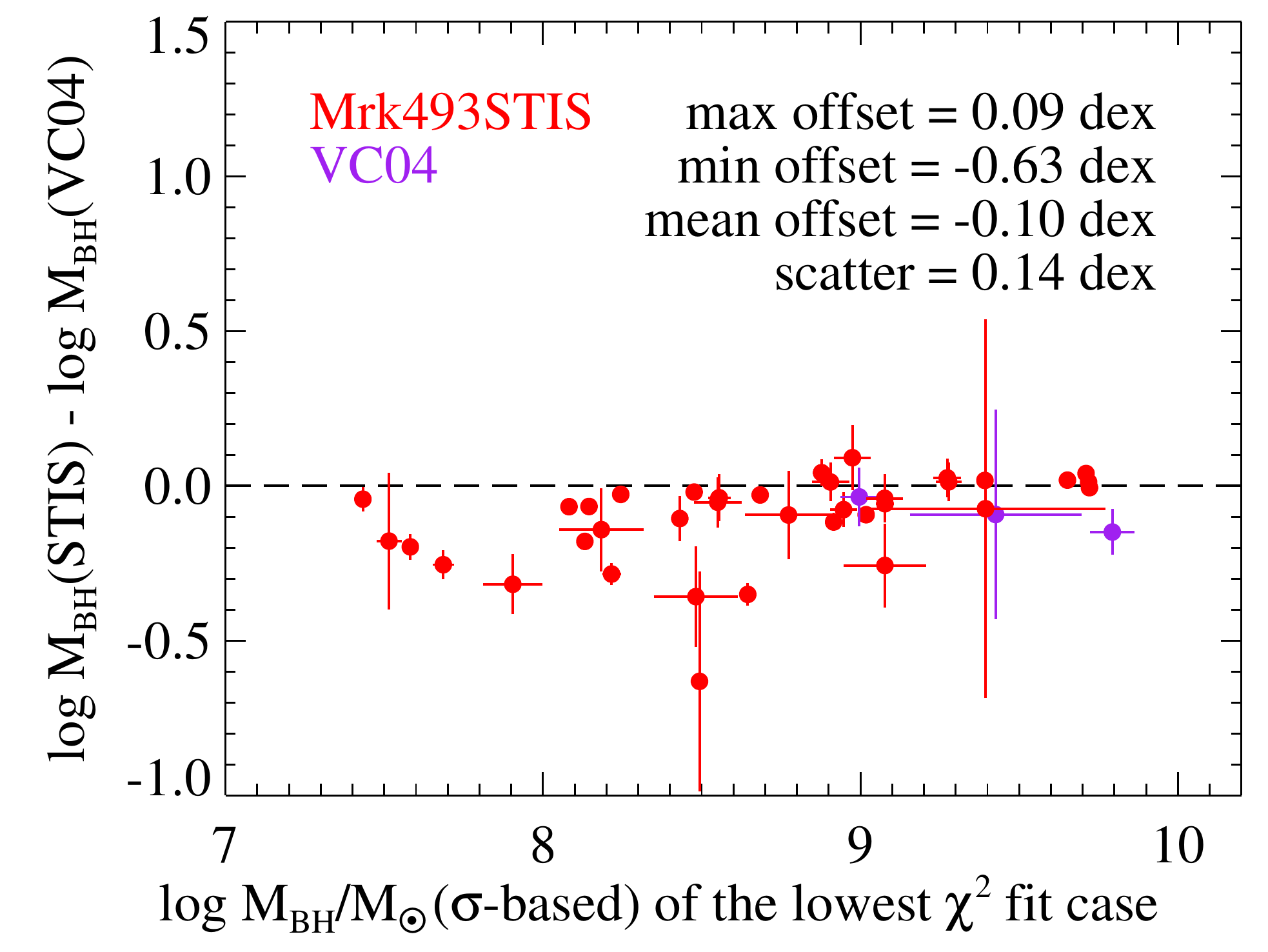}\\
	\includegraphics[width=0.33\textwidth]{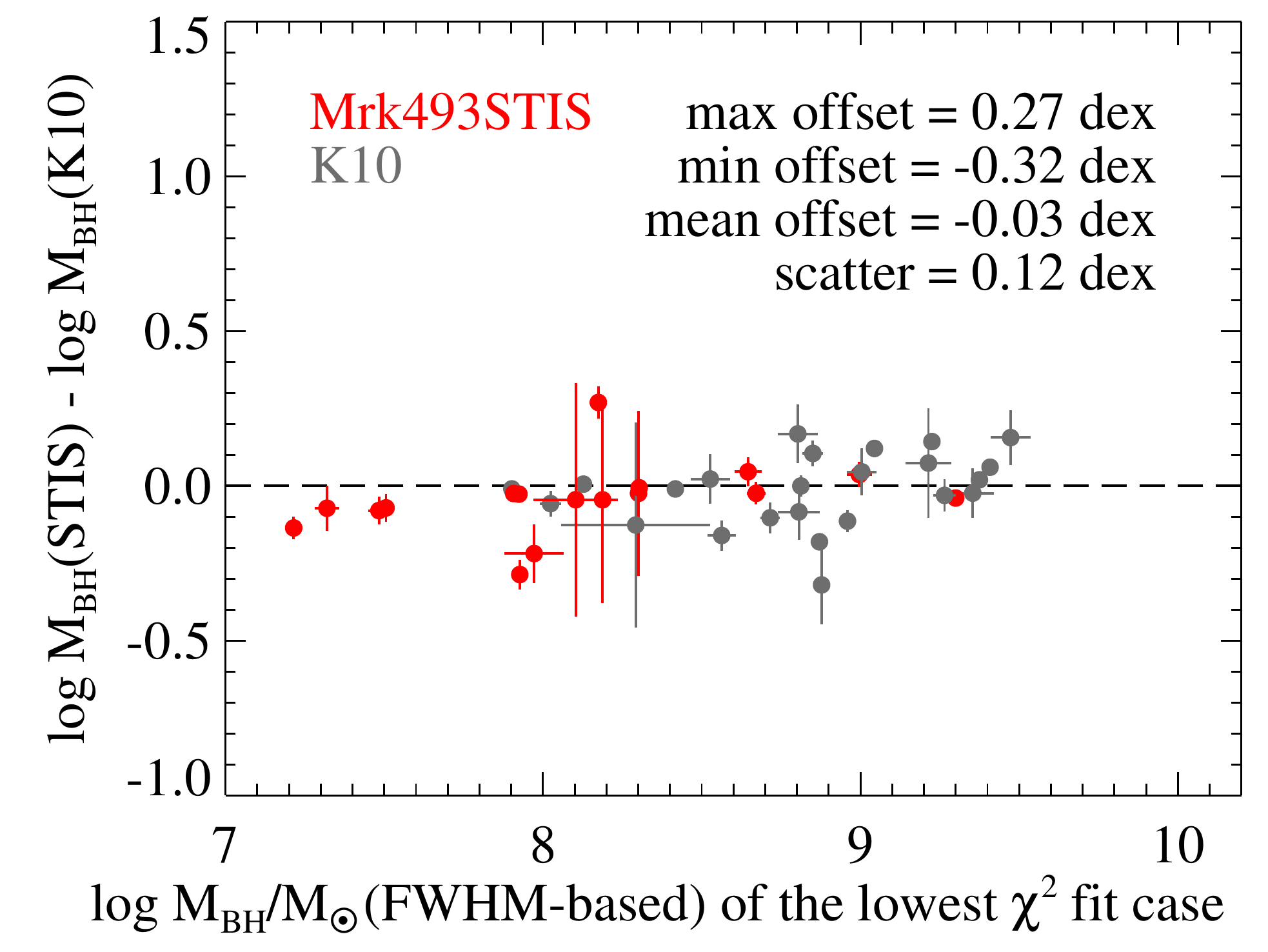}
	\includegraphics[width=0.33\textwidth]{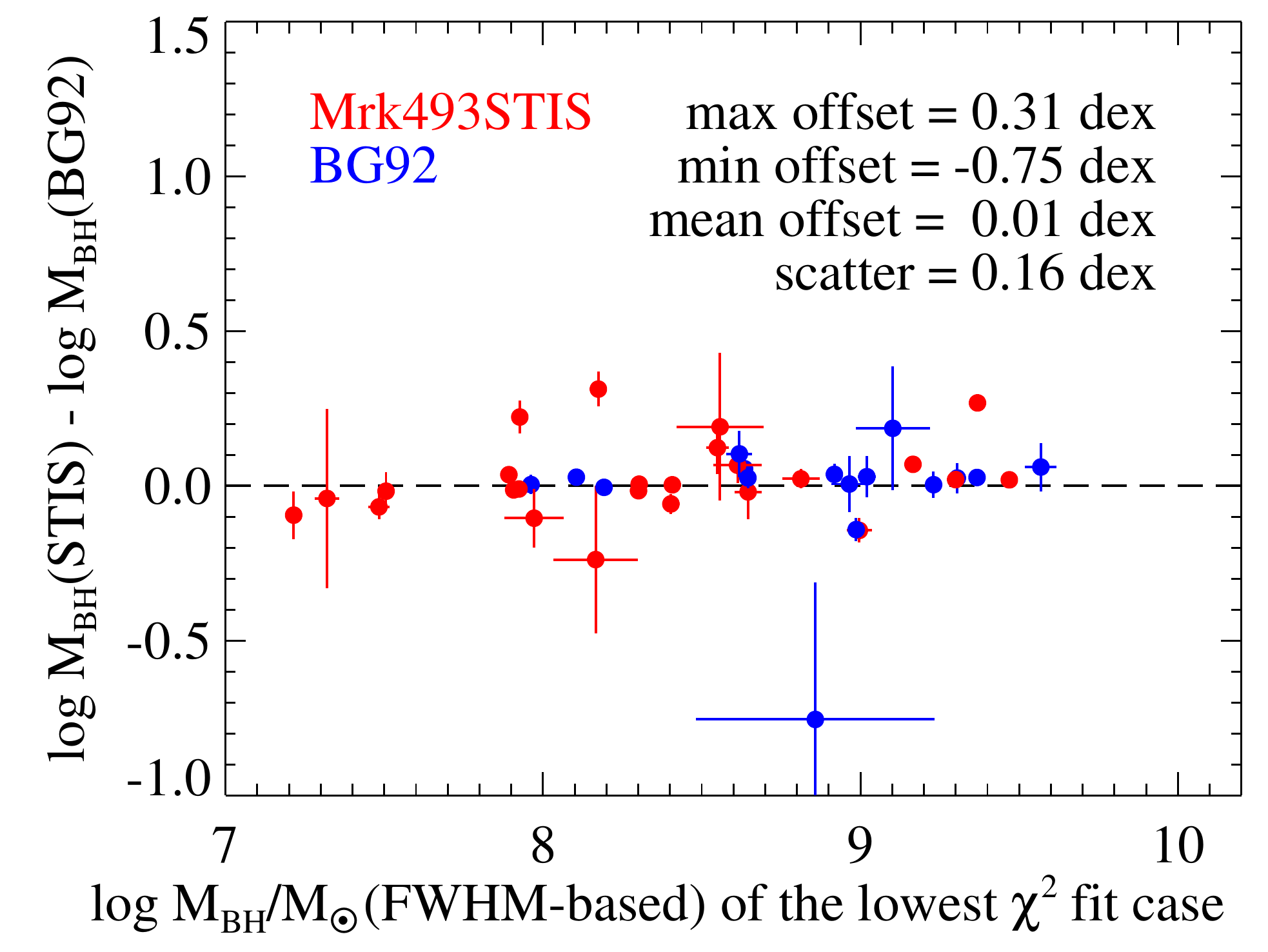}
	\includegraphics[width=0.33\textwidth]{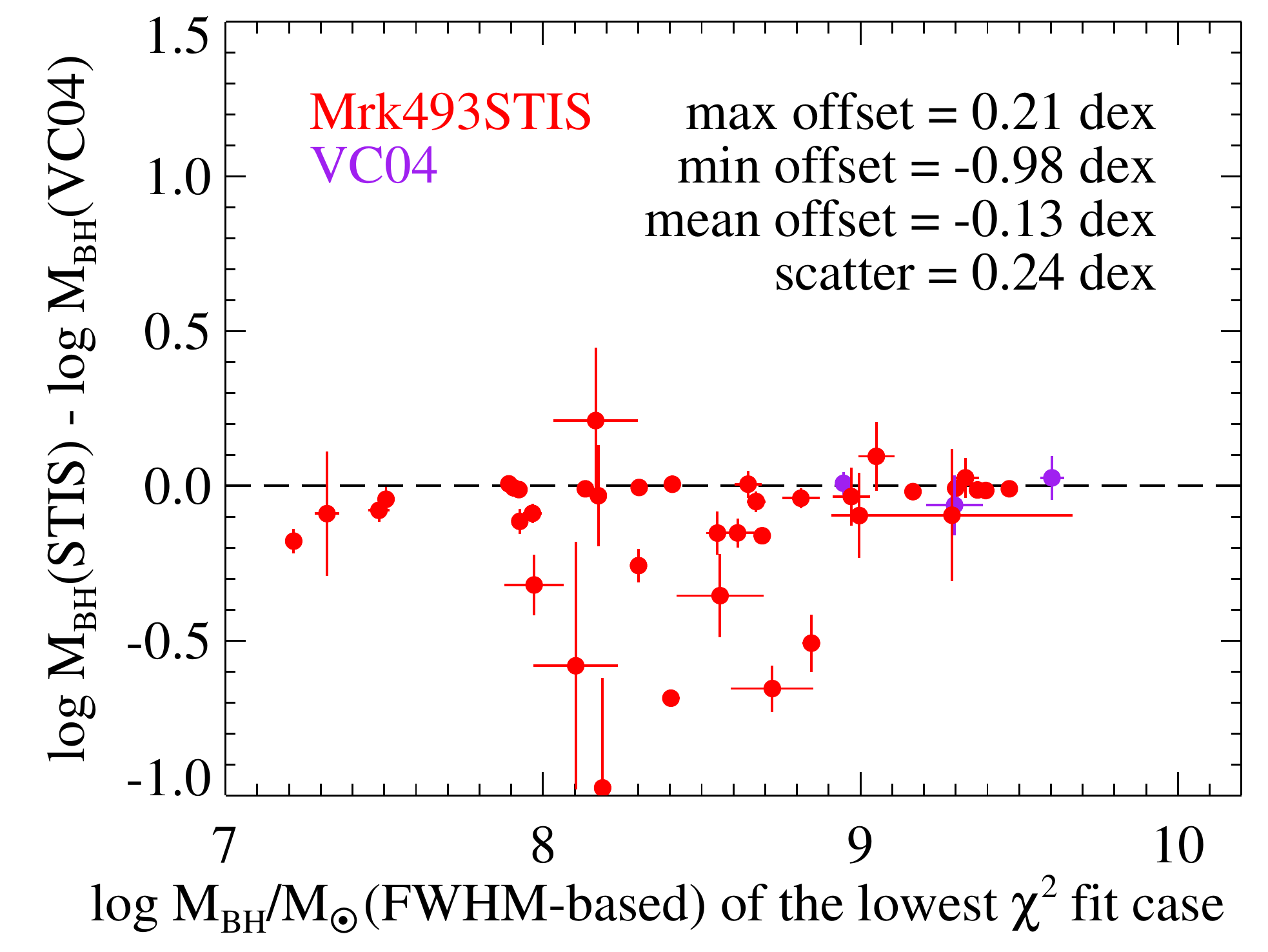}
	\caption{
		Same as Fig.~\ref{fig:SDSS_fit_EWdiff}, but for 
		\mbh\ estimates based on $\sigma_{\rm H\beta}$ (upper row) and FWHM$_{\rm H\beta}$ (lower row). The adopted symbol color for each object indicates which template yielded the lower $\chi^2$ value between the two templates compared in each panel.
	}
	\label{fig:SDSS_fit_MBHdiff}
\end{figure*}

\autoref{fig:SDSS_fit_Vdiff} shows differences between the \Hb\ broad emission line width measurements, 
$\sigma_{\rm H\beta}$ and FWHM$_{\rm H\beta}$, respectively, depending on template choice.
The measurements are on average consistent with each other showing a very small offset of $\sim0.00-0.02$ dex, except for the VC04 fits. Our tests indicate that the \Hb\ broad-line width will be on average systematically overestimated when using the VC04 template, as shown by the negative mean offset in size of $0.05-0.07$ dex for both FWHM and line dispersion, and the discrepancies appear most pronounced for AGN having narrower emission lines.

Interestingly, there is a slight systematic trend as a function of line dispersion in the offsets between \Hb\ width measurements for the Mrk493STIS template versus the K10 and BG92 fits.
A somewhat clear separation at $\sim2000$ \kms\ of $\sigma_{\rm H\beta}$ is again observed
especially in the comparison with the K10 results as already have seen in Fig.~\ref{fig:SDSS_fit_chi2DistAll}.
The line dispersion is on average over-estimated if the K10 template is used instead of our Mrk493STIS template in narrower line objects,
while over-estimation occurs on average in broader line objects if our Mrk493STIS template is used instead of the K10 template.
This trend is caused primarily by the template differences around the \Hb\ line wing regions between the K10 and our Mrk493STIS
as have seen in Fig.~\ref{fig:compare_templatesSTIS} and \ref{fig:SDSS_fit_decomp}.
The Mrk493STIS template is more complete than the K10 
in the sense that narrow iron lines and related lines are identified and included in the template,
enabling a more accurate removal of these features in the \Hb\ region. In objects with broader lines, however, the flexibility of the multicomponent K10 template works better in terms of dealing with the \Hb\ red-shelf region by fitting the red shelf flux as iron emission.

These systematic trends are much weaker in the FWHM distribution 
since the measured FWHM values are much less sensitive to the line wings than the line dispersions.
Instead, the FWHM is very sensitive to the determination of the amplitude of the line peak, which could thus be easily biased if subtraction of the \Hb\ narrow component is inaccurate.
The strong outliers in the FWHM comparisons, objects B20 and B31,
are due to the uncertainty in the \Hb\ narrow-line subtraction, and their relatively large FWHM uncertainties primarily reflect degeneracy in decomposing the narrow and broad \Hb\ components in the SDSS data. This is particularly problematic for intrinsically [\ion{O}{3}]-weak objects since the [\ion{O}{3}] line flux after subtraction of the underlying iron contribution changes significantly depending on the template choice.

\autoref{fig:SDSS_fit_MBHdiff} shows differences between the \mbh\ estimates derived from the different template fits to illustrate possible systematic offsets in inferred mass depending on iron template choice. 
The BH  mass is calculated using the line width, $\sigma_{\rm H\beta}$ or FWHM$_{\rm H\beta}$, and the
continuum luminosity at $5100$ \AA, $L_{5100\text{\AA}}$, measurements
based on the \mbh($\sigma_{\rm H\beta}$, $L_{5100\text{\AA}}$) equation adopted by \citet{Bennert+2015} 
and \mbh(FWHM$_{\rm H\beta}$, $L_{5100\text{\AA}}$) listed in \citet{ShenLiu2012} (originally from \citealt{Assef+2011}) respectively.
There is a systematic trend in the \mbh\ estimates particularly for those based on $\sigma_{\rm H\beta}$ from the K10 and BG92 template fits in comparison with the Mrk493STIS fits, while for the VC04 fits
there is an average offset by $0.10-0.13$ dex (in the sense that the VC04 template gives systematically higher \mbh\ than the Mrk493STIS template).
The trends seen in these comparisons directly reflect the systematic differences in the line width measurements as shown in Fig.~\ref{fig:SDSS_fit_Vdiff}. The $L_{5100\text{\AA}}$ estimates from the best-fit power-law continuum component are closely
consistent with each other among the different template fits with only small mean offsets of $0.01-0.03$ dex and scatter of $0.01-0.04$ dex and no systematic trend trend as a function of line width.
Thus, the choice of iron template causes a systematic trend in $\sigma_{\rm H\beta}$ measurements, 
which directly propagates into \mbh\ estimates. 

If we divide the sample into high-mass and low-mass groups using a threshold of $\log(M_{\rm BH}/M_{\odot})=8.5$ for the K10 fits using $\sigma_{\rm H\beta}$,
the mean offsets for the two sub-samples respectively are found to be both $\sim0.1$ dex, with differences for individual objects of up to $\sim0.3-0.5$ dex. This indicates that a systematic bias of $\sim0.1$ dex on average on \mbh\ would occur on either side according to the iron template choice.
However, these estimates are subject to small number statistics and possible selection biases of our test sample.
It is also worth noting that the comparison results presented above would change if different fitting methods or assumptions were adopted 
(e.g., different emission line model profiles, continuum models, or inclusion of additional lines such as \ion{He}{1} $\lambda\lambda4922,5016$).

\section{Summary and Discussion} \label{sec:summary}  

We have constructed a new empirical iron emission template in the optical wavelength range $4000-5600$ \AA, which is the primary region of interest for single-epoch BH mass estimation in low-$z$ AGN and for reverberation mapping of the \Hb\ emission line.
A major advantage of this new template is that it is based on Mrk 493, which has narrower lines, lower reddening, and a less extreme Eddington ratio value than I Zw 1. Using HST STIS data taken with an 0\farcs2 slit as the basis of the template construction provides a further advantage by almost completely eliminating any starlight contribution to the data. When used for fitting AGN spectra, our template arguably provides more accurate spectral measurements and recovery of Balmer lines with a better line profile match among those than the BG92, VC04, and K10 templates. Thus, the new Mrk493STIS template will be a useful addition to the library of available iron templates, and can provide substantial advantages for a variety of applications in AGN spectroscopy including systematic application to data from large spectroscopic surveys 
\citep[e.g.,][]{Shen+2011,Calderone+2017,Rakshit+2020}.

Based on our comparison tests using objects selected from the SDSS quasar sample to span the full range of \ion{Fe}{2} emission properties,
we showed that our Mrk493STIS template works best for AGN having line widths of $\sigma_{\rm H\beta}\lesssim2000$ \kms\ (or FWHM$_{\rm H\beta}\lesssim4000$ \kms), while the K10 template gives more precise fits for objects with $\sigma_{\rm H\beta}\gtrsim2000$ \kms\ (or FWHM$_{\rm H\beta}\gtrsim4000$ \kms). We find that there is a possible systematic bias in line width measurements ($\sigma_{\rm H\beta}$) depending on a template choice,
which in turn leads to as much as $\sim0.1$ dex offset on \mbh\ estimates on average and up to $\sim0.3-0.5$ dex offset for individual objects. 
Interestingly, the separation velocity (FWHM$_{\rm H\beta}\sim4000$ \kms) we found in this work between AGN that are best fitted by the Mrk493STIS or K10 templates
is similar to the separation threshold found by \citet{Sulentic+2000,Sulentic+2002,Sulentic+2009} between their A and B sub-populations of radio-quiet AGN
and also the transition \Hb\ FWHM above which the \CIV\ line becomes 
narrower, rather than broader, than \Hb\ \citep[see Fig.~3 in][]{BaskinLaor2005_CIV}.

Our new template will provide substantial benefits for 
systematic studies of active galaxies having strong and narrow iron emission (e.g., NLS1). In addition to its advantages for measurement of \Hb\ and \ion{Fe}{2} features, our tests show that the Mrk493STIS template will also have benefits for measurement of accurate \Hg\ and \Hd\ line profiles in AGN showing strong iron emission.

Despite the improvements achieved, there may still be shortcomings in the template construction due to the limited spectral resolution of the data, imperfect deblending of lines in the Mrk 493 spectrum, or incompleteness in the adopted line list. Such issues could lead to systematic errors or biases in the template shape and in inferences based on use of the template.
Thus, further cross-checks and tests of the template will be worthwhile, and improvements in template construction methods are still needed. Most importantly, the availability of the complete STIS UV through optical spectrum of Mrk 493 will provide a lasting reference and benchmark for future template construction and tests.

As a template object Mrk 493 has several preferable properties (e.g., narrowness and intrinsic reddening) in comparison to I Zw 1, but no monolithic template based on a single object can fully represent the diversity of \ion{Fe}{2} emission properties across all AGN.
It would be useful to carry out further tests of object-to-object differences (e.g., iron line ratios and profiles) 
using similar high-quality data for other possible template candidates from the NLS1 population. STIS data for a larger NLS1 sample could provide a ``basis set'' for construction of more flexible multi-component templates that could provide substantial further improvements in fitting AGN spectra.

Although the STIS optical data used to construct the iron template in this work is much better than any of the previously available data,
its spectral resolution ($R\sim1500$) is actually insufficient to fully resolve and unambiguously decompose all of the broad and narrow components in the blended emission complexes across this optical region. 
Improvements in the line decomposition could be obtained from optical spectra of Mrk 493 having much higher spectral resolution. Such data could also detect and resolve the elusive narrow component of iron emission lines \citep[e.g.,][]{Wang+2008,Dong+2010} and investigate the properties and kinematics of these features.

Other limitations of our template construction include the fact that all the Balmer lines in the Mrk 493 spectral decomposition were forced to have exactly the same line profile (i.e., the same velocity distribution with different total line intensities), 
which proved necessary in order to obtain a successful fit given the model complexity and limited resolution of the data. (This same assumption was used by VC04 in constructing their I Zw 1 model.)
However, the Balmer series lines are expected to have slightly different velocity distributions due to stratified BLR structure.
A single  power-law function was used to model the continuum emission when constructing the iron template, but a more complex continuum could be present due to the contributions of host-galaxy starlight and Paschen continuum from the BLR \citep[see, e.g.,][]{Vincentelli+2021},
although the former is minimized by the narrow slit width of the STIS observations
and the latter is completely degenerate with the power-law model over the limited range of our spectral fit. A more definitive decomposition of these components could be done by fitting models over a broader wavelength range.
It is also possible that our derived best-fit solution for the Mrk 493 spectral decomposition model 
using $\chi^2$ minimization (based on the Levenberg-Marquardt method)
does not represent the true global $\chi^2$ minimum, due to the unavoidable model-fitting degeneracy when optimizing a fit over such a large number of spectral components simultaneously.
We plan to revisit this optimization issue in the future
using Bayesian Markov chain Monte Carlo sampling techniques in future works.

Our new monolithic template provides better fits over almost all regions  of the EV1 plane 
than the existing empirical templates having a monolithic structure (BG92 and VC04). However, all of these monolithic templates have the drawback that they do not allow for the possibility of different relative intensities of iron lines in different objects.
The K10 multicomponent template works better than ours especially for objects with broader line widths and having significant \Hb\ red-shelf features, by adjusting individual iron line group flux levels independently.  Ultimately, the ideal iron template would thus have both the multicomponent flexibility of the K10 template while also including a larger line list and setting line profiles and relative intensities based on a fit to high-quality spectroscopic data such as ours over the entire UV-optical spectral range.

As a next step, we will construct the UV portion of the Mrk 493 iron template, following methods used for prior I Zw 1 templates by \citet{VestergaardWilkes2001,Tsuzuki+2006}. We also plan to extend the optical portion of the template to longer wavelengths, although \ion{Fe}{2} features are relatively weak longward of $\sim6000$ \AA. 
The result will be the first  empirical iron template covering the entire wavelength range from Ly$\alpha$ through H$\alpha$ constructed from a single, quasi-simultaneous spectroscopic observation over this full wavelength range. We also plan to obtain ground-based optical spectra of Mrk 493 over the \Hb\ region at significantly higher spectral resolution than the STIS data, which will aid in carrying out more detailed deblending of the complex iron emission features in this region.

\begin{acknowledgments}
Support for HST program GO-14744 was provided by NASA through a grant
from the Space Telescope Science Institute, which is operated by the
Association of Universities for Research in Astronomy, Inc., under NASA
contract NAS 5-26555.
This work was supported by the National Research Foundation of Korea (NRF) grant 
funded by the Korea government (MSIT) (No. 2020R1C1C1010802).
Research by A.J.B. was also supported by NSF grant AST-1907290.
L.C.H. was supported by the National Key R\&D Program of China (2016YFA0400702) 
and the National Science Foundation of China (11721303, 11991052).
A.L. acknowledges support by the Israel Science Foundation (grant no. 1008/18).
D.P. thanks Minjin Kim and Min-Su Shin for useful discussions.
We thank the anonymous referee for useful suggestions and comments that have improved the paper.
\end{acknowledgments}

\facility{HST (STIS)}   

\software{IDL, Python, PyRAF}

\bibliography{daeseong}	 

\begin{thebibliography}{}
\expandafter\ifx\csname natexlab\endcsname\relax\def\natexlab#1{#1}\fi
\providecommand{\url}[1]{\href{#1}{#1}}
\providecommand{\dodoi}[1]{doi:~\href{http://doi.org/#1}{\nolinkurl{#1}}}
\providecommand{\doeprint}[1]{\href{http://ascl.net/#1}{\nolinkurl{http://ascl.net/#1}}}
\providecommand{\doarXiv}[1]{\href{https://arxiv.org/abs/#1}{\nolinkurl{https://arxiv.org/abs/#1}}}

\bibitem[{{Assef} {et~al.}(2011){Assef}, {Denney}, {Kochanek}, {Peterson},
  {Koz{\l}owski}, {Ageorges}, {Barrows}, {Buschkamp}, {Dietrich}, {Falco},
  {Feiz}, {Gemperlein}, {Germeroth}, {Grier}, {Hofmann}, {Juette}, {Khan},
  {Kilic}, {Knierim}, {Laun}, {Lederer}, {Lehmitz}, {Lenzen}, {Mall}, {Madsen},
  {Mandel}, {Martini}, {Mathur}, {Mogren}, {Mueller}, {Naranjo}, {Pasquali},
  {Polsterer}, {Pogge}, {Quirrenbach}, {Seifert}, {Stern}, {Shappee}, {Storz},
  {Van Saders}, {Weiser}, \& {Zhang}}]{Assef+2011}
{Assef}, R.~J., {Denney}, K.~D., {Kochanek}, C.~S., {et~al.} 2011, \apj, 742,
  93, \dodoi{10.1088/0004-637X/742/2/93}

\bibitem[{{Bahk} {et~al.}(2019){Bahk}, {Woo}, \& {Park}}]{Bahk+2019}
{Bahk}, H., {Woo}, J.-H., \& {Park}, D. 2019, \apj, 875, 50,
  \dodoi{10.3847/1538-4357/ab100d}

\bibitem[{{Baldwin} {et~al.}(2004){Baldwin}, {Ferland}, {Korista}, {Hamann}, \&
  {LaCluyz{\'e}}}]{Baldwin+2004}
{Baldwin}, J.~A., {Ferland}, G.~J., {Korista}, K.~T., {Hamann}, F., \&
  {LaCluyz{\'e}}, A. 2004, \apj, 615, 610, \dodoi{10.1086/424683}

\bibitem[{{Barth} {et~al.}(2011){Barth}, {Pancoast}, {Thorman}, {Bennert},
  {Sand}, {Li}, {Canalizo}, {Filippenko}, {Gates}, {Greene}, {Malkan}, {Stern},
  {Treu}, {Woo}, {Assef}, {Bae}, {Brewer}, {Buehler}, {Cenko}, {Clubb},
  {Cooper}, {Diamond-Stanic}, {Hiner}, {H{\"o}nig}, {Joner}, {Kandrashoff},
  {Laney}, {Lazarova}, {Nierenberg}, {Park}, {Silverman}, {Son}, {Sonnenfeld},
  {Tollerud}, {Walsh}, {Walters}, {da Silva}, {Fumagalli}, {Gregg}, {Harris},
  {Hsiao}, {Lee}, {Lopez}, {Rex}, {Suzuki}, {Trump}, {Tytler}, {Worseck}, \&
  {Yesuf}}]{Barth+2011:mrk50}
{Barth}, A.~J., {Pancoast}, A., {Thorman}, S.~J., {et~al.} 2011, \apjl, 743,
  L4, \dodoi{10.1088/2041-8205/743/1/L4}

\bibitem[{{Barth} {et~al.}(2013){Barth}, {Pancoast}, {Bennert}, {Brewer},
  {Canalizo}, {Filippenko}, {Gates}, {Greene}, {Li}, {Malkan}, {Sand}, {Stern},
  {Treu}, {Woo}, {Assef}, {Bae}, {Buehler}, {Cenko}, {Clubb}, {Cooper},
  {Diamond-Stanic}, {H{\"o}nig}, {Joner}, {Laney}, {Lazarova}, {Nierenberg},
  {Silverman}, {Tollerud}, \& {Walsh}}]{Barth+2013}
{Barth}, A.~J., {Pancoast}, A., {Bennert}, V.~N., {et~al.} 2013, \apj, 769,
  128, \dodoi{10.1088/0004-637X/769/2/128}

\bibitem[{{Barth} {et~al.}(2015){Barth}, {Bennert}, {Canalizo}, {Filippenko},
  {Gates}, {Greene}, {Li}, {Malkan}, {Pancoast}, {Sand}, {Stern}, {Treu},
  {Woo}, {Assef}, {Bae}, {Brewer}, {Cenko}, {Clubb}, {Cooper},
  {Diamond-Stanic}, {Hiner}, {H{\"o}nig}, {Hsiao}, {Kandrashoff}, {Lazarova},
  {Nierenberg}, {Rex}, {Silverman}, {Tollerud}, \& {Walsh}}]{Barth+2015}
{Barth}, A.~J., {Bennert}, V.~N., {Canalizo}, G., {et~al.} 2015, \apjs, 217,
  26, \dodoi{10.1088/0067-0049/217/2/26}

\bibitem[{{Baskin} \& {Laor}(2005{\natexlab{a}})}]{BaskinLaor2005}
{Baskin}, A., \& {Laor}, A. 2005{\natexlab{a}}, \mnras, 358, 1043,
  \dodoi{10.1111/j.1365-2966.2005.08841.x}

\bibitem[{{Baskin} \& {Laor}(2005{\natexlab{b}})}]{BaskinLaor2005_CIV}
---. 2005{\natexlab{b}}, \mnras, 356, 1029,
  \dodoi{10.1111/j.1365-2966.2004.08525.x}

\bibitem[{{Bennert} {et~al.}(2011){Bennert}, {Auger}, {Treu}, {Woo}, \&
  {Malkan}}]{Bennert+2011}
{Bennert}, V.~N., {Auger}, M.~W., {Treu}, T., {Woo}, J.-H., \& {Malkan}, M.~A.
  2011, \apj, 726, 59, \dodoi{10.1088/0004-637X/726/2/59}

\bibitem[{{Bennert} {et~al.}(2015){Bennert}, {Treu}, {Auger}, {Cosens}, {Park},
  {Rosen}, {Harris}, {Malkan}, \& {Woo}}]{Bennert+2015}
{Bennert}, V.~N., {Treu}, T., {Auger}, M.~W., {et~al.} 2015, \apj, 809, 20,
  \dodoi{10.1088/0004-637X/809/1/20}

\bibitem[{{Bennert} {et~al.}(2018){Bennert}, {Loveland}, {Donohue}, {Cosens},
  {Lewis}, {Komossa}, {Treu}, {Malkan}, {Milgram}, {Flatland}, {Auger}, {Park},
  \& {Lazarova}}]{Bennert+2018}
{Bennert}, V.~N., {Loveland}, D., {Donohue}, E., {et~al.} 2018, \mnras, 481,
  138, \dodoi{10.1093/mnras/sty2236}

\bibitem[{{Boroson} \& {Green}(1992)}]{BorosonGreen1992}
{Boroson}, T.~A., \& {Green}, R.~F. 1992, \apjs, 80, 109,
  \dodoi{10.1086/191661}

\bibitem[{{Bruhweiler} \& {Verner}(2008)}]{BruhweilerVerner2008}
{Bruhweiler}, F., \& {Verner}, E. 2008, \apj, 675, 83, \dodoi{10.1086/525557}

\bibitem[{{Cackett} {et~al.}(2021){Cackett}, {Bentz}, \& {Kara}}]{Cackett+2021}
{Cackett}, E.~M., {Bentz}, M.~C., \& {Kara}, E. 2021, iScience, 24, 102557,
  \dodoi{10.1016/j.isci.2021.102557}

\bibitem[{{Calderone} {et~al.}(2013){Calderone}, {Ghisellini}, {Colpi}, \&
  {Dotti}}]{Calderone+2013}
{Calderone}, G., {Ghisellini}, G., {Colpi}, M., \& {Dotti}, M. 2013, \mnras,
  431, 210, \dodoi{10.1093/mnras/stt157}

\bibitem[{{Calderone} {et~al.}(2017){Calderone}, {Nicastro}, {Ghisellini},
  {Dotti}, {Sbarrato}, {Shankar}, \& {Colpi}}]{Calderone+2017}
{Calderone}, G., {Nicastro}, L., {Ghisellini}, G., {et~al.} 2017, \mnras, 472,
  4051, \dodoi{10.1093/mnras/stx2239}

\bibitem[{{Cappellari} {et~al.}(2002){Cappellari}, {Verolme}, {van der Marel},
  {Verdoes Kleijn}, {Illingworth}, {Franx}, {Carollo}, \& {de
  Zeeuw}}]{Cappellari+2002}
{Cappellari}, M., {Verolme}, E.~K., {van der Marel}, R.~P., {et~al.} 2002,
  \apj, 578, 787, \dodoi{10.1086/342653}

\bibitem[{{Coatman} {et~al.}(2016){Coatman}, {Hewett}, {Banerji}, \&
  {Richards}}]{Coatman+2016}
{Coatman}, L., {Hewett}, P.~C., {Banerji}, M., \& {Richards}, G.~T. 2016,
  \mnras, 461, 647, \dodoi{10.1093/mnras/stw1360}

\bibitem[{{Constantin} \& {Shields}(2003)}]{ConstantinShields2003}
{Constantin}, A., \& {Shields}, J.~C. 2003, \pasp, 115, 592,
  \dodoi{10.1086/374724}

\bibitem[{{Crenshaw} {et~al.}(2002){Crenshaw}, {Kraemer}, {Turner}, {Collier},
  {Peterson}, {Brandt}, {Clavel}, {George}, {Horne}, {Kriss}, {Mathur},
  {Netzer}, {Pogge}, {Pounds}, {Romano}, {Shemmer}, \&
  {Wamsteker}}]{Crenshaw+2002}
{Crenshaw}, D.~M., {Kraemer}, S.~B., {Turner}, T.~J., {et~al.} 2002, \apj, 566,
  187, \dodoi{10.1086/338058}

\bibitem[{{De Rosa} {et~al.}(2014){De Rosa}, {Venemans}, {Decarli}, {Gennaro},
  {Simcoe}, {Dietrich}, {Peterson}, {Walter}, {Frank}, {McMahon}, {Hewett},
  {Mortlock}, \& {Simpson}}]{DeRosa+2014}
{De Rosa}, G., {Venemans}, B.~P., {Decarli}, R., {et~al.} 2014, \apj, 790, 145,
  \dodoi{10.1088/0004-637X/790/2/145}

\bibitem[{{Denney} {et~al.}(2009){Denney}, {Peterson}, {Dietrich},
  {Vestergaard}, \& {Bentz}}]{Denney+2009}
{Denney}, K.~D., {Peterson}, B.~M., {Dietrich}, M., {Vestergaard}, M., \&
  {Bentz}, M.~C. 2009, \apj, 692, 246, \dodoi{10.1088/0004-637X/692/1/246}

\bibitem[{{Denney} {et~al.}(2013){Denney}, {Pogge}, {Assef}, {Kochanek},
  {Peterson}, \& {Vestergaard}}]{Denney+2013}
{Denney}, K.~D., {Pogge}, R.~W., {Assef}, R.~J., {et~al.} 2013, \apj, 775, 60,
  \dodoi{10.1088/0004-637X/775/1/60}

\bibitem[{{Denney} {et~al.}(2016){Denney}, {Horne}, {Shen}, {Brandt}, {Ho},
  {Peterson}, {Richards}, {Trump}, \& {Ge}}]{Denney+2016}
{Denney}, K.~D., {Horne}, K., {Shen}, Y., {et~al.} 2016, \apjs, 224, 14,
  \dodoi{10.3847/0067-0049/224/2/14}

\bibitem[{{Dietrich} {et~al.}(2002){Dietrich}, {Appenzeller}, {Vestergaard}, \&
  {Wagner}}]{Dietrich+2002}
{Dietrich}, M., {Appenzeller}, I., {Vestergaard}, M., \& {Wagner}, S.~J. 2002,
  \apj, 564, 581, \dodoi{10.1086/324337}

\bibitem[{{Ding} {et~al.}(2017){Ding}, {Treu}, {Suyu}, {Wong}, {Morishita},
  {Park}, {Sluse}, {Auger}, {Agnello}, {Bennert}, \& {Collett}}]{Ding+2017}
{Ding}, X., {Treu}, T., {Suyu}, S.~H., {et~al.} 2017, \mnras, 472, 90,
  \dodoi{10.1093/mnras/stx1972}

\bibitem[{{Ding} {et~al.}(2020){Ding}, {Silverman}, {Treu}, {Schulze},
  {Schramm}, {Birrer}, {Park}, {Jahnke}, {Bennert}, {Kartaltepe}, {Koekemoer},
  {Malkan}, \& {Sanders}}]{Ding+2020}
{Ding}, X., {Silverman}, J., {Treu}, T., {et~al.} 2020, \apj, 888, 37,
  \dodoi{10.3847/1538-4357/ab5b90}

\bibitem[{{Dong} {et~al.}(2008){Dong}, {Wang}, {Wang}, {Yuan}, {Zhou}, {Dai},
  \& {Zhang}}]{Dong+2008}
{Dong}, X., {Wang}, T., {Wang}, J., {et~al.} 2008, \mnras, 383, 581,
  \dodoi{10.1111/j.1365-2966.2007.12560.x}

\bibitem[{{Dong} {et~al.}(2010){Dong}, {Ho}, {Wang}, {Wang}, {Wang}, {Fan}, \&
  {Zhou}}]{Dong+2010}
{Dong}, X.-B., {Ho}, L.~C., {Wang}, J.-G., {et~al.} 2010, \apjl, 721, L143,
  \dodoi{10.1088/2041-8205/721/2/L143}

\bibitem[{{Dong} {et~al.}(2011){Dong}, {Wang}, {Ho}, {Wang}, {Fan}, {Wang},
  {Zhou}, \& {Yuan}}]{Dong+2011}
{Dong}, X.-B., {Wang}, J.-G., {Ho}, L.~C., {et~al.} 2011, \apj, 736, 86,
  \dodoi{10.1088/0004-637X/736/2/86}

\bibitem[{{Fitzpatrick}(1999)}]{Fitzpatrick1999}
{Fitzpatrick}, E.~L. 1999, \pasp, 111, 63, \dodoi{10.1086/316293}

\bibitem[{{Garcia-Rissmann} {et~al.}(2012){Garcia-Rissmann},
  {Rodr{\'\i}guez-Ardila}, {Sigut}, \& {Pradhan}}]{Garcia-Rissmann2012}
{Garcia-Rissmann}, A., {Rodr{\'\i}guez-Ardila}, A., {Sigut}, T.~A.~A., \&
  {Pradhan}, A.~K. 2012, \apj, 751, 7, \dodoi{10.1088/0004-637X/751/1/7}

\bibitem[{{Greene} \& {Ho}(2005)}]{GreeneHo2005}
{Greene}, J.~E., \& {Ho}, L.~C. 2005, \apj, 630, 122, \dodoi{10.1086/431897}

\bibitem[{{Greene} \& {Ho}(2007)}]{Greene&Ho2007}
---. 2007, \apj, 670, 92, \dodoi{10.1086/522082}

\bibitem[{{Grier} {et~al.}(2017){Grier}, {Pancoast}, {Barth}, {Fausnaugh},
  {Brewer}, {Treu}, \& {Peterson}}]{Grier+2017}
{Grier}, C.~J., {Pancoast}, A., {Barth}, A.~J., {et~al.} 2017, \apj, 849, 146,
  \dodoi{10.3847/1538-4357/aa901b}

\bibitem[{{Ho} {et~al.}(2012){Ho}, {Goldoni}, {Dong}, {Greene}, \&
  {Ponti}}]{Ho+2012}
{Ho}, L.~C., {Goldoni}, P., {Dong}, X.-B., {Greene}, J.~E., \& {Ponti}, G.
  2012, \apj, 754, 11, \dodoi{10.1088/0004-637X/754/1/11}

\bibitem[{{Hu} {et~al.}(2008){Hu}, {Wang}, {Ho}, {Chen}, {Zhang}, {Bian}, \&
  {Xue}}]{Hu+2008}
{Hu}, C., {Wang}, J.-M., {Ho}, L.~C., {et~al.} 2008, \apj, 687, 78,
  \dodoi{10.1086/591838}

\bibitem[{{Hu} {et~al.}(2015){Hu}, {Du}, {Lu}, {Li}, {Wang}, {Qiu}, {Bai},
  {Kaspi}, {Ho}, {Netzer}, {Wang}, \& {SEAMBH Collaboration}}]{Hu+2015}
{Hu}, C., {Du}, P., {Lu}, K.-X., {et~al.} 2015, \apj, 804, 138,
  \dodoi{10.1088/0004-637X/804/2/138}

\bibitem[{{Hu} {et~al.}(2021){Hu}, {Li}, {Yang}, {Yang}, {Guo}, {Bao}, {Jiang},
  {Du}, {Li}, {Xiao}, {Songsheng}, {Yu}, {Bai}, {Ho}, {Brotherton}, {Aceituno},
  {Winkler}, {Wang}, \& {Seambh Collaboration}}]{Hu+2021}
{Hu}, C., {Li}, S.-S., {Yang}, S., {et~al.} 2021, \apjs, 253, 20,
  \dodoi{10.3847/1538-4365/abd774}

\bibitem[{{Huang} {et~al.}(2019){Huang}, {Hu}, {Zhao}, {Zhang}, {Lu}, {Wang},
  {Zhang}, {Du}, {Li}, {Bai}, {Ho}, {Bian}, {Yuan}, \& {Wang}}]{Huang+2019}
{Huang}, Y.-K., {Hu}, C., {Zhao}, Y.-L., {et~al.} 2019, \apj, 876, 102,
  \dodoi{10.3847/1538-4357/ab16ef}

\bibitem[{{Kelly} \& {Shen}(2013)}]{KellyShen2013}
{Kelly}, B.~C., \& {Shen}, Y. 2013, \apj, 764, 45,
  \dodoi{10.1088/0004-637X/764/1/45}

\bibitem[{{Kova{\v c}evi{\'c}} {et~al.}(2010){Kova{\v c}evi{\'c}},
  {Popovi{\'c}}, \& {Dimitrijevi{\'c}}}]{Kovacevic+2010}
{Kova{\v c}evi{\'c}}, J., {Popovi{\'c}}, L.~{\v C}., \& {Dimitrijevi{\'c}},
  M.~S. 2010, \apjs, 189, 15, \dodoi{10.1088/0067-0049/189/1/15}

\bibitem[{{Kova{\v{c}}evi{\'c}-Doj{\v{c}}inovi{\'c}} \&
  {Popovi{\'c}}(2015)}]{KovacevicPopovic2015}
{Kova{\v{c}}evi{\'c}-Doj{\v{c}}inovi{\'c}}, J., \& {Popovi{\'c}}, L.~{\v{C}}.
  2015, \apjs, 221, 35, \dodoi{10.1088/0067-0049/221/2/35}

\bibitem[{{Kurk} {et~al.}(2007){Kurk}, {Walter}, {Fan}, {Jiang}, {Riechers},
  {Rix}, {Pentericci}, {Strauss}, {Carilli}, \& {Wagner}}]{Kurk+2007}
{Kurk}, J.~D., {Walter}, F., {Fan}, X., {et~al.} 2007, \apj, 669, 32,
  \dodoi{10.1086/521596}

\bibitem[{{Laor} {et~al.}(1997){Laor}, {Jannuzi}, {Green}, \&
  {Boroson}}]{Laor+1997}
{Laor}, A., {Jannuzi}, B.~T., {Green}, R.~F., \& {Boroson}, T.~A. 1997, \apj,
  489, 656, \dodoi{10.1086/304816}

\bibitem[{{Li} {et~al.}(2018){Li}, {Songsheng}, {Qiu}, {Hu}, {Du}, {Lu},
  {Huang}, {Bai}, {Bian}, {Yuan}, {Ho}, \& {Wang}}]{Li+2018}
{Li}, Y.-R., {Songsheng}, Y.-Y., {Qiu}, J., {et~al.} 2018, \apj, 869, 137,
  \dodoi{10.3847/1538-4357/aaee6b}

\bibitem[{{Lu} {et~al.}(2019){Lu}, {Bai}, {Zhang}, {Du}, {Hu}, {Kim}, {Wang},
  {Ho}, {Li}, {Bian}, {Yuan}, {Xiao}, {Feng}, {Wang}, {Xu}, {Ding}, {Yu},
  {Xin}, {Ye}, {Wang}, {Lun}, {Zhang}, {Zhang}, {Ji}, {Fan}, \&
  {Chang}}]{Lu+2019}
{Lu}, K.-X., {Bai}, J.-M., {Zhang}, Z.-X., {et~al.} 2019, \apj, 887, 135,
  \dodoi{10.3847/1538-4357/ab5790}

\bibitem[{{Lu} {et~al.}(2021){Lu}, {Wang}, {Zhang}, {Huang}, {Xu}, {Xin}, {Yu},
  {Ding}, {Wang}, \& {Feng}}]{Lu+2021}
{Lu}, K.-X., {Wang}, J.-G., {Zhang}, Z.-X., {et~al.} 2021, arXiv e-prints,
  arXiv:2106.10589.
\newblock \doarXiv{2106.10589}

\bibitem[{{Marinello} {et~al.}(2016){Marinello}, {Rodr{\'\i}guez-Ardila},
  {Garcia-Rissmann}, {Sigut}, \& {Pradhan}}]{Marinello+2016}
{Marinello}, M., {Rodr{\'\i}guez-Ardila}, A., {Garcia-Rissmann}, A., {Sigut},
  T.~A.~A., \& {Pradhan}, A.~K. 2016, \apj, 820, 116,
  \dodoi{10.3847/0004-637X/820/2/116}

\bibitem[{{Markwardt}(2009)}]{Markwardt2009}
{Markwardt}, C.~B. 2009, in Astronomical Society of the Pacific Conference
  Series, Vol. 411, Astronomical Data Analysis Software and Systems XVIII, ed.
  D.~A. {Bohlender}, D.~{Durand}, \& P.~{Dowler}, 251.
\newblock \doarXiv{0902.2850}

\bibitem[{{Marziani} {et~al.}(2003){Marziani}, {Sulentic}, {Zamanov},
  {Calvani}, {Dultzin-Hacyan}, {Bachev}, \& {Zwitter}}]{Marziani2003}
{Marziani}, P., {Sulentic}, J.~W., {Zamanov}, R., {et~al.} 2003, \apjs, 145,
  199, \dodoi{10.1086/346025}

\bibitem[{{McGill} {et~al.}(2008){McGill}, {Woo}, {Treu}, \&
  {Malkan}}]{McGill+2008}
{McGill}, K.~L., {Woo}, J.-H., {Treu}, T., \& {Malkan}, M.~A. 2008, \apj, 673,
  703, \dodoi{10.1086/524349}

\bibitem[{{McLure} \& {Dunlop}(2004)}]{McLureDunlop2004}
{McLure}, R.~J., \& {Dunlop}, J.~S. 2004, \mnras, 352, 1390,
  \dodoi{10.1111/j.1365-2966.2004.08034.x}

\bibitem[{{Mej{\'\i}a-Restrepo} {et~al.}(2016){Mej{\'\i}a-Restrepo},
  {Trakhtenbrot}, {Lira}, {Netzer}, \& {Capellupo}}]{Mejia-Restrepo+2016}
{Mej{\'\i}a-Restrepo}, J.~E., {Trakhtenbrot}, B., {Lira}, P., {Netzer}, H., \&
  {Capellupo}, D.~M. 2016, \mnras, 460, 187, \dodoi{10.1093/mnras/stw568}

\bibitem[{{Osterbrock} \& {Pogge}(1985)}]{Osterbrock1985}
{Osterbrock}, D.~E., \& {Pogge}, R.~W. 1985, \apj, 297, 166,
  \dodoi{10.1086/163513}

\bibitem[{{Pancoast} {et~al.}(2014){Pancoast}, {Brewer}, {Treu}, {Park},
  {Barth}, {Bentz}, \& {Woo}}]{Pancoast+2014-II}
{Pancoast}, A., {Brewer}, B.~J., {Treu}, T., {et~al.} 2014, \mnras, 445, 3073,
  \dodoi{10.1093/mnras/stu1419}

\bibitem[{{Pancoast} {et~al.}(2018){Pancoast}, {Barth}, {Horne}, {Treu},
  {Brewer}, {Bennert}, {Canalizo}, {Gates}, {Li}, {Malkan}, {Sand}, {Schmidt},
  {Valenti}, {Woo}, {Clubb}, {Cooper}, {Crawford}, {H{\"o}nig}, {Joner},
  {Kandrashoff}, {Lazarova}, {Nierenberg}, {Romero-Colmenero}, {Son},
  {Tollerud}, {Walsh}, \& {Winkler}}]{Pancoast+2018}
{Pancoast}, A., {Barth}, A.~J., {Horne}, K., {et~al.} 2018, \apj, 856, 108,
  \dodoi{10.3847/1538-4357/aab3c6}

\bibitem[{{Park} {et~al.}(2017){Park}, {Barth}, {Woo}, {Malkan}, {Treu},
  {Bennert}, {Assef}, \& {Pancoast}}]{Park+2017}
{Park}, D., {Barth}, A.~J., {Woo}, J.-H., {et~al.} 2017, \apj, 839, 93,
  \dodoi{10.3847/1538-4357/aa6a53}

\bibitem[{{Park} {et~al.}(2015){Park}, {Woo}, {Bennert}, {Treu}, {Auger}, \&
  {Malkan}}]{Park+2015}
{Park}, D., {Woo}, J.-H., {Bennert}, V.~N., {et~al.} 2015, \apj, 799, 164,
  \dodoi{10.1088/0004-637X/799/2/164}

\bibitem[{{Park} {et~al.}(2013){Park}, {Woo}, {Denney}, \& {Shin}}]{Park+2013}
{Park}, D., {Woo}, J.-H., {Denney}, K.~D., \& {Shin}, J. 2013, \apj, 770, 87,
  \dodoi{10.1088/0004-637X/770/2/87}

\bibitem[{{Park} {et~al.}(2012){Park}, {Woo}, {Treu}, {Barth}, {Bentz},
  {Bennert}, {Canalizo}, {Filippenko}, {Gates}, {Greene}, {Malkan}, \&
  {Walsh}}]{Park+2012}
{Park}, D., {Woo}, J.-H., {Treu}, T., {et~al.} 2012, \apj, 747, 30,
  \dodoi{10.1088/0004-637X/747/1/30}

\bibitem[{{Pei} {et~al.}(2017){Pei}, {Fausnaugh}, {Barth}, {Peterson}, {Bentz},
  {De Rosa}, {Denney}, {Goad}, {Kochanek}, {Korista}, {Kriss}, {Pogge},
  {Bennert}, {Brotherton}, {Clubb}, {Dalla Bont{\`a}}, {Filippenko}, {Greene},
  {Grier}, {Vestergaard}, {Zheng}, {Adams}, {Beatty}, {Bigley}, {Brown},
  {Brown}, {Canalizo}, {Comerford}, {Coker}, {Corsini}, {Croft}, {Croxall},
  {Deason}, {Eracleous}, {Fox}, {Gates}, {Henderson}, {Holmbeck}, {Holoien},
  {Jensen}, {Johnson}, {Kelly}, {Kim}, {King}, {Lau}, {Li}, {Lochhaas}, {Ma},
  {Manne-Nicholas}, {Mauerhan}, {Malkan}, {McGurk}, {Morelli}, {Mosquera},
  {Mudd}, {Muller Sanchez}, {Nguyen}, {Ochner}, {Ou-Yang}, {Pancoast}, {Penny},
  {Pizzella}, {Poleski}, {Runnoe}, {Scott}, {Schimoia}, {Shappee}, {Shivvers},
  {Simonian}, {Siviero}, {Somers}, {Stevens}, {Strauss}, {Tayar}, {Tejos},
  {Treu}, {Van Saders}, {Vican}, {Villanueva}, {Yuk}, {Zakamska}, {Zhu},
  {Anderson}, {Ar{\'e}valo}, {Bazhaw}, {Bisogni}, {Borman}, {Bottorff},
  {Brandt}, {Breeveld}, {Cackett}, {Carini}, {Crenshaw}, {De
  Lorenzo-C{\'a}ceres}, {Dietrich}, {Edelson}, {Efimova}, {Ely}, {Evans},
  {Ferland}, {Flatland}, {Gehrels}, {Geier}, {Gelbord}, {Grupe}, {Gupta},
  {Hall}, {Hicks}, {Horenstein}, {Horne}, {Hutchison}, {Im}, {Joner}, {Jones},
  {Kaastra}, {Kaspi}, {Kelly}, {Kennea}, {Kim}, {Kim}, {Klimanov}, {Lee},
  {Leonard}, {Lira}, {MacInnis}, {Mathur}, {McHardy}, {Montouri}, {Musso},
  {Nazarov}, {Netzer}, {Norris}, {Nousek}, {Okhmat}, {Papadakis}, {Parks},
  {Pott}, {Rafter}, {Rix}, {Saylor}, {Schn{\"u}lle}, {Sergeev}, {Siegel},
  {Skielboe}, {Spencer}, {Starkey}, {Sung}, {Teems}, {Turner}, {Uttley},
  {Villforth}, {Weiss}, {Woo}, {Yan}, {Young}, \& {Zu}}]{Pei+2017}
{Pei}, L., {Fausnaugh}, M.~M., {Barth}, A.~J., {et~al.} 2017, \apj, 837, 131,
  \dodoi{10.3847/1538-4357/aa5eb1}

\bibitem[{{Peterson}(1993)}]{Peterson1993}
{Peterson}, B.~M. 1993, \pasp, 105, 247, \dodoi{10.1086/133140}

\bibitem[{{Popovic} {et~al.}(2013){Popovic}, {Kovacevic}, \&
  {Dimitrijevic}}]{Popovic+2013}
{Popovic}, L.~C., {Kovacevic}, J., \& {Dimitrijevic}, M.~S. 2013, ArXiv
  e-prints.
\newblock \doarXiv{1301.6941}

\bibitem[{{Popovi{\'c}} {et~al.}(2019){Popovi{\'c}},
  {Kova{\v{c}}evi{\'c}-Doj{\v{c}}inovi{\'c}}, \&
  {Mar{\v{c}}eta-Mandi{\'c}}}]{Popovic+2019}
{Popovi{\'c}}, L.~{\v{C}}., {Kova{\v{c}}evi{\'c}-Doj{\v{c}}inovi{\'c}}, J., \&
  {Mar{\v{c}}eta-Mandi{\'c}}, S. 2019, \mnras, 484, 3180,
  \dodoi{10.1093/mnras/stz157}

\bibitem[{{Rakshit} {et~al.}(2020){Rakshit}, {Stalin}, \&
  {Kotilainen}}]{Rakshit+2020}
{Rakshit}, S., {Stalin}, C.~S., \& {Kotilainen}, J. 2020, \apjs, 249, 17,
  \dodoi{10.3847/1538-4365/ab99c5}

\bibitem[{{Rakshit} {et~al.}(2019){Rakshit}, {Woo}, {Gallo}, {Hodges-Kluck},
  {Shin}, {Jeon}, {Bae}, {Baldassare}, {Cho}, {Cho}, {Foord}, {Kang}, {Kang},
  {Karouzos}, {Kim}, {Kim}, {Le}, {Park}, {Park}, {Son}, {Sung}, {Bennert}, \&
  {Malkan}}]{Rakshit+2019}
{Rakshit}, S., {Woo}, J.-H., {Gallo}, E., {et~al.} 2019, \apj, 886, 93,
  \dodoi{10.3847/1538-4357/ab49fd}

\bibitem[{{Sarkar} {et~al.}(2021){Sarkar}, {Ferland}, {Chatzikos},
  {Guzm{\'a}n}, {van Hoof}, {Smyth}, {Ramsbottom}, {Keenan}, \&
  {Ballance}}]{Sarkar+2021}
{Sarkar}, A., {Ferland}, G.~J., {Chatzikos}, M., {et~al.} 2021, \apj, 907, 12,
  \dodoi{10.3847/1538-4357/abcaa6}

\bibitem[{{Schlafly} \& {Finkbeiner}(2011)}]{Schlafly+2011}
{Schlafly}, E.~F., \& {Finkbeiner}, D.~P. 2011, \apj, 737, 103,
  \dodoi{10.1088/0004-637X/737/2/103}

\bibitem[{{Schulze} {et~al.}(2018){Schulze}, {Silverman}, {Kashino}, {Akiyama},
  {Schramm}, {Sanders}, {Kartaltepe}, {Daddi}, {Rodighiero}, {Renzini},
  {Arimoto}, {Nagao}, {Puglisi}, {Trakhtenbrot}, {Civano}, \&
  {Suh}}]{Schulze+2018}
{Schulze}, A., {Silverman}, J.~D., {Kashino}, D., {et~al.} 2018, \apjs, 239,
  22, \dodoi{10.3847/1538-4365/aae82f}

\bibitem[{{Science Software Branch at STScI}(2012)}]{PyRAF2012}
{Science Software Branch at STScI}. 2012, {PyRAF: Python alternative for IRAF}.
\newblock \doeprint{1207.011}

\bibitem[{{Shapovalova} {et~al.}(2012){Shapovalova}, {Popovi{\'c}}, {Burenkov},
  {Chavushyan}, {Ili{\'c}}, {Kova{\v{c}}evi{\'c}}, {Kollatschny},
  {Kova{\v{c}}evi{\'c}}, {Bochkarev}, {Valdes}, {Torrealba},
  {Le{\'o}n-Tavares}, {Mercado}, {Ben{\'\i}tez}, {Carrasco}, {Dultzin}, \& {de
  la Fuente}}]{Shapovalova+2012}
{Shapovalova}, A.~I., {Popovi{\'c}}, L.~{\v{C}}., {Burenkov}, A.~N., {et~al.}
  2012, \apjs, 202, 10, \dodoi{10.1088/0067-0049/202/1/10}

\bibitem[{{Shen}(2013)}]{Shen2013}
{Shen}, Y. 2013, Bulletin of the Astronomical Society of India, 41, 61.
\newblock \doarXiv{1302.2643}

\bibitem[{{Shen} {et~al.}(2008){Shen}, {Greene}, {Strauss}, {Richards}, \&
  {Schneider}}]{Shen+2008}
{Shen}, Y., {Greene}, J.~E., {Strauss}, M.~A., {Richards}, G.~T., \&
  {Schneider}, D.~P. 2008, \apj, 680, 169, \dodoi{10.1086/587475}

\bibitem[{{Shen} \& {Ho}(2014)}]{ShenHo2014}
{Shen}, Y., \& {Ho}, L.~C. 2014, \nat, 513, 210, \dodoi{10.1038/nature13712}

\bibitem[{{Shen} \& {Kelly}(2012)}]{ShenKelly2012}
{Shen}, Y., \& {Kelly}, B.~C. 2012, \apj, 746, 169,
  \dodoi{10.1088/0004-637X/746/2/169}

\bibitem[{{Shen} \& {Liu}(2012)}]{ShenLiu2012}
{Shen}, Y., \& {Liu}, X. 2012, \apj, 753, 125,
  \dodoi{10.1088/0004-637X/753/2/125}

\bibitem[{{Shen} {et~al.}(2011){Shen}, {Richards}, {Strauss}, {Hall},
  {Schneider}, {Snedden}, {Bizyaev}, {Brewington}, {Malanushenko},
  {Malanushenko}, {Oravetz}, {Pan}, \& {Simmons}}]{Shen+2011}
{Shen}, Y., {Richards}, G.~T., {Strauss}, M.~A., {et~al.} 2011, \apjs, 194, 45,
  \dodoi{10.1088/0067-0049/194/2/45}

\bibitem[{{Sigut} \& {Pradhan}(1998)}]{SigutPradhan1998}
{Sigut}, T.~A.~A., \& {Pradhan}, A.~K. 1998, \apjl, 499, L139,
  \dodoi{10.1086/311369}

\bibitem[{{Sigut} \& {Pradhan}(2003)}]{SigutPradhan2003}
---. 2003, \apjs, 145, 15, \dodoi{10.1086/345498}

\bibitem[{{Storey} \& {Zeippen}(2000)}]{StoreyZeippen2000}
{Storey}, P.~J., \& {Zeippen}, C.~J. 2000, \mnras, 312, 813,
  \dodoi{10.1046/j.1365-8711.2000.03184.x}

\bibitem[{{Sulentic} {et~al.}(2002){Sulentic}, {Marziani}, {Zamanov}, {Bachev},
  {Calvani}, \& {Dultzin-Hacyan}}]{Sulentic+2002}
{Sulentic}, J.~W., {Marziani}, P., {Zamanov}, R., {et~al.} 2002, \apjl, 566,
  L71, \dodoi{10.1086/339594}

\bibitem[{{Sulentic} {et~al.}(2009){Sulentic}, {Marziani}, \&
  {Zamfir}}]{Sulentic+2009}
{Sulentic}, J.~W., {Marziani}, P., \& {Zamfir}, S. 2009, \nar, 53, 198,
  \dodoi{10.1016/j.newar.2009.06.001}

\bibitem[{{Sulentic} {et~al.}(2000){Sulentic}, {Zwitter}, {Marziani}, \&
  {Dultzin-Hacyan}}]{Sulentic+2000}
{Sulentic}, J.~W., {Zwitter}, T., {Marziani}, P., \& {Dultzin-Hacyan}, D. 2000,
  \apjl, 536, L5, \dodoi{10.1086/312717}

\bibitem[{{Trakhtenbrot} \& {Netzer}(2012)}]{TrakhtenbrotNetzer2012}
{Trakhtenbrot}, B., \& {Netzer}, H. 2012, \mnras, 427, 3081,
  \dodoi{10.1111/j.1365-2966.2012.22056.x}

\bibitem[{{Tsuzuki} {et~al.}(2006){Tsuzuki}, {Kawara}, {Yoshii}, {Oyabu},
  {Tanab{\'e}}, \& {Matsuoka}}]{Tsuzuki+2006}
{Tsuzuki}, Y., {Kawara}, K., {Yoshii}, Y., {et~al.} 2006, \apj, 650, 57,
  \dodoi{10.1086/506376}

\bibitem[{{Valdes} {et~al.}(2004){Valdes}, {Gupta}, {Rose}, {Singh}, \&
  {Bell}}]{Valdes+2004}
{Valdes}, F., {Gupta}, R., {Rose}, J.~A., {Singh}, H.~P., \& {Bell}, D.~J.
  2004, \apjs, 152, 251, \dodoi{10.1086/386343}

\bibitem[{{van der Marel} \& {Franx}(1993)}]{vanderMarelFranx1993}
{van der Marel}, R.~P., \& {Franx}, M. 1993, \apj, 407, 525,
  \dodoi{10.1086/172534}

\bibitem[{{van Dokkum}(2001)}]{vanDokkum2001}
{van Dokkum}, P.~G. 2001, \pasp, 113, 1420, \dodoi{10.1086/323894}

\bibitem[{{Vanden Berk} {et~al.}(2001){Vanden Berk}, {Richards}, {Bauer},
  {Strauss}, {Schneider}, {Heckman}, {York}, {Hall}, {Fan}, {Knapp},
  {Anderson}, {Annis}, {Bahcall}, {Bernardi}, {Briggs}, {Brinkmann}, {Brunner},
  {Burles}, {Carey}, {Castander}, {Connolly}, {Crocker}, {Csabai}, {Doi},
  {Finkbeiner}, {Friedman}, {Frieman}, {Fukugita}, {Gunn}, {Hennessy},
  {Ivezi{\'c}}, {Kent}, {Kunszt}, {Lamb}, {Leger}, {Long}, {Loveday}, {Lupton},
  {Meiksin}, {Merelli}, {Munn}, {Newberg}, {Newcomb}, {Nichol}, {Owen}, {Pier},
  {Pope}, {Rockosi}, {Schlegel}, {Siegmund}, {Smee}, {Snir}, {Stoughton},
  {Stubbs}, {SubbaRao}, {Szalay}, {Szokoly}, {Tremonti}, {Uomoto}, {Waddell},
  {Yanny}, \& {Zheng}}]{VandenBerk+2001}
{Vanden Berk}, D.~E., {Richards}, G.~T., {Bauer}, A., {et~al.} 2001, \aj, 122,
  549, \dodoi{10.1086/321167}

\bibitem[{{Verner} {et~al.}(2004){Verner}, {Bruhweiler}, {Verner}, {Johansson},
  {Kallman}, \& {Gull}}]{Verner+2004}
{Verner}, E., {Bruhweiler}, F., {Verner}, D., {et~al.} 2004, \apj, 611, 780,
  \dodoi{10.1086/422303}

\bibitem[{{Verner} {et~al.}(1999){Verner}, {Verner}, {Korista}, {Ferguson},
  {Hamann}, \& {Ferland}}]{Verner+1999}
{Verner}, E.~M., {Verner}, D.~A., {Korista}, K.~T., {et~al.} 1999, \apjs, 120,
  101, \dodoi{10.1086/313171}

\bibitem[{{V{\'e}ron} {et~al.}(2002){V{\'e}ron}, {Gon{\c{c}}alves}, \&
  {V{\'e}ron-Cetty}}]{Veron+2002}
{V{\'e}ron}, P., {Gon{\c{c}}alves}, A.~C., \& {V{\'e}ron-Cetty}, M.~P. 2002,
  \aap, 384, 826, \dodoi{10.1051/0004-6361:20020072}

\bibitem[{{V{\'e}ron-Cetty} {et~al.}(2004){V{\'e}ron-Cetty}, {Joly}, \&
  {V{\'e}ron}}]{Veron-Cetty2004}
{V{\'e}ron-Cetty}, M.-P., {Joly}, M., \& {V{\'e}ron}, P. 2004, \aap, 417, 515,
  \dodoi{10.1051/0004-6361:20035714}

\bibitem[{{Vestergaard} \& {Wilkes}(2001)}]{VestergaardWilkes2001}
{Vestergaard}, M., \& {Wilkes}, B.~J. 2001, \apjs, 134, 1,
  \dodoi{10.1086/320357}

\bibitem[{{Vincentelli} {et~al.}(2021){Vincentelli}, {McHardy}, {Cackett},
  {Barth}, {Horne}, {Goad}, {Korista}, {Gelbord}, {Brandt}, {Edelson},
  {Miller}, {Pahari}, {Peterson}, {Schmidt}, {Baldi}, {Breedt}, {Hern{\'a}ndez
  Santisteban}, {Romero-Colmenero}, {Ward}, \& {Williams}}]{Vincentelli+2021}
{Vincentelli}, F.~M., {McHardy}, I., {Cackett}, E.~M., {et~al.} 2021, \mnras,
  504, 4337, \dodoi{10.1093/mnras/stab1033}

\bibitem[{{Wang} {et~al.}(2009){Wang}, {Dong}, {Wang}, {Ho}, {Yuan}, {Wang},
  {Zhang}, {Zhang}, \& {Zhou}}]{Wang+2009}
{Wang}, J.-G., {Dong}, X.-B., {Wang}, T.-G., {et~al.} 2009, \apj, 707, 1334,
  \dodoi{10.1088/0004-637X/707/2/1334}

\bibitem[{{Wang} {et~al.}(2014){Wang}, {Du}, {Hu}, {Netzer}, {Bai}, {Lu},
  {Kaspi}, {Qiu}, {Li}, {Wang}, \& {SEAMBH Collaboration}}]{Wang+2014}
{Wang}, J.-M., {Du}, P., {Hu}, C., {et~al.} 2014, \apj, 793, 108,
  \dodoi{10.1088/0004-637X/793/2/108}

\bibitem[{{Wang} {et~al.}(2020){Wang}, {Shen}, {Jiang}, {Grier}, {Horne},
  {Homayouni}, {Peterson}, {Trump}, {Brandt}, {Hall}, {Ho}, {Li}, {Hernandez
  Santisteban}, {Kinemuchi}, {McGreer}, \& {Schneider}}]{Wang+2020}
{Wang}, S., {Shen}, Y., {Jiang}, L., {et~al.} 2020, \apj, 903, 51,
  \dodoi{10.3847/1538-4357/abb36d}

\bibitem[{{Wang} {et~al.}(2008){Wang}, {Dai}, \& {Zhou}}]{Wang+2008}
{Wang}, T., {Dai}, H., \& {Zhou}, H. 2008, \apj, 674, 668,
  \dodoi{10.1086/525242}

\bibitem[{{Williams} {et~al.}(2018){Williams}, {Pancoast}, {Treu}, {Brewer},
  {Barth}, {Bennert}, {Buehler}, {Canalizo}, {Cenko}, {Clubb}, {Cooper},
  {Filippenko}, {Gates}, {Hoenig}, {Joner}, {Kandrashoff}, {Laney}, {Lazarova},
  {Li}, {Malkan}, {Rex}, {Silverman}, {Tollerud}, {Walsh}, \&
  {Woo}}]{Williams+2018}
{Williams}, P.~R., {Pancoast}, A., {Treu}, T., {et~al.} 2018, \apj, 866, 75,
  \dodoi{10.3847/1538-4357/aae086}

\bibitem[{{Wills} {et~al.}(1999){Wills}, {Laor}, {Brotherton}, {Wills},
  {Wilkes}, {Ferland}, \& {Shang}}]{Wills+1999}
{Wills}, B.~J., {Laor}, A., {Brotherton}, M.~S., {et~al.} 1999, \apjl, 515,
  L53, \dodoi{10.1086/311980}

\bibitem[{{Wills} {et~al.}(1985){Wills}, {Netzer}, \& {Wills}}]{Wills+1985}
{Wills}, B.~J., {Netzer}, H., \& {Wills}, D. 1985, \apj, 288, 94,
  \dodoi{10.1086/162767}

\bibitem[{{Woo} {et~al.}(2008){Woo}, {Treu}, {Malkan}, \& {Bland
  ford}}]{Woo+2008}
{Woo}, J.-H., {Treu}, T., {Malkan}, M.~A., \& {Bland ford}, R.~D. 2008, \apj,
  681, 925, \dodoi{10.1086/588804}

\bibitem[{{Woo} {et~al.}(2007){Woo}, {Treu}, {Malkan}, {Ferry}, \&
  {Misch}}]{Woo+2007}
{Woo}, J.-H., {Treu}, T., {Malkan}, M.~A., {Ferry}, M.~A., \& {Misch}, T. 2007,
  \apj, 661, 60, \dodoi{10.1086/516564}

\end{thebibliography}

\clearpage
\appendix

\section{Line lists} \label{app:linelist}

\startlongtable
\begin{deluxetable}{lcc}
	\tablenum{A.1}
	\tablecolumns{3}
	\tablewidth{0pt}
	\tablecaption{All broad emission lines identified in Mrk 493}
	\tablehead{
		\colhead{Line} &
		\colhead{Vacuum Wavelength} &
		\colhead{Line Flux} \\
		\colhead{(name)} &
		\colhead{(\AA)} &
		\colhead{($\rm 10^{-15}\ erg\ s^{-1}\ cm^{-2}$)}
	}
	\startdata
	He I 18          &  $4027.50$  &  $   2.95$  \\
	Fe II            &  $4094.82$  &  $   1.37$  \\
	H$\delta$        &  $4102.89$  &  $  12.99$  \\
	Fe II 28         &  $4259.36$  &  $   1.70$  \\
	Fe II 32         &  $4279.30$  &  $   3.38$  \\
	Fe II 28         &  $4297.78$  &  $   5.38$  \\
	Fe II 32         &  $4315.50$  &  $   5.74$  \\
	H$\gamma$        &  $4341.62$  &  $  22.72$  \\
	Fe II 28         &  $4370.63$  &  $   2.29$  \\
	Fe II            &  $4404.27$  &  $   6.09$  \\
	Fe II 27         &  $4418.06$  &  $   0.32$  \\
	Fe II 37         &  $4474.18$  &  $   8.60$  \\
	Fe II 38         &  $4509.54$  &  $   6.30$  \\
	Fe II 37         &  $4557.17$  &  $   0.00$  \\
	Ti II 50         &  $4565.04$  &  $   5.45$  \\
	Fe II 38         &  $4577.61$  &  $   5.45$  \\
	Ti II 50         &  $4591.27$  &  $   6.45$  \\
	Fe II 38         &  $4596.97$  &  $   1.03$  \\
	Fe II] 43        &  $4602.70$  &  $   0.98$  \\
	Fe II 38         &  $4621.80$  &  $   1.76$  \\
	Fe II] 43        &  $4658.27$  &  $   2.77$  \\
	Fe II 37         &  $4668.06$  &  $   6.19$  \\
	He II            &  $4687.02$  &  $   7.14$  \\
	Fe II] 26        &  $4714.50$  &  $   0.65$  \\
	Fe II] 54        &  $4721.47$  &  $   1.92$  \\
	H$\beta$         &  $4862.66$  &  $  56.48$  \\
	Fe II] 54        &  $4888.29$  &  $   0.74$  \\
	Fe II 42         &  $4925.29$  &  $   1.54$  \\
	Fe II 42         &  $5019.84$  &  $   2.01$  \\
	Fe II            &  $5032.04$  &  $   3.02$  \\
	Si II 5          &  $5057.76$  &  $   1.35$  \\
	Fe II] 35        &  $5134.10$  &  $   2.82$  \\
	Fe II] 35        &  $5155.84$  &  $   2.62$  \\
	Fe II 42         &  $5170.47$  &  $   3.72$  \\
	Fe II 49         &  $5199.02$  &  $   8.38$  \\
	Fe II 49         &  $5236.08$  &  $   4.15$  \\
	Fe II 41         &  $5258.35$  &  $   1.72$  \\
	Fe II 48         &  $5266.27$  &  $   4.30$  \\
	Fe II 41         &  $5285.56$  &  $   1.17$  \\
	Fe II 49         &  $5318.09$  &  $   6.57$  \\
	Fe II 48         &  $5339.19$  &  $   1.74$  \\
	Fe II 48         &  $5364.35$  &  $   2.92$  \\
	Fe II 48         &  $5415.59$  &  $   3.01$  \\
	Fe II] 55        &  $5434.49$  &  $   2.44$  \\
	Fe II            &  $5506.78$  &  $   0.85$  \\
	Fe II] 55        &  $5536.37$  &  $   1.93$  \\
	\enddata
	\label{tab:BLRlinelist}
	\tablecomments{Measurements are given in the AGN rest frame.}
\end{deluxetable}

\startlongtable
\begin{deluxetable}{lcc}
	\tablenum{A.2}
	\tablecolumns{3}
	\tablewidth{0pt}
	\tablecaption{All narrow emission lines identified in Mrk 493}
	\tablehead{
		\colhead{Line} &
		\colhead{Vacuum Wavelength} &
		\colhead{Line Flux} \\
		\colhead{(name)} &
		\colhead{(\AA)} &
		\colhead{($\rm 10^{-15}\ erg\ s^{-1}\ cm^{-2}$)}
	}
	\startdata
	Cr II 194        &  $4004.46$  &  $   0.69$  \\
	Ni II 12         &  $4016.64$  &  $   0.20$  \\
	Fe II 126        &  $4034.09$  &  $   0.67$  \\
	Ti II 87         &  $4054.96$  &  $   0.73$  \\
	Cr II 19         &  $4065.09$  &  $   0.75$  \\
	$[$S II] 1F      &  $4069.77$  &  $   1.55$  \\
	$[$S II] 1F      &  $4077.37$  &  $   0.36$  \\
	H$\delta$        &  $4102.89$  &  $   3.28$  \\
	$[$Fe II] 23F    &  $4115.63$  &  $   1.97$  \\
	Fe II 28         &  $4123.80$  &  $   1.19$  \\
	Fe II 27         &  $4129.89$  &  $   2.32$  \\
	$[$Fe II] 21F    &  $4147.82$  &  $   0.78$  \\
	Ti II 105        &  $4164.81$  &  $   0.21$  \\
	Fe II] 149       &  $4168.86$  &  $   0.69$  \\
	Fe II 27         &  $4174.63$  &  $   2.37$  \\
	Fe II] 21        &  $4178.88$  &  $   3.16$  \\
	Fe II 28         &  $4180.04$  &  $   0.20$  \\
	Fe II] 21        &  $4184.38$  &  $   1.28$  \\
	Fe II]           &  $4192.14$  &  $   0.37$  \\
	Fe II]           &  $4205.66$  &  $   0.62$  \\
	Fe II] 45        &  $4228.36$  &  $   0.78$  \\
	Fe II 27         &  $4234.36$  &  $   4.49$  \\
	Fe II]           &  $4238.76$  &  $   0.61$  \\
	$[$Fe II] 21F    &  $4245.17$  &  $   0.82$  \\
	$[$Fe II] 21F    &  $4246.01$  &  $   1.26$  \\
	Fe II 220        &  $4260.52$  &  $   0.91$  \\
	$[$Fe II] 7F     &  $4288.60$  &  $   0.41$  \\
	Fe II 28         &  $4297.78$  &  $   1.70$  \\
	Fe II 27         &  $4304.38$  &  $   2.00$  \\
	Fe II 32         &  $4315.50$  &  $   0.55$  \\
	$[$Fe II] 21F    &  $4320.83$  &  $   0.43$  \\
	Fe II] 20        &  $4328.26$  &  $   1.36$  \\
	H$\gamma$        &  $4341.62$  &  $   5.74$  \\
	$[$Fe II] 21F    &  $4348.07$  &  $   1.64$  \\
	$[$Fe II] 21F    &  $4354.00$  &  $   4.01$  \\
	$[$Fe II] 21F    &  $4359.58$  &  $   0.61$  \\
	Fe II] 202       &  $4360.35$  &  $   1.23$  \\
	$[$Fe II] 7F     &  $4360.57$  &  $   0.08$  \\
	$[$O III]        &  $4364.44$  &  $   1.18$  \\
	$[$Fe II] 21F    &  $4373.66$  &  $   2.10$  \\
	$[$Fe II] 6F     &  $4383.97$  &  $   0.89$  \\
	Fe II 27         &  $4386.61$  &  $   2.90$  \\
	Ti II 19         &  $4396.26$  &  $   0.61$  \\
	$[$Fe II] 7F     &  $4415.02$  &  $   1.07$  \\
	$[$Fe II] 6F     &  $4417.51$  &  $   2.90$  \\
	Ti II 93         &  $4423.19$  &  $   0.76$  \\
	$[$Fe II] 6F     &  $4433.69$  &  $   0.85$  \\
	Ti II 19         &  $4445.05$  &  $   1.31$  \\
	Ti II 19         &  $4451.74$  &  $   1.22$  \\
	$[$Fe II] 7F     &  $4476.16$  &  $   0.28$  \\
	$[$Fe II] 6F     &  $4490.01$  &  $   2.89$  \\
	$[$Fe II] 6F     &  $4493.89$  &  $   1.79$  \\
	Fe II 38         &  $4509.54$  &  $   2.20$  \\
	Fe II 37         &  $4516.61$  &  $   3.56$  \\
	Fe II 37         &  $4521.49$  &  $   2.37$  \\
	Fe II 38         &  $4523.90$  &  $   3.67$  \\
	$[$Fe II] 6F     &  $4529.65$  &  $   1.08$  \\
	Fe II 37         &  $4535.44$  &  $   2.89$  \\
	Fe II 38         &  $4542.79$  &  $   2.65$  \\
	Fe II 38         &  $4550.75$  &  $   4.20$  \\
	Fe II 37         &  $4557.17$  &  $   3.44$  \\
	Fe II 38         &  $4585.11$  &  $   4.08$  \\
	Cr II 44         &  $4617.93$  &  $   1.71$  \\
	Fe II 38         &  $4621.80$  &  $   1.82$  \\
	Fe II 37         &  $4630.64$  &  $   5.82$  \\
	Fe II 25         &  $4635.90$  &  $   2.43$  \\
	$[$Fe II] 4F     &  $4640.97$  &  $   0.11$  \\
	N II 5           &  $4644.39$  &  $   1.92$  \\
	$[$Fe II] 4F     &  $4665.75$  &  $   0.02$  \\
	Fe II 25         &  $4671.48$  &  $   1.89$  \\
	Cr II            &  $4698.91$  &  $   0.40$  \\
	Fe II            &  $4703.86$  &  $   0.88$  \\
	Fe II] 43        &  $4732.76$  &  $   1.54$  \\
	$[$Fe II] 20F    &  $4776.06$  &  $   0.02$  \\
	Fe II]           &  $4803.95$  &  $   0.36$  \\
	$[$Fe II] 20F    &  $4815.88$  &  $   0.98$  \\
	Cr II 30         &  $4825.48$  &  $   2.23$  \\
	H$\beta$         &  $4862.66$  &  $  14.26$  \\
	Cr II 30         &  $4877.84$  &  $   1.88$  \\
	$[$Fe II] 4F     &  $4890.99$  &  $   1.46$  \\
	$[$Fe II]        &  $4899.98$  &  $   1.19$  \\
	$[$Fe II] 20F    &  $4906.71$  &  $   0.60$  \\
	Ti II 114        &  $4912.57$  &  $   1.02$  \\
	Fe II 42         &  $4925.29$  &  $   6.79$  \\
	Fe II]           &  $4929.69$  &  $   0.33$  \\
	N I 9            &  $4936.41$  &  $   0.67$  \\
	Fe II 36         &  $4948.71$  &  $   0.87$  \\
	$[$O III]        &  $4960.28$  &  $   3.94$  \\
	$[$Fe II] 20F    &  $4974.78$  &  $   0.36$  \\
	Fe II 36         &  $4994.74$  &  $   2.70$  \\
	$[$Fe II] 20F    &  $5006.91$  &  $   0.38$  \\
	$[$O III]        &  $5008.20$  &  $  11.85$  \\
	Fe II 42         &  $5019.84$  &  $   5.38$  \\
	$[$Fe II] 20F    &  $5021.63$  &  $   0.63$  \\
	$[$Fe II] 20F    &  $5044.93$  &  $   0.87$  \\
	$[$Fe II]        &  $5061.49$  &  $   0.25$  \\
	Ti II 113        &  $5073.69$  &  $   0.90$  \\
	Fe II            &  $5103.22$  &  $   0.54$  \\
	$[$Fe II] 18F    &  $5109.36$  &  $   0.27$  \\
	$[$Fe II] 18F    &  $5159.44$  &  $   0.87$  \\
	Fe II 42         &  $5170.47$  &  $   5.03$  \\
	$[$Fe II] 18F    &  $5183.39$  &  $   0.50$  \\
	$[$Fe II] 19F    &  $5186.24$  &  $   0.22$  \\
	Fe II 49         &  $5199.02$  &  $   0.42$  \\
	$[$N I]          &  $5201.71$  &  $   0.89$  \\
	$[$Fe II] 19F    &  $5221.51$  &  $   0.24$  \\
	Fe II 49         &  $5236.08$  &  $   3.07$  \\
	$[$Fe II] 19F    &  $5263.08$  &  $   0.67$  \\
	Fe II 48         &  $5266.27$  &  $   0.35$  \\
	$[$Fe II] 18F    &  $5274.82$  &  $   2.05$  \\
	Fe II 49         &  $5277.46$  &  $   3.07$  \\
	Fe II 41         &  $5285.55$  &  $   1.48$  \\
	$[$Fe II] 19F    &  $5298.30$  &  $   0.43$  \\
	Fe II 49         &  $5318.09$  &  $   4.11$  \\
	$[$Fe II] 19F    &  $5335.13$  &  $   1.13$  \\
	Fe II 48         &  $5364.35$  &  $   2.13$  \\
	$[$Fe II] 19F    &  $5377.95$  &  $   0.03$  \\
	Fe II            &  $5380.53$  &  $   0.83$  \\
	Fe II 49         &  $5426.78$  &  $   0.91$  \\
	$[$Fe II] 34F    &  $5478.76$  &  $   0.07$  \\
	Fe II] 55        &  $5536.37$  &  $   0.96$  \\
	Fe II]           &  $5542.02$  &  $   0.46$  \\
	Fe II]           &  $5575.53$  &  $   0.13$  \\
	\enddata
	\label{tab:NLRlinelist}
	\tablecomments{Measurements are given in the AGN rest frame.}
\end{deluxetable}

\section{Fits to the AGN Test Sample} \label{app:allSDSSfits}

\begin{figure}[ht!]
	\figurenum{B.1}
	\centering
	\includegraphics[width=0.40\textwidth]{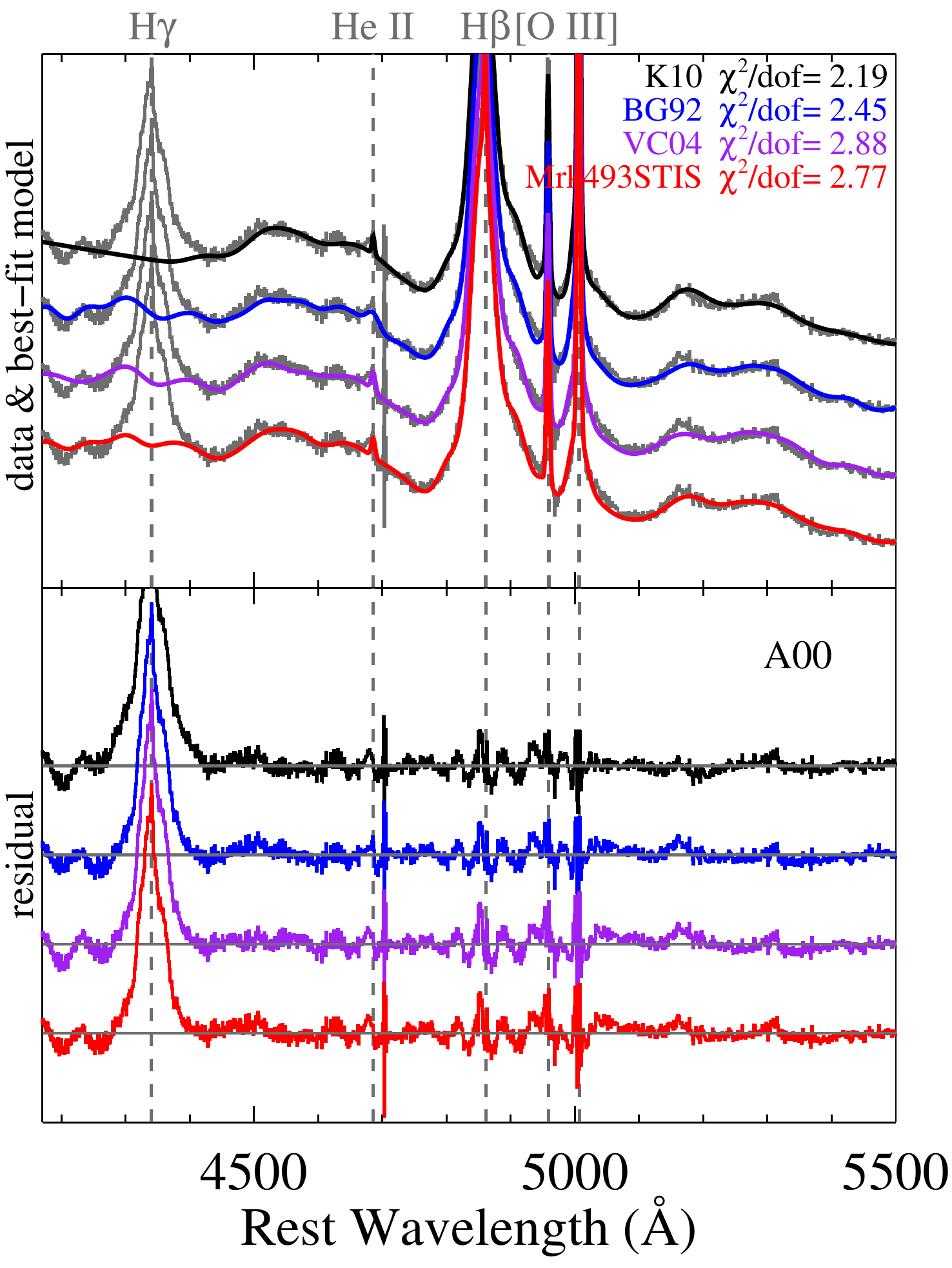}
	\caption{
		Spectral fitting results for the test sample of 41 AGN using the four different iron templates.	
		The observed data (dark gray) and best-fit models (black for K10, blue for BG92, purple for VC04, red for Mrk493STIS) 
		with the resulting reduced $\chi^2$ values are shown in the upper panel in each graph.
		The residuals, representing the difference between the data and the best-fit model, are shown in the lower panel in each graph.
		The four spectra in each panel are shifted vertically by an arbitrary amount to facilitate comparison between the fits with the different iron templates.
		The full content of this figure (a total of 41 AGN) is given in the electronic version of the Journal.
		A portion is shown here for guidance regarding its form and content.
	}
	\label{AppFig:allSDSS_fits1}
\end{figure}

\end{CJK*}
\end{document}